\documentclass[12pt]{article}
\usepackage{lscape}
\usepackage{geometry}
\usepackage[onehalfspacing]{setspace}
\usepackage{amssymb}
\usepackage{amsfonts}
\usepackage{graphicx}
\usepackage{amsmath}
\usepackage{epstopdf}
\usepackage{natbib}
\usepackage{bibunits}
\usepackage{morefloats}
\usepackage{float}
\usepackage{comment}
\usepackage{pdflscape}
\usepackage{adjustbox}
\usepackage{subcaption}
\usepackage{longtable}
\usepackage{apalike}
\usepackage{xfrac}
\usepackage[dvipsnames,table]{xcolor}
\usepackage{neuralnetwork}
\usepackage{listings}
\usepackage{subcaption} 

\usepackage{qtree}
\usepackage{multicol}
\usepackage{booktabs}
\usepackage{threeparttable}
\usepackage{tabularx}
\usepackage{footnote}
\usepackage{wrapfig}

\usepackage{dcolumn}
\newcolumntype{d}[1]{D..{#1}} 

\usepackage{siunitx}
\usepackage{mathpazo} 
\usepackage[scaled]{helvet} 
\usepackage{courier} 
\normalfont
\usepackage[T1]{fontenc}
\usepackage{listings}
\usepackage{epigraph}
\def\sym#1{\ifmmode^{#1}\else\(^{#1}\)\fi}

\definecolor{DarkerPineGreen}{RGB}{0, 90, 80} 
\definecolor{darkpowderblue}{rgb}{0.0, 0.04, 0.4}

\usepackage[pdftex,bookmarks,colorlinks]{hyperref}
\hypersetup{
  colorlinks,
  citecolor=DarkerPineGreen,
  linkcolor=red,
  urlcolor=blue}
\usepackage{cleveref}

\usepackage{tikz}
\usetikzlibrary{positioning}

\usepackage{etoolbox}

\renewenvironment{abstract}
 {\small
  \begin{center}
  \bfseries \abstractname\vspace{-.5em}\vspace{0pt}
  \end{center}
  \list{}{
    \setlength{\leftmargin}{1.35cm}    \setlength{\rightmargin}{\leftmargin}  }  \item\relax}
 {\endlist}
\begin{document}
\sloppy
\title{\vspace*{-0.01cm} \LARGE \textbf{\color{darkpowderblue} A Neural Phillips Curve and a Deep Output Gap} \\ \phantom{.}} 
\author{Philippe Goulet Coulombe\thanks{%
Departement des Sciences Économiques, \href{mailto:p.gouletcoulombe@gmail.com}{\texttt{goulet\_coulombe.philippe@uqam.ca}}. First,  I am grateful to Hugo Couture for outstanding research assistance.  For helpful discussions and/or  comments, I would like to thank Frank Diebold,  Mikael Frenette, Alain Guay, Maximilian Goebel,  Guillaume Poulin-Bellisle,  Josefine Quast, Dalibor Stevanovic,  David Wigglesworth,  Boyuan Zhang,  as well as three anonymous referees.  I also thank participants at the UQAM weekly seminar,  at at the CIRANO Montreal Macro Workshop,  the SKEMA/CERGY Inflation Forecasting Workshop 2021,  the Bank of Canada Macro Brownbag Seminar,  the CFE 2021,  the International Symposium on Forecasting 2022,   the Bank of England Advanced Analytics Workshop 2022,  TimeWorld 2022,  the Bocconi Finance Brownbag Seminar,  the RCEA Big Data \& Machine Learning Conference,  the 35th SUERF Colloquium and 49th OeNB Annual Economic Conference 2022,  the Symposium of the Society for Nonlinear Dynamics \& Econometrics,  the Cleveland Fed Real-Time Data Conferences, and 4th conference on Non-traditional Data, Machine Learning, and Natural Language Processing in Macroeconomics. }}
\date{\vspace{-0.4cm}
Université du Québec à Montréal \\[2ex]%
\small
\small
}
\maketitle

 
\begin{abstract}

\noindent Many problems plague empirical Phillips curves (PCs). Among them is the hurdle that the two key components, inflation expectations and the output gap, are both unobserved. Traditional remedies include proxying for the absentees or extracting them via assumptions-heavy filtering procedures. I propose an alternative route: a Hemisphere Neural Network (HNN) whose architecture yields a final layer where components can be interpreted as latent states within a Neural PC.  First, HNN conducts the supervised estimation of nonlinearities that arise when translating a high-dimensional set of observed regressors into latent states. Second, forecasts are economically interpretable. Among other findings, the contribution of real activity to inflation appears understated in traditional PCs. In contrast, HNN captures the 2021 upswing in inflation and attributes it to a large positive output  gap starting from late 2020. The unique path of HNN’s gap comes from dispensing with unemployment and GDP in favor of an amalgam of nonlinearly processed alternative tightness indicators.

\end{abstract}

\thispagestyle{empty}







\clearpage


\clearpage 
\setcounter{page}{1}

\newgeometry{left=1.8 cm, right= 1.8 cm, top=1.9 cm, bottom=1.8 cm}

\section{Introduction}

Few equations are as central to modern macroeconomics and current monetary policy debates as the Phillips Curve (PC) -- and its modern incarnation, the New Keynesian Phillips Curve (NKPC).  Yet, many problems plague its estimation and thus, our understanding of how increasing economic activity translates into higher pressures on the price level suffers.  Similarly,  our understanding of how inflation expectations influence current inflation is also compromised.  



This paper focuses on a predictive Phillips curve -- building an equation that uses, among other things, some measures of real activity and expectations to \textit{forecast} inflation.  It provides a new solution to a pervasive problem in empirical inflation modeling and economics research in general.  Namely,  the two key components of the NKPC,  inflation expectations ($\mathcal{E}_t$) and the output gap ($g_t$),  are both unobserved.  Instantly,  this opens the gates to a zoo of proxies.  Which gap to choose? Which inflation expectations at what horizon,  and from whom? Those are crucial empirical choices on which theory is mostly silent.   $\mathcal{E}_t$ and $g_t$ are necessary to produce \textit{and} understand inflation forecasts,  both of which are needed to guide monetary policy action.  


{\noindent \sc \textbf{A {Hemisphere} Neural Network.}}  Taking a step back,  what basic macroeconomic theory suggests is that two sufficient statistics summarizing different groups of economic indicators should predict inflation reasonably well.  More precisely,  we know that (i) there should exist some abstract output gap, or in other words,  a possibly nonlinear combination of variables related to the state of the economy (labor markets, industrial production, national accounts) that influence inflation, and (ii) some combination of price variables (past CPI values and several others) and other measures of inflation expectations also impact inflation directly.  I make this vision operational by developing a new Deep Neural Network (DNN) architecture coined Hemisphere Neural Network (HNN).  The DNN is restricted so that its final inflation prediction is the sum of components composed from groups of predictors. These group of predictors are separated at the entrance of the network into different hemispheres, defined as modules or sub-networks operating in isolation until their outputs are combined in a final layer.  This particular structure allows the interpretation of the top layer's cells output as key macroeconomic latent states in a linear equation -- a Phillips curve.  Moreover,  the estimation of time-varying PC coefficients and the key latent states can be performed within a single model.  

While HNN's development is motivated from  inflation predictive modeling, its applicability extends to the various problems in economics where the link between theoretical variables and actual variables in our databases is not crystal clear.  Examples  include the neutral interest rate,  Taylor rules inputs, term premium, and of great interest recently,  "financial conditions" \citep{adrian2019}.  This also extends to poorly measured observed explanatory variables.   From an econometric perspective, this paper introduces a novel approach, leveraging modern deep learning to address mismeasurement issues impeding empirical macroeconometrics.  Obviously,  HNN is by no means the first methodology dealing with latent state extraction or attenuation bias.  But, when compared to the older generation of methods, its empirical merits will be decisive.   As such,  this paper sits at the intersection of three literatures: (i) estimating the output gap and Phillips curves,  (ii) the application of deep learning methods in macroeconomic forecasting,  and (iii) interpretable artificial intelligence (AI).  I now review those concisely and discuss how HNN may improve upon existing methods.








{\sc \noindent \textbf{Output Gap and Phillips Curves}.}  By the virtues of being a supervised learning problem,  HNN improves over methods where $g_t$ is extracted  from an economic activity series and then, optionally,  evaluated in a "second stage" PC regression.  That is,  ${g}_t^{\text{HNN}}$ is \textit{by construction} the most relevant summary statistic of real activity  to explain inflation.  The unsupervised approach  -- filtering either GDP,  unemployment,  or related indicators in a certain way -- is by far the most widespread practice in central banks.    There are now alternative filtering tools \citep{hamilton2018you}, and others that incorporates additional (timely) information \citep{berger2020nowcasting,de2017real}.  Taking a step back,  there is the deeper question of whether this filtered $g_t$ is what we should be after at all, especially that its explanatory power for inflation seems to be vanishing quickly \citep{blanchard2015inflation}.  Little prevents the true $g_t$ from falling outside the range of what trend-cycle decompositions can extract.  HNN surely has its own set of assumptions -- like that real activity should impact inflation through a linear reduced-form equation --  but I will argue throughout that they are more palatable.

With respect to econometric methods that included some mild form of supervision in the estimation of $g_t$  \citep{kichian1999measuring,blanchard2015inflation,chan2016,chan2018,hasenzagl2018model,jarocinski2018inflation}, HNN improves by dispensing with restrictive law of motion assumptions inherent to a state-space methodology.  This is not innocuous,  as the smoothness and composition assumptions plausibly drive results to an appreciable extent, leading us back,  at least in part,  towards unsupervised methods.  Additionally,  with respect to the aforementioned approaches,  HNN easily handles a high-dimensional group of inputs for both $g_t$ and $\mathcal{E}_t$  and performs computations quickly using standard highly optimized deep learning software.  Nonlinearities in how activity variables translate into $g_t$ or $\mathcal{E}_t$ (and ultimately $\pi_t$) are allowed through a deep and wide network architecture with over 2 million parameters.  In that sense,  HNN fully embraces the implications from the widely documented {double descent} phenomenon \citep{belkin2019reconciling} by being overtly overparametrized and yet providing good \textit{non-overfitting} results.  Nonlinearities are in fact a necessary feature given the accumulating evidence that the PC might be nonlinear with respect to traditional slack indicators \citep{linde2019resolving,MRFjae,forbes2021low,benigno2023s}.


{\sc \noindent \textbf{Macroeconomic Forecasting and Interpretable AI}.}  Regarding applications of modern neural networks to inflation forecasting (such as \cite{hauzenberger2020real} and \cite{paranhos2021predicting}), HNN is an inherently interpretable model, tailored to inflation by incorporating minimal "theoretical" restrictions. These restrictions enable the outputs of the last layer to be understood as economic states. In its simplest form,  HNN is a generalized additive model  \citep{hastie2017generalized} where more than one regressor is allowed to enter each linearly separated nonparametric function,  and all functions are learned simultaneously through a gradient-based approach.  By that,  HNN fits within what \cite{hothorn2010model} defines as structure-based additive models or group-level structured neural networks in \cite{bianchi2021}.  Closely related,  \cite{agarwal2020neural} and others develop architectures inspired from generalized additive models to enhance interpretability in deep networks for generic tasks.  While these articles certainly tackle some of the opacity issues coming from fully nonparametric estimation, none address those that are inherent to non-sparse high-dimensional (even linear) regression.  This also represents a bottleneck for post-hoc interpretability methods, such as Shapley values, which typically require some form of input pre-selection to yield economically meaningful results \citep{buckmann2022interpretable}.  The key observation behind HNN's design is that grouping variables into hemispheres and combining their outputs according to theory enables the interpretation of the high-dimensional nonlinear black box as a sparse linear unobserved components model.  Thus, more generally,  HNN is related to interpretable AI approaches aiming for user-centered explanations, addressing how AI systems outputs can be communicated effectively to (human) decision makers \citep{Gunning2019,Miller2019}.


{\sc \noindent \textbf{Results}.}  Two main variants of HNN are proposed.   They primarily differ in their handling of time-variation in the model.  The first one (\textbf{HNN}) is less restrictive on how exogenous time-variation mixes with other nonlinearities. The caveat is that only the gap's \textit{contribution} to inflation can be extracted from the model.  The second architecture (\textbf{HNN-F},  the flagship model) has a built-in \textit{factorization}  which disentangles $g_t$'s estimates from its exogenously time-varying coefficient.

Many new insights are obtained.  \textbf{First},  forecasts are better than traditional PC-based forecasts.  For the post-2008 period, this can be partly attributed to HNN-F's gap -- projected \textit{out-of-sample} -- closing much faster than traditional ones,  then slipping back gently into negative territory in the mid-2010s.  HNN also captures the 2021 upswing in inflation and attributes it to a strongly positive output gap.  While both peaks are comparable in size to the 1970s,  the components show much less persistence than they did four decades ago---in line with the stop-and-go nature of economic constraints of the pandemic era.  \textbf{Second}, throughout the whole sample and for both architectures,   the contribution of the output gap component is shown to be significantly higher than what is reported from time-varying PC regressions with traditional gap measures.  Conversely, the effect of the expectations component is found to be milder overall, with the notable exception of 2021, where it moves upward quickly while traditional estimates remain flat.  \textbf{Third}, the Phillips curve coefficient in HNN-F decreased in the early 1980s and experienced a revival starting from the 2000s. This contrasts with traditional PC regressions, which suggest a very weak relationship in recent years.  As a result,  HNN-F -- through its positive gap and alive-and-well PC coefficient -- forecasts the inflation awakening of 2021.  \textbf{Fourth},  important input variables driving $g_t$ are found to be hours worked and vacancies.  While the former points towards the importance of the intensive margins, giving indication why an unemployment-based gap might underperform, the prominent role of vacancies  highlights that a better leading indicator appears to be the demand side rather than the supply side of the labor market.  A surrogate model analysis shows that the shape of the linkage between those two indicators and $g_t$ is highly nonlinear -- reminiscent for vacancies of the hockey stick shape in \cite{benigno2023s}.


{\sc \noindent \textbf{Outline,  Supplementary Material,  and Extensions}.} Section \ref{sec:HNN} introduces HNN,  motivates its structure,  and discusses practical aspects.   Section \ref{sec:emp} conducts the empirical analysis and Section \ref{sec:fcast} presents forecasting results.  Section \ref{sec:con} concludes.  Supplementary material  includes a simulation study as well as many robustness checks and variations on the main specification.  It also contain vignettes highlighting the wide applicability of the method.  The first extension leverages \cite{sims2019four}'s 4-equation NK model to extract a more focused indicator of latent "credit conditions".  The second  creates a supervised composite from a panel of international GDP growth data to evaluate the effect of global economic conditions on US inflation.  Those extensions show the wide applicability of the method when theory implies testing with empirically ambiguous indicators.  Lastly,  a more significant extension is considered in follow-up work:  \cite{GCFK} obtain and evaluate density forecasts  by introducing an additional "volatility" hemisphere and altering the loss function used to train the model.

\section{The Architecture}\label{sec:HNN}
    
This section discusses the motivation behind the newly proposed network architecture.   Basic New Keynesian theory tells us that current inflation  ($\pi_{t}$) is the sum of the contributions of inflation expectations for the next  period ($E_t(\pi_{t+1})$) and the output gap  ($g_t$).   \cite{stock2008phillips} define Phillips curve forecasts as those which are structured so to feature (at least) both of these components.   Defining expectations less stringently as $\mathcal{E}_t$,   the simple $s$-steps ahead predictive problem would be formulated as
    \begin{align}\label{nkpc1_2}
\pi_{t+s} = \theta \mathcal{E}_t + \gamma g_t + \nu_{t+s}.
\end{align}   
where $\nu_{t+s}$ is unpredictable noise.  Before even thinking about the definition of $\mathcal{E}_t $ and $g_t $,  \eqref{nkpc1_2} needs to be refined to include two features that are known to matter when bringing this equation to the data.  First,  coefficients may evolve through time ($\theta_t$ and $\gamma_t$),  which has become an empirical necessity to accurately describe inflation in most advanced economies \citep{blanchard2015inflation}.  Second,  commodity price cycles (particularly energy) play an important role in driving short-run inflation fluctuations, having a direct impact on it (i.e.,  not channeled through the real activity component).  In fact,  the typical worry is that omitting it, using  \cite{hasenzagl2018model}'s words,  may "overpower" the Phillips curve.  While this is certainly more of a worry when setting the modeling problem as a contemporaneous one, we would not want the omission to lead to, e.g., an overstatement of the importance of real economic activity.  The equation featuring all those empirically-desirable  features is
\begin{align}\label{nkpc3}
\pi_{t+s} = \underbrace{\theta_t \mathcal{E}_t}_{h_{t,\mathcal{E}}} \,\,\, + \,\,\, \underbrace{\gamma_t g_t}_{h_{t,g}} \,\,\, + \,\,\, \underbrace{\zeta_t c_t}_{h_{t,c}} \,\,\, + \,\,\, \nu_{t+s}.
\end{align}

\begin{figure}[ht!]
\begin{center} 
\hspace*{-0.15cm}\includegraphics[width=0.9\textwidth]{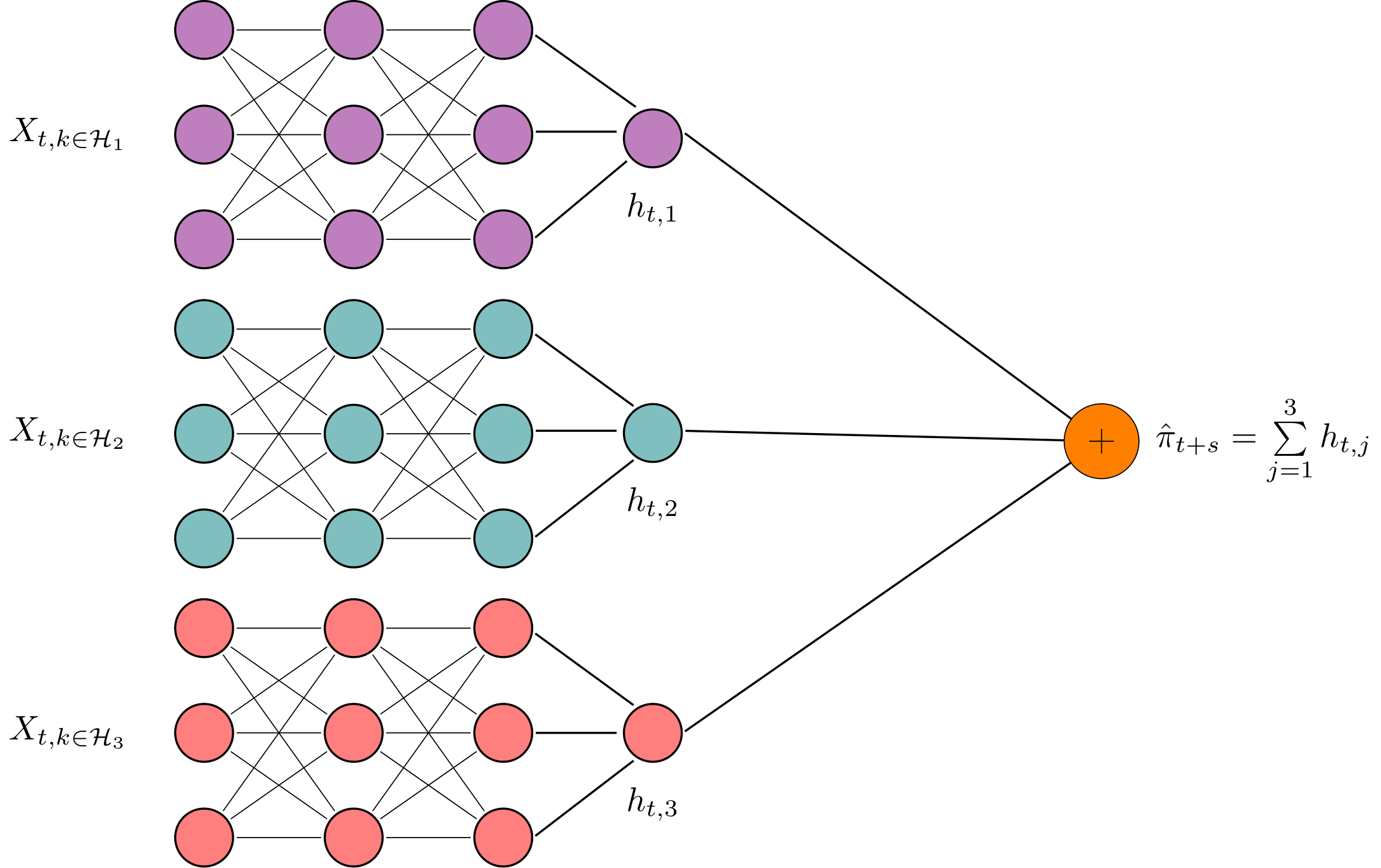}
\caption{Hemisphere Neural Network's Architecture,  with 3 hemispheres, 3 regressors per hemisphere, and (for each hemisphere) 3 hidden layers of 3 neurons each.} \label{hnnarchi}
\end{center}
\end{figure}

\noindent Essentially, this is a 3-factor model where $h_{t,\mathcal{E}}$,  $h_{t,g}$,  and $h_{t,c}$ are contributions.  Thus, let $\mathcal{H}_g$, $\mathcal{H}_\mathcal{E}$, and $\mathcal{H}_c$ be the expectations,  real activity,  and commodities prices hemispheres, respectively.\footnote{The terms "gap" and "output gap" are used throughout the paper in a loose fashion,  meaning they refer to a generic latent indicator of economic non-slack.  That is,  it refers to an abstract gap between aggregate demand and aggregate supply conditions, not a deviation from the trend of a particular observed measure of economic output.   That being said,  there is no claim that $g_t$ is the "true" wedge between aggregate demand and aggregate supply, but it is a \textit{useful indicator} that is optimized to match the implications we would hope such a true $g_t$ to have.} To make this operational,  I impose some restrictions on a  fully connected NN so that its $h_t$'s will carry economic meaning.  A shallow and narrow (for visual convenience) HNN architecture for the three hemispheres case is displayed in Figure \ref{hnnarchi}.

Some remarks are in order.  First, HNN's architecture is trivially extendable to more than 3 hemispheres.\footnote{ I use the term hemisphere throughout, but modules or sub-network could be equally adequate.   The appellation "hemisphere" is very loosely inspired from neurons in the AI model being allocated to different parts of a brain assembling outputs to produce a final decision,  although here it is detached from its etymological  sense of being two parts of a sphere.} This makes it convenient for splitting some hemispheres into sub-hemispheres (like expectations into short-run vs. long-run).    It also makes it a flexible testing ground for theories claiming the NKPC should be augmented with something, but that something is not clearly defined in terms of what is in our databases.  Such extensions are considered in Appendix \phantomsection{\ref{sec:ext}}.


Second,  HNN does not give us $g_t$ nor $\gamma_t$, but their product $h_{t,g}$.  This is \textit{not} the neural network's doing,  but rather the design of the problem.  With $g_t$ and $\gamma_t$ both unobserved and possibly time-varying,  they cannot be separately identified without additional assumptions on how  $g_t$ and $\gamma_t$ should or should not evolve through time.   HNN-F,  developed and motivated in Section \ref{sec:hsplit},  will propose such restrictions and leverage them to separate $g_t$ and $\gamma_t$.  The more general point being made is that the plain HNN architecture provides $h_{t,g}$ as the most sensible output given the econometric conditions, but nothing prevents a researcher from splitting it in $g_t$ and $\gamma_t$ using whichever assumptions deemed reasonable.  Nonetheless,  for policy purposes,  a crucial use of $g_t$ is to inform us on how real activity \textit{contributes} to $\pi_t$ -- and that is the HNN's direct output.  Finally, this does not prevent from comparing HNN results with other methodologies since their gap's contribution to inflation can easily be calculated from the PC regression (see Section \ref{sec:emp}).


Third,  a comment on the group-wise network structure and the "separability assumption".   Precisely, by separability, it is meant that $h_{t,j}$'s are the product of mostly non-overlapping (they share $t$ in common) groups of predictors.  Of course,  it is possible that the interaction of the prices group and the real activity group influences inflation.  On the other hand, some level of separability is crucial for interpretability in this high-dimensional environment. It is the separation, as suggested by the (linearized) PC, that gives $h_{t,j}$ its meaning.  While there is nothing sacred about linearized PCs,  it is noteworthy that the proposed separation is not new to HNN at all.  It is inherent to almost any linear PC estimation (there is a block of lags, and an output gap, all separated and typically non-interacting).  As a side note,  some overlap between the contents of $\mathcal{H}$'s is absolutely possible if the definition of $h_{t,j}$'s calls for it.   Finally,  $h_{t,j}$ need not be orthogonal since they are obtained from what is ultimately a (supervised) semi-nonparametric regression which dispenses with most of the traditional identification problems inherent unsupervised learning and factors models.

      \subsection{Data and Defining $\mathcal{H}$'s for Benchmark Model}\label{data}


The baseline estimation is at the quarterly frequency using the dataset FRED-QD \citep{mccracken2020fred}. It is \href{https://research.stlouisfed.org/wp/more/2020-005}{publicly available} at the Federal Reserve of St-Louis' website and contains 248 US macroeconomic and financial aggregates observed from 1960Q1. The target considered in the main analysis is CPI Inflation (thus $\pi_{t+1} = \Delta log(\text{CPI}_{t+1})$).  Forecasting and some robustness checks on $g_t$ are conducted using core inflation ($s=1$) and year-over-year (YoY) headline CPI four quarters ahead ($s=4$).  The transformations to induce stationarity for predictors are indicated in \cite{mccracken2020fred}.  

\begin{table}[h!]
\small
\centering
\rowcolors{2}{white}{gray!15}
\setlength{\tabcolsep}{0.96em}
\caption{Defining $\mathcal{H}$'s}
\label{hdefine}
\vspace*{-0.25em}
\begin{tabularx}{0.99\textwidth}{|l|X|}
\toprule
\toprule

$\mathcal{H}$  & Content   \\ 
\midrule
\textbf{$\mathcal{E}_t^{\text{LR}}$}   &  $t$ (exogenous time trend) \\
\textbf{$\mathcal{E}_t^{\text{SR}}$}   &  Inflation expectations from SPF/Michigan Survey,  lags of $\pi_t$ and price series in FRED-QD, $t$  \\
\textbf{$g_t$}   & Labor Market Variables, Industrial Production Variables, National Accounts, $t$   \\
\textbf{$c_t$}   & Oil and Gas price series from FRED-QD,  Metals PPI, $t$  \\
\bottomrule
\bottomrule
\end{tabularx}
\end{table}
\vspace*{-0.75em}

Our empirical baseline model comprises 4 hemispheres.  It consists of the 3 described in Section \ref{sec:HNN}, with one of them being split in two sub-hemispheres.  Precisely,  to examine them separately,  I split expectations into two \textit{additive} components: long-run/exogenous ($\mathcal{E}_t^{\text{LR}}$),  and short-run ($\mathcal{E}_t^{\text{SR}}$).  The remaining two $\mathcal{H}$'s are real activity and commodity/energy prices.  For each variable $X_{t,k}$,   I include 4 lags of it and 3 moving averages of order 2,  4,  and 8.  This is motivated by \cite{MDTM}'s so-called Moving Average Rotation of X (MARX) transformation, developed to alter the implicit prior of certain machine learning (ML) algorithms when applied to time series data -- without recoding them.  $\mathcal{H}$'s composition details are in Table \ref{hdefine} and the complete list of FRED-QD mnemonics is available in Appendix \ref{sec:mne}.


In addition to FRED-QD's price series,  I include inflation expectations/forecasts from the Survey of Professional Forecasters and the Michigan Survey of consumers. Proxying directly for inflation expectations with survey-based data emerged as a popular alternative to fully rational expectations \citep{coibion2015phillips,coibion2018formation}. The downside is that theory provides little to no guidance about the provenance of expectations that should be used.   \cite{meeks2023heterogeneous} use a functional principal component approach to summarize the distributional aspect of the expectations from the Michigan survey of consumers (among others) and finds that the additional information reduces the role of inflation persistence.  It is worth noting that this line of papers almost universally take the unemployment/output gap as given.  This paper, for simplicity and to maximize the length of the historical period being studied,  opts for very standard series of inflation expectations as inputs, like the average expectations from professional forecasters and consumers surveyed by the University of Michigan,  as well as lags from price indices.   From a methodological and practical standpoint,  nothing prevents the inclusion of a much richer and heterogeneous set of beliefs -- these would be additional regressors in $\mathcal{H}_{\mathcal{E}^\text{SR}}$.  By construction,  the HNN procures the optimal "summary statistic" of such expectations because the nonlinear information compression parameters are estimated in a supervised fashion.  Thus,  in future work,  HNN could digest larger expectations information sets (like the whole cross-section dimension of a survey,  or many quantiles of it).

{\sc \noindent \textbf{On the Separation of Expectations}.} An important question for the interpretation of HNN outputs as plausible estimates of the latent concepts "economic slack" and "inflation expectations" is what is an appropriate composition for the expectations hemisphere,  as many series are reasonable candidates to characterize consumers' information set.  The motivation from using lags of the target is obviously to proxy for backward-looking expectations in a way that is traditionally done in tightly specified linear Phillips curves.  Then,   the argument for including many price sub-indices is that consumers typically pay different levels of attention to individual prices to form expectations,  and the overall weighting likely differs from headline consumption basket weights (being used implicitly through lags of $\pi_t$).   However,  this very inclusive set of predictors may be a threat to HNN separability assumptions, necessary for the economic interpretation of results.  Appendix \ref{sec:rob} reports, among other things,  results where the $\mathcal{H}_{\mathcal{E}^\text{SR}}$ is cleared of anything but survey-based expectations series.  We will see that results on real activity remain consistent across alternative expectations specifications whereas results on expectations inevitably differ in certain historical episodes.  The data-rich $\mathcal{H}_{\mathcal{E}^\text{SR}}$ in the baseline HNN specification is found to be more timely, with the resulting series rising from its bed faster post-2020 than specifications only using survey-based information. 


{\sc \noindent \textbf{Timing Assumptions.}} The baseline HNN specification is set up as a one-step-ahead direct forecasting problem, in the spirit of \cite{stock2008phillips}'s Phillips curve inflation forecasts and the widespread use of such equations in central banking to build an inflation outlook \citep{yellen2017inflation}. Beyond the prism of forecasting, HNN, being a direct forecasting problem, can also be interpreted as the first step of a local projection studying the dynamic impact of real activity and expectations on inflation. This is similar to studies that examine the multi-horizon impacts of real activity shocks, considering that it takes time for real activity conditions to fully materialize in inflation numbers \citep{del2020s}.  The practical merits of the "forecasting" approach are evident, such as directly creating forward-looking metrics useful for predicting inflation, rather than performing a two-step process as in, e.g., \cite{banbura2021inflation}.  However, the timing assumptions embedded in the baseline HNN are not entirely aligned with the macroeconomic theory that motivated it. The NKPC yields a contemporaneous regression ($\pi_t$ regressed on $E_t(\pi_{t+1})$ and $g_t$), as opposed to using $\pi_{t+1}$ as the dependent variable.  Results are reported in Appendix \ref{sec:rob}, where an HNN specification is built using a more theory-consistent timing of the target variable.

     
 \subsection{Extracting the Output Gap and its Coefficient with HNN-F}\label{sec:hsplit}        

As a consequence of sparing HNN from the numerous assumptions typically associated with output gap extractions, the procedure only produces  $h_{t,g}$, the gap's contribution to inflation,  rather than $g_t$ itself.  It was discussed that splitting $h_{t,g}$ into $g_t$ and $\gamma_t$ can be done if the researcher is willing to use additional assumptions on  $g_t$ and $\gamma_t$.  Denote $\mathcal{H}_j \setminus k$ as the restricted  set where variable $k$ is excluded from the set of predictors originally included in the hemisphere $j$.  One possible factorization is $\gamma_t=h_{\gamma}(t)$ and $g_t=h_{g}(\mathcal{H}_g \setminus t)$.   The factorization coerces the PC coefficient to move exogenously and slowly -- like what is assumed by random walk coefficients in {\color{PineGreen} Chan, Koop, and Potter (2016)} (henceforth CKP) and many others.  This is merely an interpretation device because what we can say about $g_t$ and $\gamma_t$ depends perfectly on what we assume they can be.  For instance, a convex PC is ruled out by $\gamma_t=h_{\gamma}(t)$ but residual "convexity" will be mechanically relegated to $g_t$.   Nonetheless,  what HNN-F provides is a $g_t$ which definition (i.e.,  its composition out of real activity data)  is constant over time.  The coefficient is a slow-moving scaling coefficient ($\gamma_t$) -- which can be assumed fixed for short- and medium-run forecasting horizons.  In a way,  HNN-F tries to build a pair of $g_t$ and $\gamma_t$ which are most convenient from a macroeconomic monitoring perspective: a fixed definition of the gap which passthrough to inflation is very stable.  Most of the "complexity" is channeled through the definition of the gap itself. 


Implementing  $h_{t,g}=h_{\gamma}(t)h_{g}(\mathcal{H}_g \setminus t)$ is easy within HNN and the \texttt{PyTorch} (\texttt{Python}) or \texttt{Torch} (\texttt{R}) environments.  First, an additional hemisphere containing only $t$ is created.  Then,  in the final layer, rather than summing 3 or 4 $h_{t,j}$'s as in Figure \ref{hnnarchi},  some last layer outputs will be multiplied together. Namely, the output of the hemisphere containing only $t$ will be multiplied with that of $\mathcal{H}_g \setminus t$ and the product will be added to the rest of the sum constituting the neural PC. For consistency,  this intuitive factorization is forced on each component. Thus, using the notation established in \eqref{nkpc3}, the final layer in \textbf{HNN-F} (F for factorized) will be
\begin{align}\label{splitgap}
\hat{\pi}_{t+s} = h_{\mathcal{E}^{\text{LR}}}(t) + h_{\theta}(t)  h_{\mathcal{E}_{\text{SR}}}(\mathcal{H}_{\mathcal{E}^{\text{SR}}} \setminus t) + h_{\gamma}(t) h_g (\mathcal{H}_g \setminus t) +h_{\zeta}(t)  h_{c}(\mathcal{H}_{c} \setminus t) \, .
\end{align}
Clearly,  the various $h_t$'s of \eqref{splitgap} are not scale- and sign-identified, except for $ h_{\mathcal{E}^{\text{LR}}}(t) $ since it is not multiplied with any other component.  To identify the relevant $h_t$'s,   time-varying coefficient hemisphere outputs $\theta_t$, $\gamma_t$ and $\zeta_t$ are all forced to be non-negative by feeding them forward through an absolute value layer before they enter the final layer above.  This prevents the gap from being the symmetrical opposite of what it is expected to be.  Moreover,  the scale identification problem is fixed in estimation by setting the mean (over all $t$'s) of time-varying coefficients to 1.  One can rescale gaps and coefficients after estimation.   In Section \ref{sec:emp}, $g_t$'s standard deviation is set to that of Congressional Budget Office's (CBO) gap to facilitate comparison.  By estimating $g_t$ flexibly and allowing for $\gamma_t$ to vary over time,  HNN-F allows for an investigation of the declining link between real activity and inflation with a lessened worry that a declining $\gamma_t$ is due to a mismeasured $g_t$.  This hypothesis has often been ruled out when using traditional gaps or economic indicators in the context of \textit{linear} econometric methods \citep{stock2019slack,del2020s}. 



 \subsection{Estimation and Tuning}\label{sec:archi}

Within \textit{each} $\mathcal{H}$,  we have a standard feed-forward fully connected network.   We set $\texttt{layers} = 5$ and $\texttt{neurons} = 400$.  For HNN, I maximize efficiency by enabling weight sharing \citep{nowlan1992simplifying,bender2020can} across hemispheres.  In other words,  nonlinear processing parameters are forced to be identical across hemispheres.  In HNN-F, I relax that constraint and the states hemispheres are given $\texttt{neurons} = 400$ and $\texttt{layers} = 3$ while the coefficients hemispheres (with only input being $t$) have $\texttt{neurons} = 100$ and $\texttt{layers} = 3$.  More layers or neurons beyond that point visibly increase what is apparent noise in the components, and not improve out-of-bag MSE.    

The maximal number of epochs (optimizer steps) is fixed at 500. The activation functions are all \textit{ReLU} ($\operatorname{ReLU}(x)=\max \{0, x\}$) and the \texttt{learning rate} is 0.005 for HNN and 0.05 for HNN-F.   The adequacy of such values can be assessed by examining optimization curves (along epochs).  65\% of the training sample is used to estimate the parameters and the MSE of the remaining 35\% is used to determine when to optimally stop optimization -- early stopping being known to perform a form of ridge regularization on network weights  \citep{raskutti2014early}.  This random shuffling of data is done through shuffling blocks of 8 quarters for quarterly data.  Users only interested in point estimates (and not, e.g.,  getting bands around $g_t$) can increase the subsampling rate to higher values such as 85\%.  The batch size is the whole sample and the optimizer is Adam.  For forecasting,  I do 50 random 65-35 allocations of data and ensemble resulting predictions. This is beneficial in two aspects.  First, it stabilizes the optimal early stopping point choice.  Second, it is known that ensembling overfitting ("interpolating") networks can give a performance similar to that of very large yet computationally costly networks, by among other things, integrating out noise coming from network weights initialization \citep{d2020double}.  Finally,  I perform a mild form of dropout by setting the \texttt{dropout rate} to 0.2.  The typical z-score normalization of inputs in neural networks needs to be adjusted to the HNN case and such details can be found in Appendix \ref{sec:VIdet}.  The R and Python codes are available on \href{https://github.com/philgoucou/hemisphere}{Github}.  They feature the functions used throughout this paper and a replicating example code of the main empirical results.

While the large total number of neurons may seem like an ill-advised choice given the lasting threat of overfitting,  there is a now a wide literature documenting the benefits of overparametrization in such models \citep{belkin2019reconciling}.  Moreover,  their documented stability in optimization (and with respect to initialization of weights) will prove useful when constructing bands from the ensemble.  The simulation study in Appendix \ref{sec:simul} documents that a large HNN architecture is indeed more apt at recovering precisely "true" latent states that are nonlinear functions of inputs. 




\subsection{Quantifying Uncertainty}\label{sec:inference}


Ensembling requirements are more demanding to quantify $h_{t,j}$'s uncertainty than to obtain point estimates and forecasts.  $B$, the total number of runs,  is set to 300 and  blocks of 2 years are subsampled (without replacement) with a subsampling rate of 0.65.   This takes about an hour to run on an M1 MacBook Air.  

Since any DNN can easily fit the training data much better than it actually does on the test data,  it is wiser to opt for an out-of-bag strategy in order to calculate $h_{t,j}$'s \textit{in-sample} as well as their quantiles.  Such a strategy was deployed in \cite{MRFjae} for time-varying coefficients obtained from a Random Forest (RF).  Given that HNN also uses dropout to a mild extent and is optimally early-stopped to maximize hold-out sample performance,  this additional precaution may not appear necessary at first sight. However,  it is the object of a burgeoning literature of its own that best-performing DNNs out-of-sample can very well overfit in-sample \citep{belkin2019reconciling}.  This obviously complicates things for in-sample analysis of historical estimates,  and considering out-of-bag estimates is a natural solution to that problem inspired from metrics traditionally reported for RF. 

The out-of-bag calculations proceed as follows.  Assume we have a sample of size 100.  I estimate HNN using data points from 1 to 65, and project it "out-of-bag" on the 35 observations not used in training. This gives us ${h}_{65:100,b}$ for a single allocation $b$ while ${h}_{1:65,b}$ are still \texttt{NA}s. By considering many such random (block) allocations where "bag" and "out-of-bag" roles are interchanged, I obtain the final $h_{t,j}$'s by averaging over $B$ at each $t$ such that 
\begin{align}
{h}_{t,j} = \frac{1}{(1-0.65) \times B} \sum_{b=1}^B I({h}_{t,j,b}\neq  \texttt{NA}){h}_{t,j,b}.
\end{align}
This constitutes an approximation to a Block Bayesian Bootstrap by replacing the posterior tree functional $\mathcal{T}$ in \cite{MRFjae} by HNN. Thus, ${h}_{t,j,b}$ draws can be used to compute credible regions. This relies on the connection between Bagging and the Bayesian Bootstrap put forward for Random Forest  by \cite{taddy2015bayesian}.  More recently, \cite{newton2021weighted} develop a weighted Bayesian Bootstrap, derive theoretical guarantees, and show its applicability to trend filtering,  deep learning, and other problems.







How should we think of the statistical adequacy of HNN and HNN-F's key outputs? There are a number of proofs of DNN's nonparametric consistency for generic architectures -- for instance \cite{farrell2021deep}.  HNN and HNN-F are restricted DNNs,  or, alternatively,  semiparametric models.   If restrictions are approximately true (like the separability in HNN, and the factorization in HNN-F),  then we can be confident our $h_{t,j}$'s are close to true latent states.  Those restrictions can be implicitly tested by fitting a fully connected DNN with the same data and comparing predictive performance out-of-sample or out-of-bag. Thus,  if HNN increases bias much less than it curbs variance,  it will supplant the plain DNN. It is interesting to note that the restrictions' benefits are twofold: they reduce variance \textit{and} provide interpretability.  



The proposed strategy to quantify uncertainty is computationally economical since it directly recycles the ensemble,  in the spirit of \cite{MRFjae} for Macro Random Forest.   However,  it is not devoid of possible deficiencies.  First,  in an ideal world with more data,   the block size should be higher than 2 years to accommodate for highly persistent predictors in HNN's information set.  Second,  one needs to think carefully about how initialization randomness ends up artificially increasing the reported bands.  The good news are that while small networks estimation results depend significantly on randomly drawn weight initialization values,  that of large networks do not \citep{d2020double}.   Thereby,  quantifying uncertainty directly through the ensemble will provide a reasonable and indicative measure of estimation uncertainty only for large networks (which is being used in this paper).  Additionally,  HNN uses full batches of data (the whole subsampled datasets for each $b$) rather than \textit{stochastic} gradient descent.  Thus,  for each $b$,  randomness comes from the estimation subsample,  with a minimal contribution from initialization,  and little to no contribution from optimization itself.   Appendix \ref{sec:simul} features a simulation study with a data-generating process (DGP) plausibly resembling that of the empirical work.   It is found that the uncertainty quantification strategy delivers suitable estimates for the kind of (large) architecture considered in Section \ref{sec:emp}.   



 \section{Analysis}\label{sec:emp}

I first analyze contributions series obtained from HNN and HNN-F.  Then,  I focus on HNN-F which provides a breakdown of contributions in terms of gaps and coefficients.  Lastly,  I investigate which original inputs contribute most to HNN-F's $g_t$,  and how they do so.

 \subsection{Looking at Contributions}\label{sec:emp_comp}

\begin{figure}[t!]
\begin{center} 
\hspace*{-0.05cm}\includegraphics[trim={0cm 1.75cm 0cm 0cm},clip,width=0.9\textwidth]{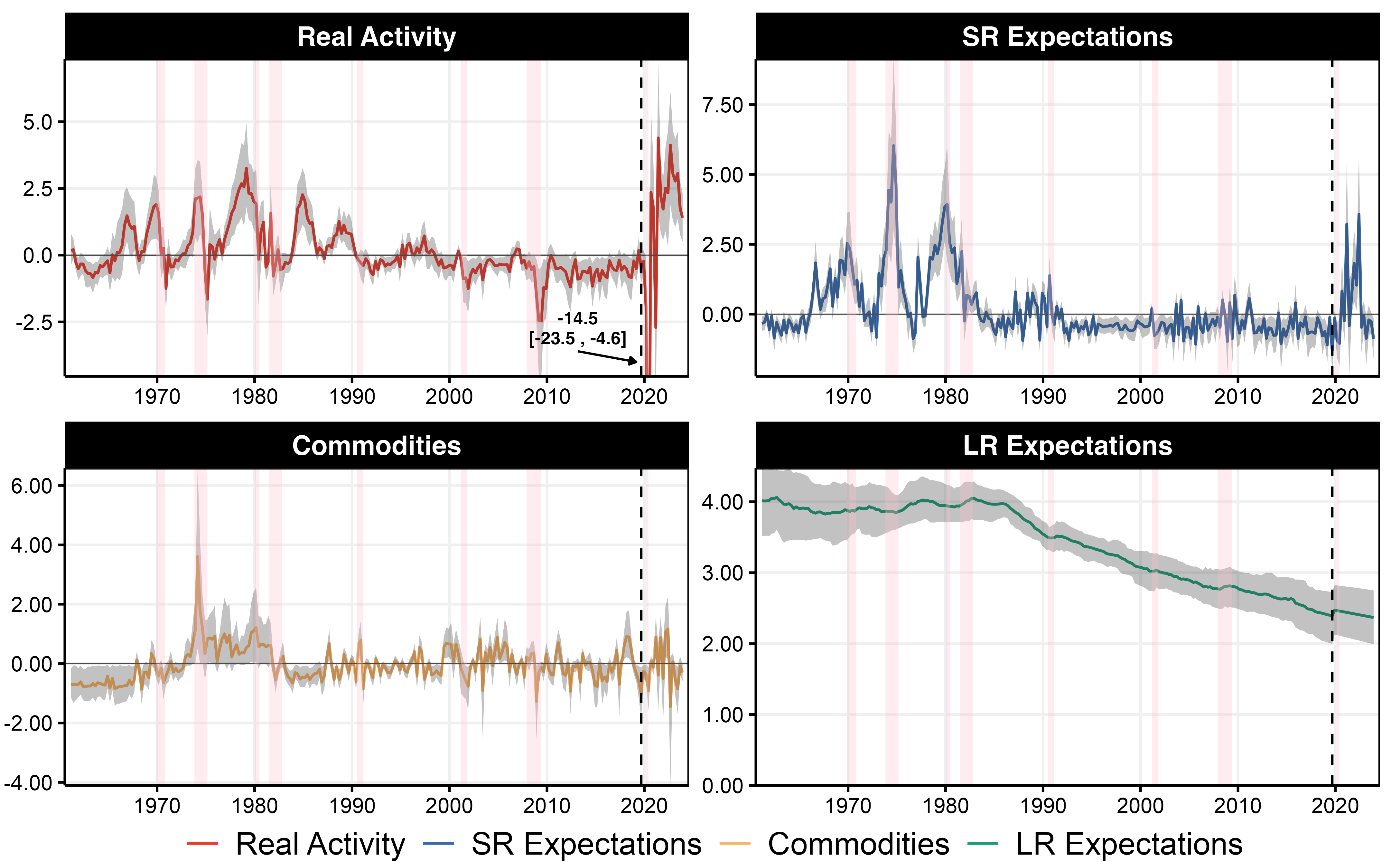}
\caption{\footnotesize Contributions ($h_{t,j}$'s) from HNN-F.  Notes: Dashed line is the beginning of the out-of-sample.  NBER recessions are in pink shadowing.  Gray shading represents the 14-86\% quantiles from the out-of-bag ensemble. }\label{hs_2019}
\end{center}
\end{figure}

\begin{figure}[t!]
  \begin{subfigure}[b]{0.5\textwidth}
\includegraphics[trim={0cm 0cm 0cm 0cm},clip,width=0.995\textwidth]{{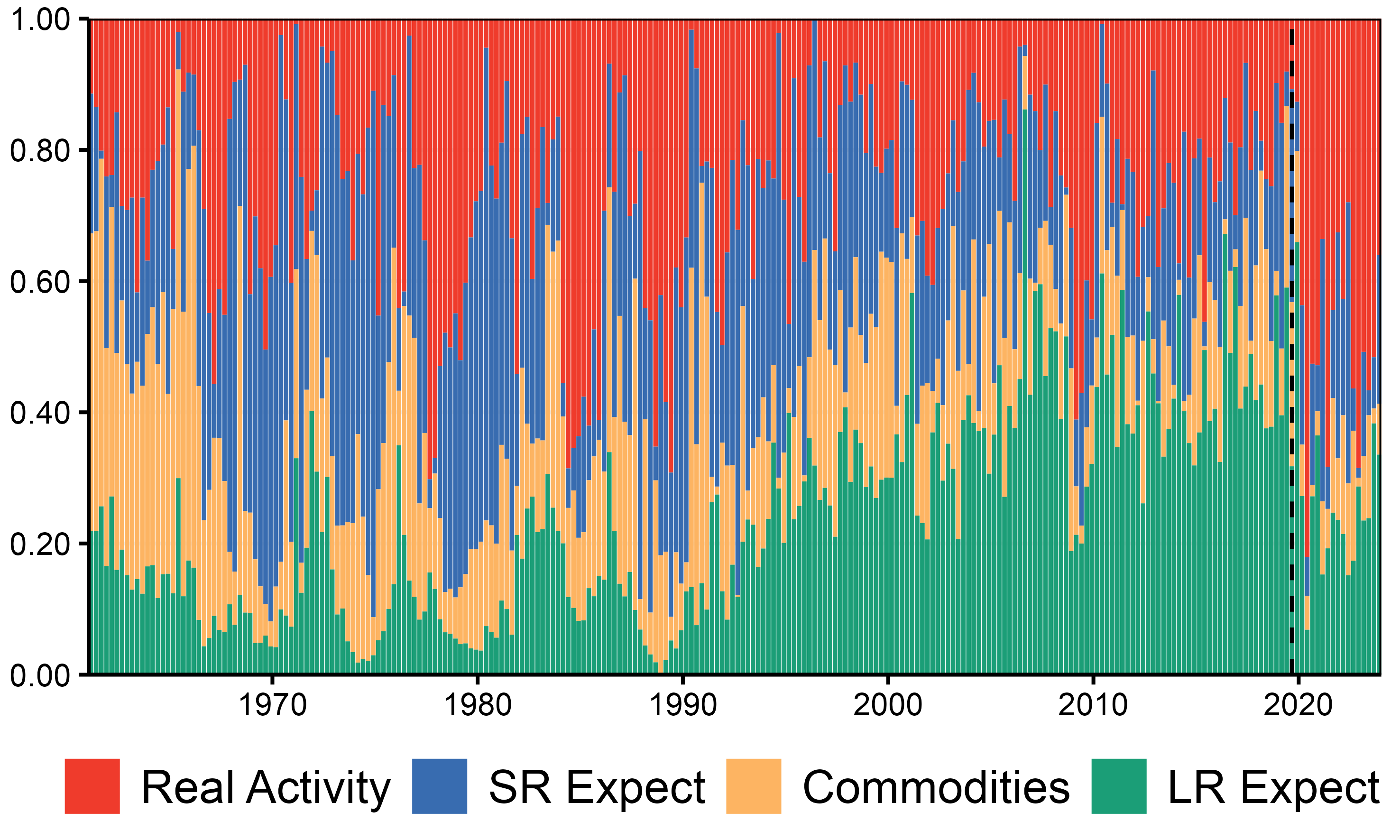}}
\caption{\footnotesize Absolute Contributions to $\hat{\pi}_t$}\label{shares}
      \end{subfigure}
  \hspace{0.2em}
  \begin{subfigure}[b]{0.5\textwidth}
\includegraphics[trim={0cm 0cm 0cm 0cm},clip,width=0.995\textwidth]{{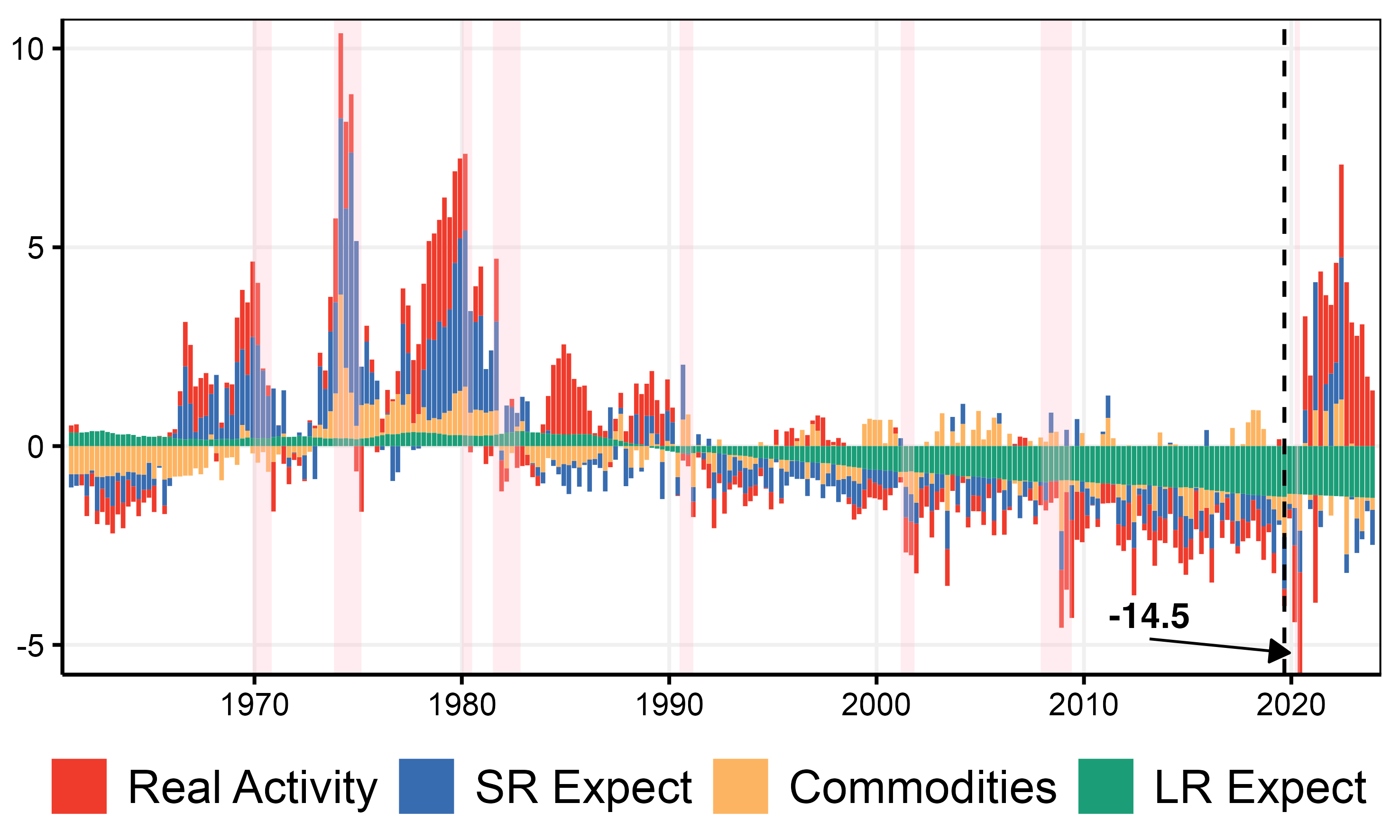}}
\caption{\footnotesize $\hat{\pi}_t$ decomposition}\label{decomp}
      \end{subfigure}
  \caption{\footnotesize Summed contributions from the four $\mathcal{H}$'s through time.  Notes:  estimation ends in 2019.   Pink shading is NBER recessions.  }\label{inf_ds}
\end{figure}

As starting point,  hemisphere outputs are displayed in Figure \ref{hs_2019} for a training sample ending in 2019Q4.  Figure \ref{hs_2007} (Appendix) reports largely unchanged estimates from using a training sample ending in 2007.   Lastly,  Figure \ref{inf_ds} reports inflation shares in two ways.

First,  we observe large positive contributions of $h_{t,g}$ in the 1970s and 1980s which have been much more muted since then,  in line with the declining PC narrative (this will be formally assessed when looking at $\gamma_t$ in Figure \ref{og_coef}).  But that was before the pandemic.  HNN-F and HNN (Figure \ref{hs_2019_hnn}) both report an extremely high positive contribution from $g_t$ to  $\pi_{t+1}$ starting from late 2020--as projected from a fixed structure estimated up until 2019. As a result, HNN-F's (and HNN as well) are forecasting annualized headline inflation consistently above 4.5\% starting from 2020Q4 (see Figures \ref{decomp} and \ref{fcast_q_2007}).   HNN's estimation of $h_{t,g}$ seems to align with the growing evidence that the PC is nonlinear \textit{in traditional economic indicators space} and that the steep part of it has simply been unsolicited in recent decades \citep{linde2019resolving,MRFjae,forbes2021low}.  This will be more formally assessed in Section \ref{sec:nlplus}.  Results also cast some doubts on methodologies forcing smoothness through laws of motion.  Those typically require  potential output to trend upward slowly (a random walk, or local-level process) whereas $h_{t,g}$ has been subject to many upward/downward movements that are much faster than these  methods typically allow for.

HNN's successes and failures in forecasting post-2019 inflation can be further understood from Figure \ref{decomp}.  We see that the "overkill" downswing that will later materialize in 2020 inflation  forecasts is entirely due to the real activity component.  This is because HNN — and Phillips curve regressions in general — are not supplied with external information about the exceptional nature of this recession (a forced partial shutdown of the economy).  This makes inference and extrapolation from previous recessions difficult and empirical results for 2020Q1-2020Q3 reflect that.  When it comes to the  increase in the first half of 2021 is due to a pattern very similar to the 1970s being replicated, that is, a gentle positive impulse from $g_t$ followed by upward pressure  from $\mathcal{E}_t^{\text{SR}}$ which eventually settles whereas the forcing from $g_t$ remains.


Contribution of the $\mathcal{E}_t^{\text{SR}}$ component was extremely strong during the 1970s and has been dormant since the beginning of Paul Volker's chairmanship up until early 2021.    In 2021,  the hibernating $h_{\mathcal{E}^{\text{SR}},t}$ wakes up, and captures the early consequences of supply chain disruptions and the expected consequences of it.  It appears that the main reason why inflation forecasts did not climb to 1970s levels in late 2021 is that the overall $h_{\mathcal{E}^{\text{SR}},t}$,  despite its earlier spike,  shows much less persistence than 4-5 decades ago.  Said differently,  expectations are still relatively well-anchored,  by not deviating persistently from the long-run ones.    Additionally,  $h_{\mathcal{E}^{\text{LR}},t}$ helps by being  almost 2  \%-points lower than in the 1970s.   Indeed,  $h_{\mathcal{E}^{\text{LR}},t}$ is found to be slowly decreasing from the 1980s onward, as expected.\footnote{Note that the overall level of $h_{\mathcal{E}^{\text{LR}},t}$ is not identified separately from $h_{\mathcal{E}^{\text{SR}},t}$  and here it was set by normalizing the other three components to have mean zero over the sample. }  Finally,  the commodity group (with oil being naturally its most influential member) contributed strongly from the first oil crisis of the 1970s, through the second oil shock, and ends after the second of the twin recessions.

 In Figure \ref{shares} shows the decline of the overall influence of $\mathcal{E}_t^{\text{SR}}$ in favor of $\mathcal{E}_t^{\text{LR}}$,  with the emergence of trend inflation dominance in the mid-1990s.   $\mathcal{E}_t^{\text{SR}}$'s contribution  peaks during  the 3 inflation spirals of the 1970s.   The share of $h_{t,g}$ appears more stable than what is typically reported by PC regressions although it seems to be milder between 2000 and 2020.  Lasting effects of energy and commodity prices appear to be gently declining.  Figure \ref{decomp} makes clear that key historical increases are always due in large part to $\mathcal{E}_t^{\text{SR}}$ up until 2020,  when $h_{t,g}$ takes the center stage.

\begin{figure}[t!]
  \begin{subfigure}[b]{0.5\textwidth}
\includegraphics[trim={0cm 0cm 0cm 0cm},clip,width=0.995\textwidth]{{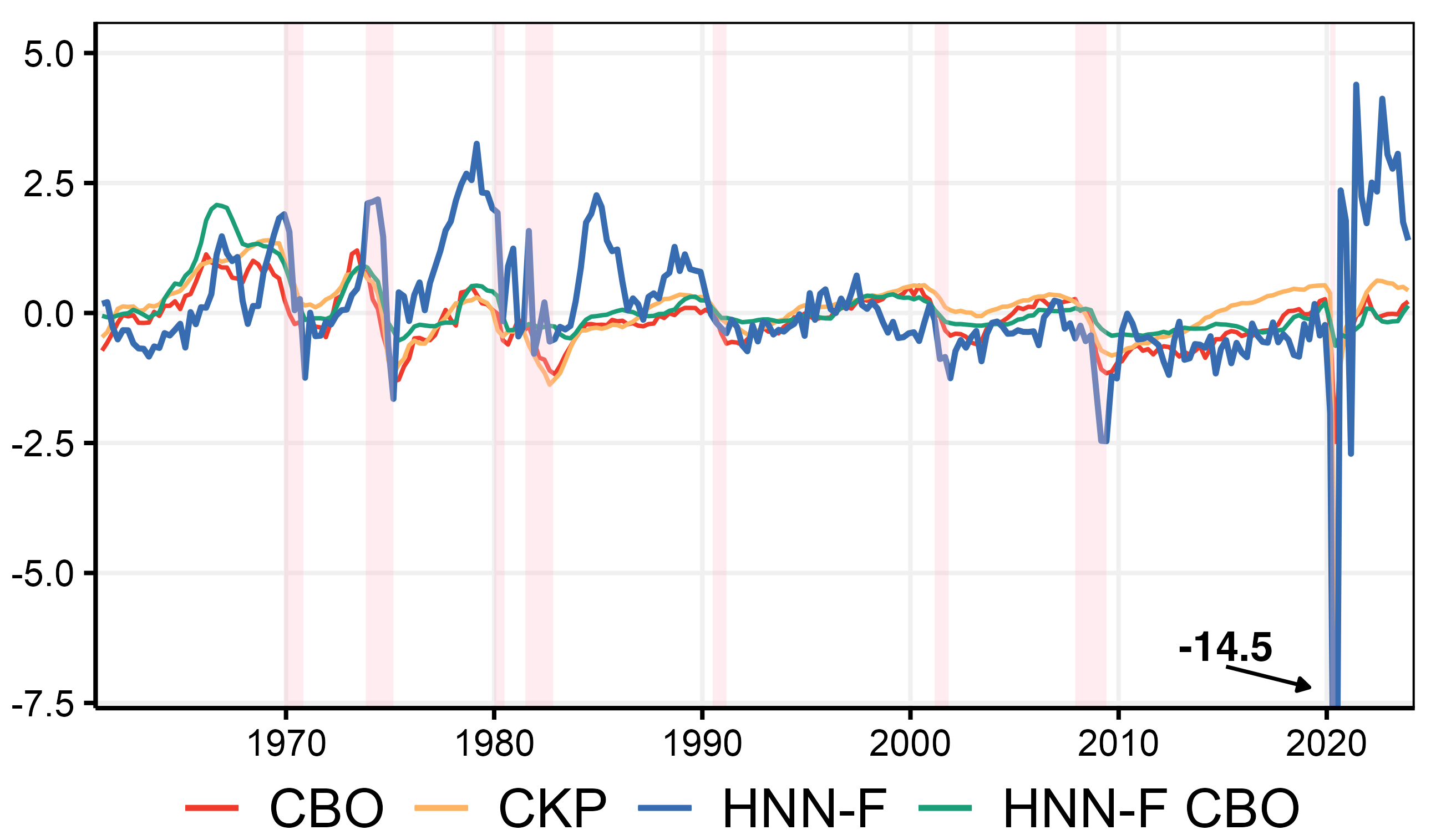}}
\caption{\footnotesize Contributions of economic activity to inflation. }\label{cbo_compa_g}
      \end{subfigure}
  \begin{subfigure}[b]{0.5\textwidth}
\includegraphics[trim={0cm 0cm 0cm 0cm},clip,width=0.995\textwidth]{{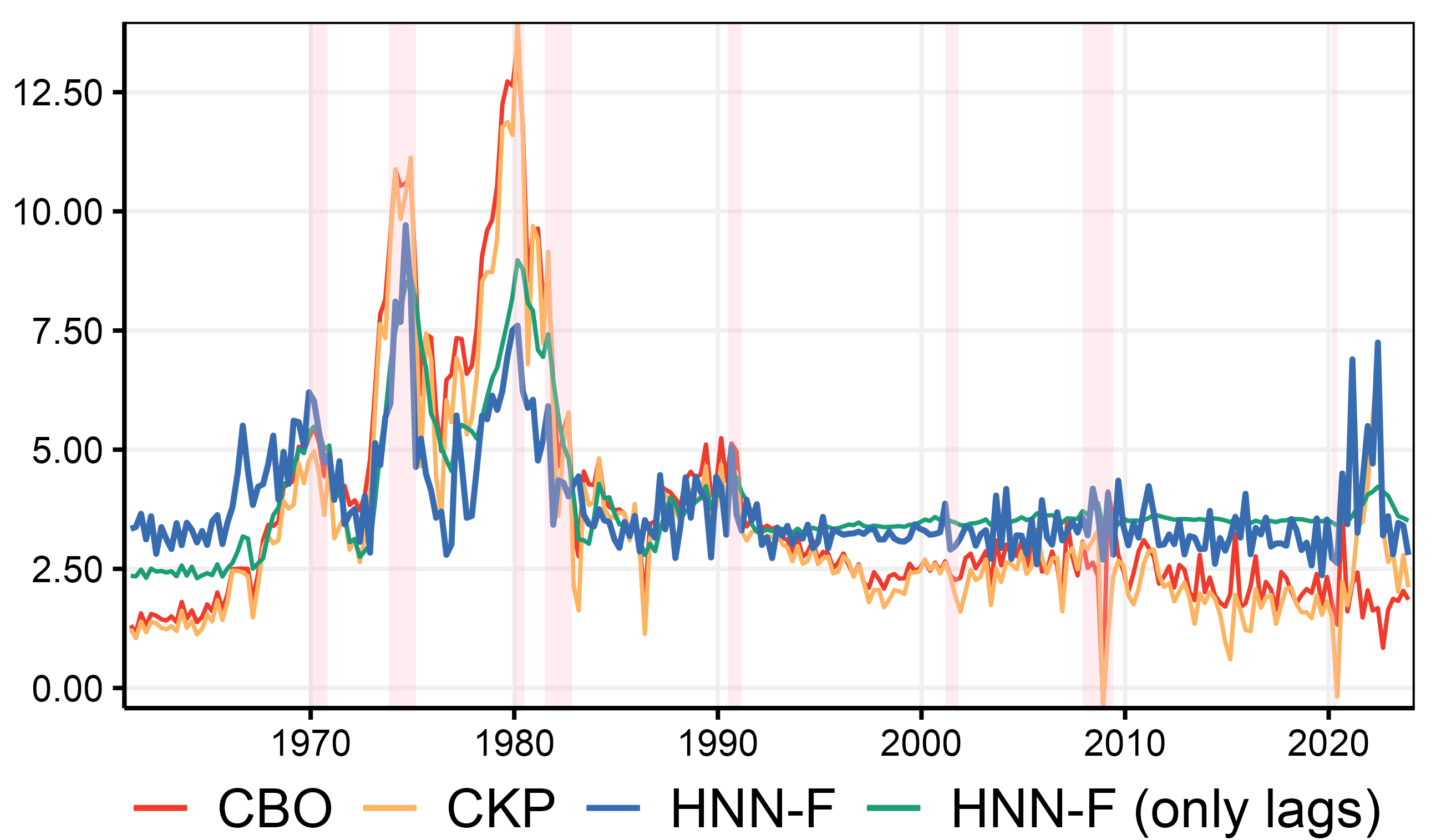}}
\caption{\footnotesize Contributions of expectations to inflation. }\label{cbo_compa_p}
      \end{subfigure}
  \caption{\footnotesize Comparing estimated contributions.  Notes: all estimations end in 2019 and are projected using the out-of-sample data from there.   Pink shading is NBER recessions.  "CBO" is the estimated contribution from a PC regression using the CBO output gap as forcing variable.   CKP means the estimated contribution as extracted from  \cite{chan2016}'s model.   "HNN-F CBO" means that all the data entering $\mathcal{H}_g$ was replaced by the CBO gap (and keeping $t$) and an HNN was rerun.  "HNN-F (only lags)" means that all the data entering $\mathcal{H}_\mathcal{E}$ was replaced by lags of $\pi_t$ (and keeping $t$) and an HNN was rerun.   In the case of expectations,  $\mathcal{E}_t^{\text{LR}}+\mathcal{E}_t^{\text{SR}}$ is plotted for HNN and the plot for "CBO" is the sum of the two lags plus the time-varying intercept.  Analogous calculations are carried for CKP.}
    \label{cbo_compa}
\end{figure}

{\noindent \sc \textbf{Comparison with Traditional Phillips Curves.}} Since gaps themselves rather than contributions are what is typically reported,  Figure \ref{cbo_compa} presents contributions from canonical PC regressions for comparison purposes.  In the case of "CBO",  those are constructed from a traditional PC specification (including 2 lags of $\pi_t$ and the CBO output gap) with time-varying coefficients obtained from \cite{GC2019} two-steps ridge regression approach.  Contributions are interesting in their own right because, unlike gaps and coefficients, they are completely identified and expressed in "inflation units".  The difference between HNN-F and alternatives is striking for $h_{t,g}$, with the latter giving real activity much less weight in driving inflation than what the former reports.  This is especially true in the 1970s and 1980s, but also for recent years.  From an ocular spectral analysis standpoint, it is clear that ${h}_{t,g}^{\text{HNN}}$ includes much higher frequencies than raditional gaps/contributions.  ${h}_{t,g}^{\text{HNN}}$ is prone to rapid spikes that the alternatives completely forego (e.g.,  the mid-1980s, the years preceding the 1990s recession,and the mid 1990s).  


"HNN-F CBO",  which replaces all the activity data in $\mathcal{H}_g$ by the CBO gap itself -- thus keeping all the other modeling ingredients from HNN  -- partly helps in understanding this wedge.  Indeed,  the green and red line follow each other closely except following the 1981-1982 and 2008-2009 recessions.  "HNN-F CBO" seems to use nonlinearities to avoid the two very negative contributions from the PC regression.  Nonetheless, it is clear that a key difference between HNN and the canonical PC regression is the nonlinear processing \textit{of a rich real activity data set}.  Finally,  only the classic PC regression and \cite{chan2016}'s gap (CKP) suggests a deep and lasting negative output gap following 2008.  "HNN-F CBO" circumvent this by interacting with a small implicit $\gamma_t$ (to be made explicit in Section \ref{sec:emp_gapcoef}).  HNN -F follows a very different pattern where the gap closes rapidly (as early as 2011) but remains gently in negative territory at least until 2018.  Finally,  from the early 1990s up until the Great Recession (GR),  both "CBO" and "CKP" contributions are practically 0 whereas HNN-F sees a mild downward contribution from real activity in the early 2000s.

HNN's post-2020 estimates differ even more from that of standard techniques.  CKP's gap in Figure \ref{og_coef} behaves like most unemployment filtering methods do.  It reports strong overheating in the late 2010s\footnote{This is because the trend has been adjusted downward by then.  Estimates including COVID-19 observations make this even more pronounced.} and a gently positive gap contribution from late 2021 onward.  As we will see in the forecasting results of Section \ref{sec:fcast},  this will be largely insufficient as an upward forcing to obtain well-centered forecasts during 2021.  This is no surprise: this approach yields an output gap which is  negative through 2021 and the PC coefficient is small.   While persistent gaps are the norm,  even in more modern incarnations like \cite{hasenzagl2018model},  they appear to be difficult to map into the inflation observational record,  and provide the well-known assessment of very low passtrough from real activity to inflation.

In Figure \ref{cbo_compa_p},  the results for HNN-F and its altered version suggest that the information contained in $\mathcal{H}_\mathcal{E}$ beyond lags of $\pi_t$ only seldom makes a difference --- except  for the latest inflation upswing.  It is also obvious that baseline HNNs allocate a smaller fraction of inflation to expectations,  which is particularly visible from the 1970s inflation spirals (mostly the second) and the 1980s.  One way to explain this is that a suboptimal \( g_t \) places an excessive burden of explanation on the lagged values of \( \pi_t \).




{\noindent \sc \textbf{Robustness to Alternative Separation and Timing Assumptions.}}  Appendix \ref{sec:rob} evaluates various deviations in the specification of the HNN model to assess if the key finding—that real economic activity has a stronger effect on inflation than captured by standard Phillips curves—still holds.  First,  by specifying short-run expectations in a tighter way (using only official expectations data), the study found that the more cleanly separated HNN yields very similar results on the importance of $g_t$.  Second,  the timing of target variables is changed to align with the theoretical NKPC (a \textit{contemporaneous} relationship) and design in studies such as \cite{blanchard2015inflation} and \cite{coibion2015phillips}.  Although $h_{t,g}$ show some quantitative differences in the post-2020 sample,  the historical path,  the resulting post-pandemic narrative,  and its overall significance remains consistent with baseline results.  Third,   a different specification of long-run expectations ($\mathcal{E}_t^{\text{LR}}$) also maintains the core findings.  This suggests that while compositional assumptions and data adjustments can influence the interpretation of short-run and/or long-run expectations, they do not significantly impact the estimated role of real activity in driving inflation.  

 \subsection{Gaps and Coefficients}\label{sec:emp_gapcoef}

\begin{figure}[t!]
\begin{center} 
\hspace*{-0.05cm}\includegraphics[trim={0cm 0cm 0cm 0cm},clip,width=0.86\textwidth]{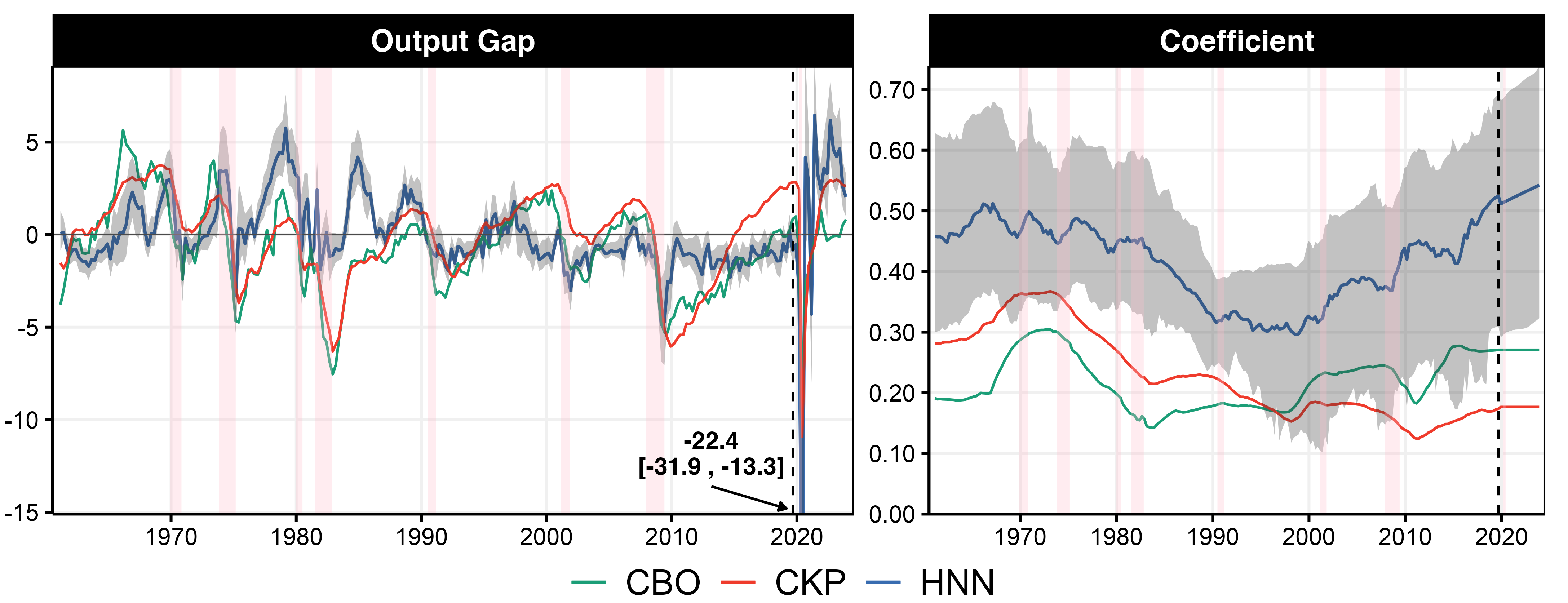}
\caption{\footnotesize HNN-F's gap ($g_t$) and associated coefficient ($\gamma_t$).  Notes: Dashed line is the beginning of the out-of-sample.  NBER recessions are in pink shadowing.  Gray shading represents the 14-86\% quantiles from the out-of-bag ensemble. See Figure \ref{cbo_compa} for specifications details.}\label{og_coef}
\end{center}
\end{figure}

So far, the focus has been on $h_{t,g}$. As discussed in Section \ref{sec:hsplit}, HNN-F, at the cost of additional assumptions,  allows for a separate inspection of $g_t$ and $\gamma_t$.  Figure \ref{og_coef} reports them for estimation ending in 2019Q4.\footnote{Figure \ref{leftover_coefs} reports the other two coefficients, which have heterogeneous shapes,  implying HNN-F is not merely "smoothing" through inflation. } Unlike recessions that preceded it, the GR is characterized by a rapid yet incomplete closing of the gap.  Interestingly,  this mildly negative gap lasting for a decade coincides in part with the so-called missing inflation era.  This observation -- a rapidly closing gap followed by a long slightly negative one -- is found whether I estimate the model using data up to today,  or end estimation in 2007.   Thus, there is no indication for a reverse engineering of $g_t$ to fit the post-GR inflation data.   Moreover, the rapid closing of \(g_t\) following the Great Recession is not observed for the early 1990s and 2000s recessions. This observation suggests that the fast closing of \(g_t\) after the GR is not a mechanical feature of HNN.





Let us now consider \(\gamma_t\), the widely studied evolving coefficient of the Phillips curve.  The evidence in Figure \ref{og_coef} is in partial agreement with the recent literature on the matter \citep{blanchard2016phillips,gali2015monetary,del2020s} in the sense that the exogenously time-varying $\gamma_t$ has been decreasing starting from the 1980s.   However, there are many notable differences.  First,  there seem to be a break around 1980,  in the midst of Volker's disinflation, where $\gamma_t$'s decline substantially accelerates.  Second,  unlike results from standard approaches,  $\gamma_t$  is not found to decline further following 2008, but rather to slowly come back to its former value  starting from the 2000s.  These observations differ from those in studies such as \cite{blanchard2016phillips}, \cite{stock2019slack}, and \cite{hazell2020slope}. Despite their diverse research designs, these studies commonly use forcing variables that are closely correlated with filtered unemployment.  Given how different HNN-F's $g_t$ is with respect to traditional indicators,   $\gamma_t^{\text{HNN}}$ atypical vivacity is not entirely surprising -- and turns out to be in line with other recent works.   \cite{bonam2021time} report that using a survey-based labor shortage indicator is a better predictor of wage inflation in many European countries and leads to stable or steepening PCs over the last decade.  Recently,  \cite{domash2022tight} report results in agreement with this view using US state-level data.

For space considerations,   many results have been relegated to the appendix.  In Appendix \ref{app:rewrite},  I report $g_t$ for various estimation windows and find that it is desirably stable across those.   In Appendix \ref{sec:altsup},  I report gaps and coefficients using two alternative supervisor variables (Core CPI and Yearly Inflation).

\subsection{What are the gap and expectations made of?}\label{sec:VI}


By construction,  $g_t$ and $\mathcal{E}_t^{\text{SR}}$ are combinations of thousands of parameters nonlinearly processing many regressors.  Consequently,  directly looking at network weights is inevitably meaningless.  More productively,  I investigate which $X_{t,k}\in \mathcal{H}_g$ seems to matter most by designing a variable importance (VI) exercise very much inspired from what \cite{MRFjae} studied for "generalized time-varying parameters" in a Random Forest context -- which is itself inspired from traditional variable importance measures for tree ensembles \textit{predictions}.  For the VI calculations,  I focus on groups of variable $k$,  meaning we will evaluate the overall effect of all transformations and lags of variable $k$.  Implementation details can be found in Appendix \ref{sec:VIdet}.

\begin{figure}[t!]
  \centering
  \begin{subfigure}[b]{0.48\textwidth}
\hspace{-0.25cm}\includegraphics[trim={0cm 1.5cm 0cm 0cm},clip,width=0.995\textwidth]{{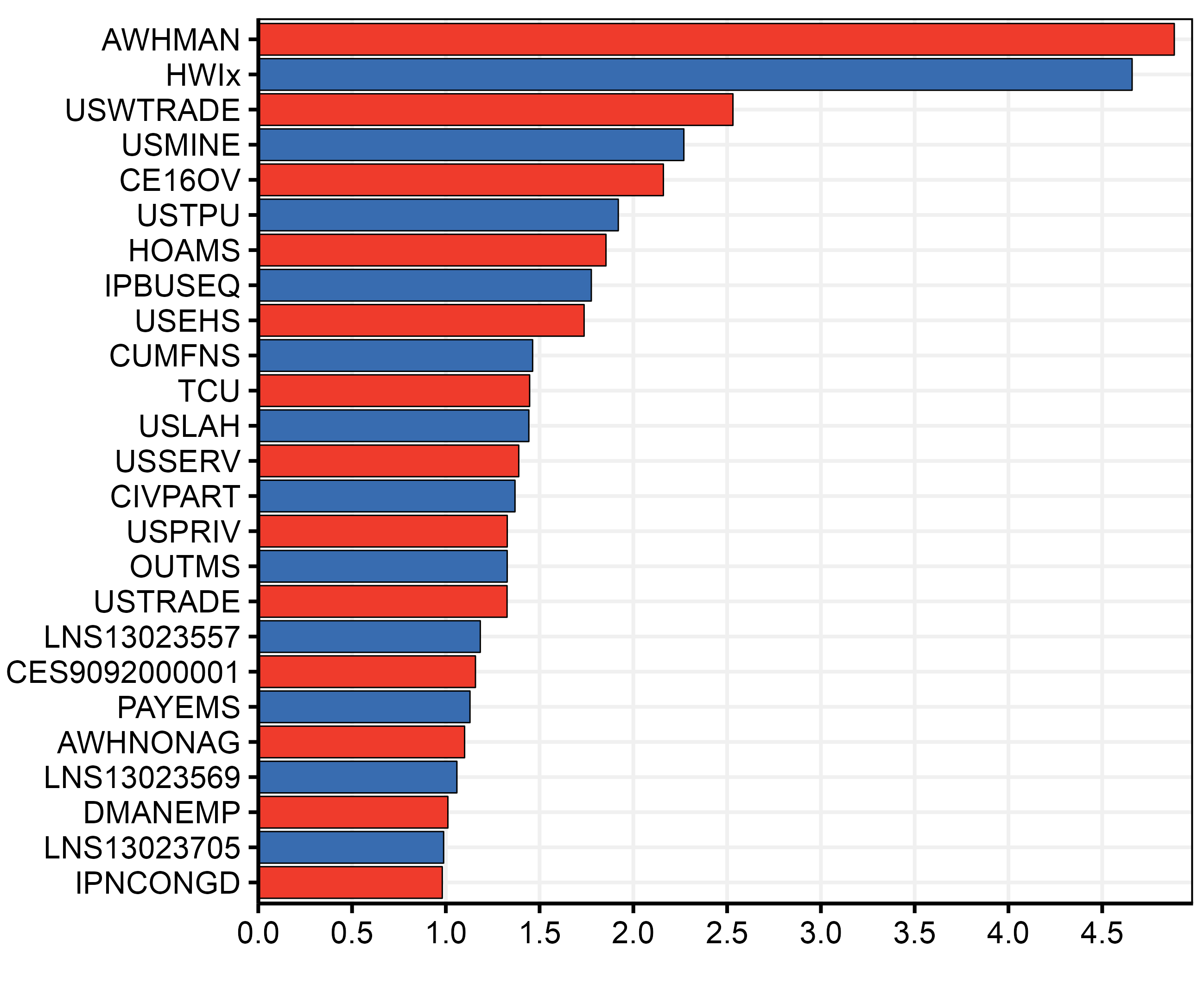}} 
\caption{\footnotesize  Gap ($g_t$)}\label{vi_g}
      \end{subfigure}
  \begin{subfigure}[b]{0.48\textwidth}
\includegraphics[trim={0cm 1.5cm 0cm 0cm},clip,width=0.995\textwidth]{{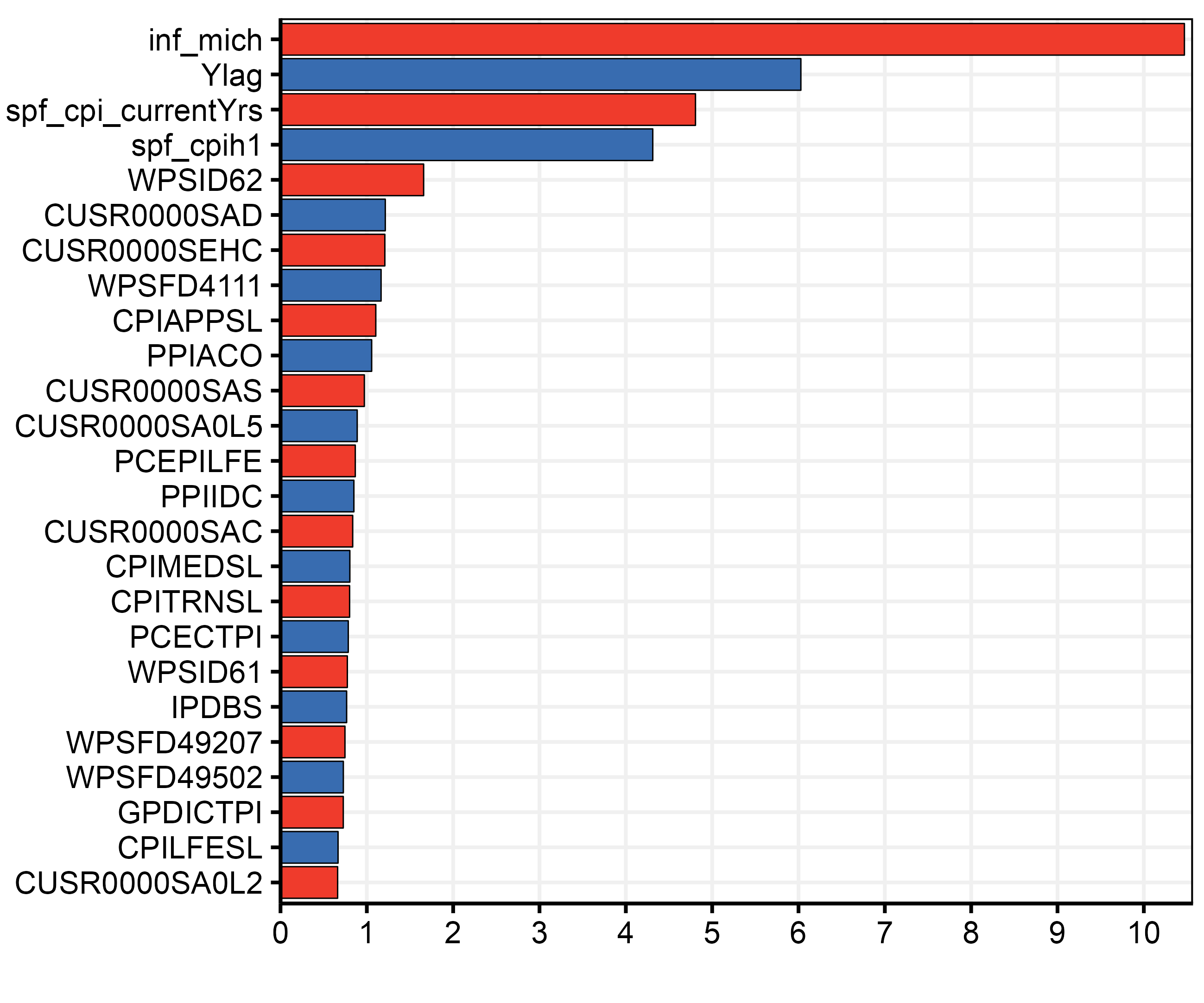}}
\caption{\footnotesize Short-Run Expectations ($\mathcal{E}_t^\text{SR}$)}\label{vi_p}
      \end{subfigure}
        \vspace{0.0001em}
  \caption{\footnotesize VI Results for benchmark HNN-F specification, with training ending in 2019Q4.  Mnemonics are those of FRED-QD \citep{mccracken2020fred}.}\label{vi_bench}
\end{figure}

VI results are reported in Figure \ref{vi_bench}.  Here are key observations for $\text{VI}^g$.  First,  \texttt{AWHMAN}'s (average weekly hours in the manufacturing sector) predominance for $g_t$ suggests an important role for the \textit{intensive} margin, whereas typical labor-based gap measures are mostly about extensive margin (like filtered unemployment).  Recently,  \cite{bulligan2019adjustments} find the former can complement the latter as forcing variables in linear PCs for the Euro area (but not the US).  Second,  the composite Help-Wanted Index (\texttt{HWIx}) of \cite{barnichon2010building} --which \cite{mccracken2020fred} splice earlier in the sample with the original Conference Board product for "print" job postings -- is shown to play an important role.  Intuitively,  the index, by construction,  characterizes increased labor demand (and perhaps shortages) which is expected, by economic theory, to translate into higher wages, and eventually, higher aggregate prices.  This partly explains the very positive gap in Figure \ref{og_coef} since  \texttt{HWIx} is effectively skyrocketing as of late 2021 despite being largely stagnant in 2018 and 2019.\footnote{ \cite{domash2022tight} also find that indicators associated with the demand side of the labor market have better predictive power for wage inflation.  Similarly,  \cite{bonam2021time} document that a survey-based labor shortage indicator is a better predictor of wage inflation in many European countries than an unemployment gap.} However, this not the whole story: nonlinear neural processing of \texttt{HWIx} seems essential as reported in the ablation study (Appendix \ref{sec:abla}).  In Figure \ref{ur_hwi} (Appendix),  we see that, at times,  the unemployment rate and \texttt{HWIx} were closely related,  like during 1990s and the 2000s.  But other times they were not,  like the 1970s and during recent years.   Moreover, their acceleration rate can differ in key recession and expansion episodes.  By betting on some transformation of \texttt{HWIx},  HNN leveraged historical patterns to rely on more potent (and timely) forcing variables.  Third,  GDP and associated measures seem unimportant,  so does the unemployment rate.  The only traditional gap measure appearing in the lower end of $\text{VI}^g$'s top 25 is total capacity utilization (\texttt{TCU}).


Turning to $\text{VI}^{\mathcal{E}_\text{SR}}$, we see that a handful of very familiar variables dominate the top 25.  First,  the obvious preponderance of the University of Michigan Survey of Consumer Inflation Expectations (\texttt{inf\_mich}) strengthens the case for the increasingly popular practice of using survey expectations in PC regressions \citep{binder2015whose,coibion2015phillips,coibion2018formation,meeks2023heterogeneous}.  It also completes the explanation as to why HNN-F forecasts did not call for lasting disinflation following the GR. That is, as suggested by \cite{coibion2015phillips},  proxying expectations using survey expectations rather than, say,  lags of the CPI,  procures more accurate post-2008 predictions.  HNN learned that prior to 2007 by putting a high weight on \texttt{inf\_mich}.  Nonetheless,  $\text{VI}^{\mathcal{E}_\text{SR}}$ suggests mixing in expectations from different economic agents and formulated for different horizons seems more appropriate,  which is in line with recent results for simpler regression models in \cite{banbura2021inflation}.   There is also a minor role for "backward-looking expectations" or "inflation persistence" as emphasized by the presence of lags of the CPI (\texttt{Ylag}) in the top 4.

\subsection{On the Nature of Nonlinearities}\label{sec:nlplus}

The black box can be opened further.  Now that we identified important inputs for $g_t$  and $\mathcal{E}_t^{\text{SR}}$,  we can use a surrogate model approach as in \cite{MRFjae} to map back the latent series of interest into pre-identified key inputs.   The surrogate model approach,  which consists of fitting an auxiliary interpretable model to a pre-estimated function output often using a focused set of inputs \citep{molnar2019interpretable}, is an economical way to asses the shape of nonlinearities.  The surrogate model used below is a bivariate Random Forest  featuring, for both $g_t$  and $\mathcal{E}_t^{\text{SR}}$,  the two most salient inputs identified by VI.  While RF is typically considered a black box model,  its fit is easy to visualize if the inputs are 2-dimensional.  Moreover,  unlike single trees (often used as surrogate devices),  it can approximate hard-thresholding \textit{and} smooth relationships.   Implementation details can be found in Appendix  \ref{sec:VIdet}. 

\begin{figure}[h!]
  \centering
  \begin{subfigure}[b]{0.4\textwidth}
    \centering
    \includegraphics[trim={0cm 0cm 0cm 0cm},clip,width=\textwidth]{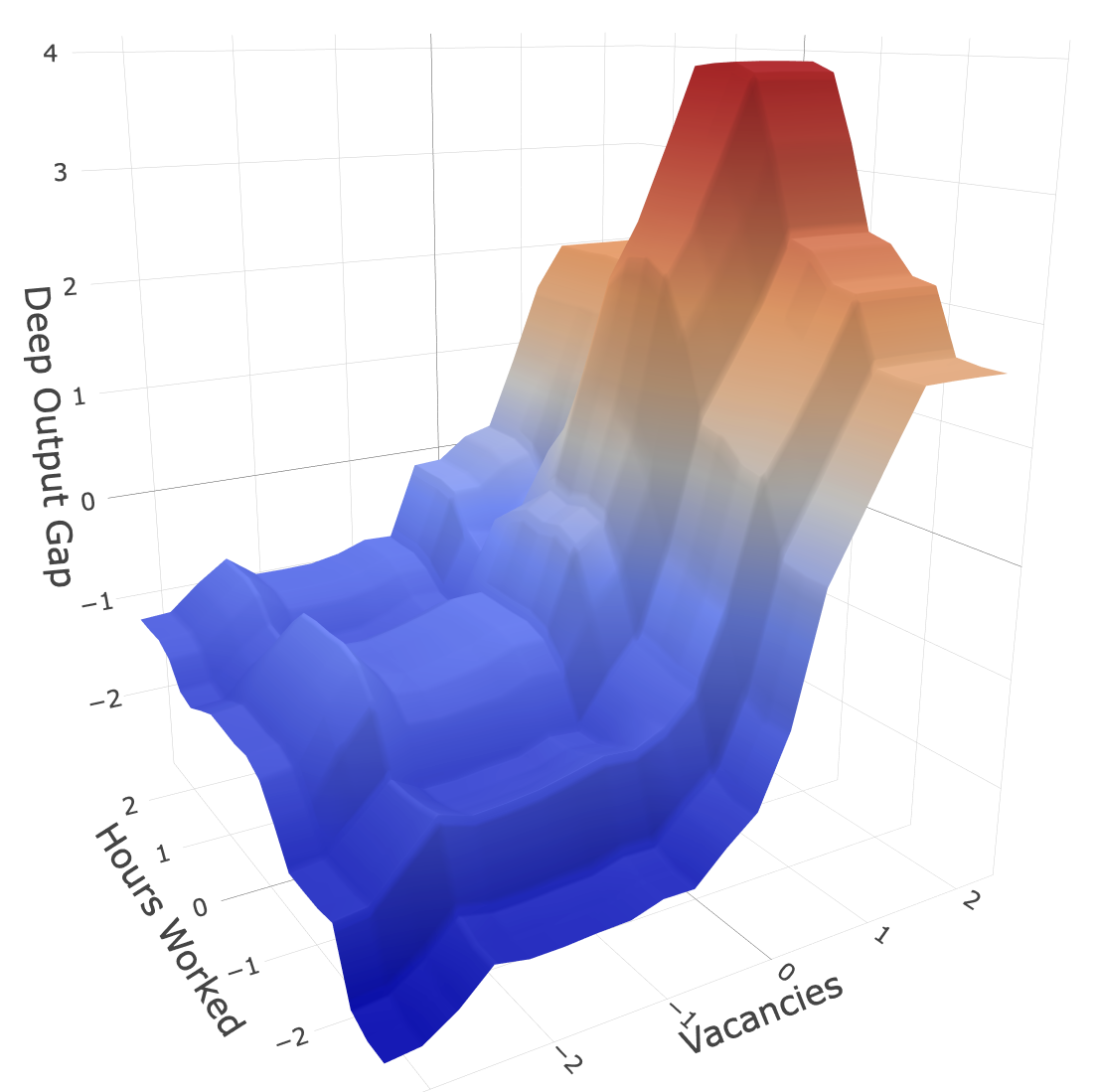}
    \caption{\footnotesize Gap ($g_t$)}
  \end{subfigure}
  \hspace{1.5em}
  \begin{subfigure}[b]{0.4\textwidth}
    \centering
    \includegraphics[trim={0cm 0cm 0cm 0cm},clip,width=\textwidth]{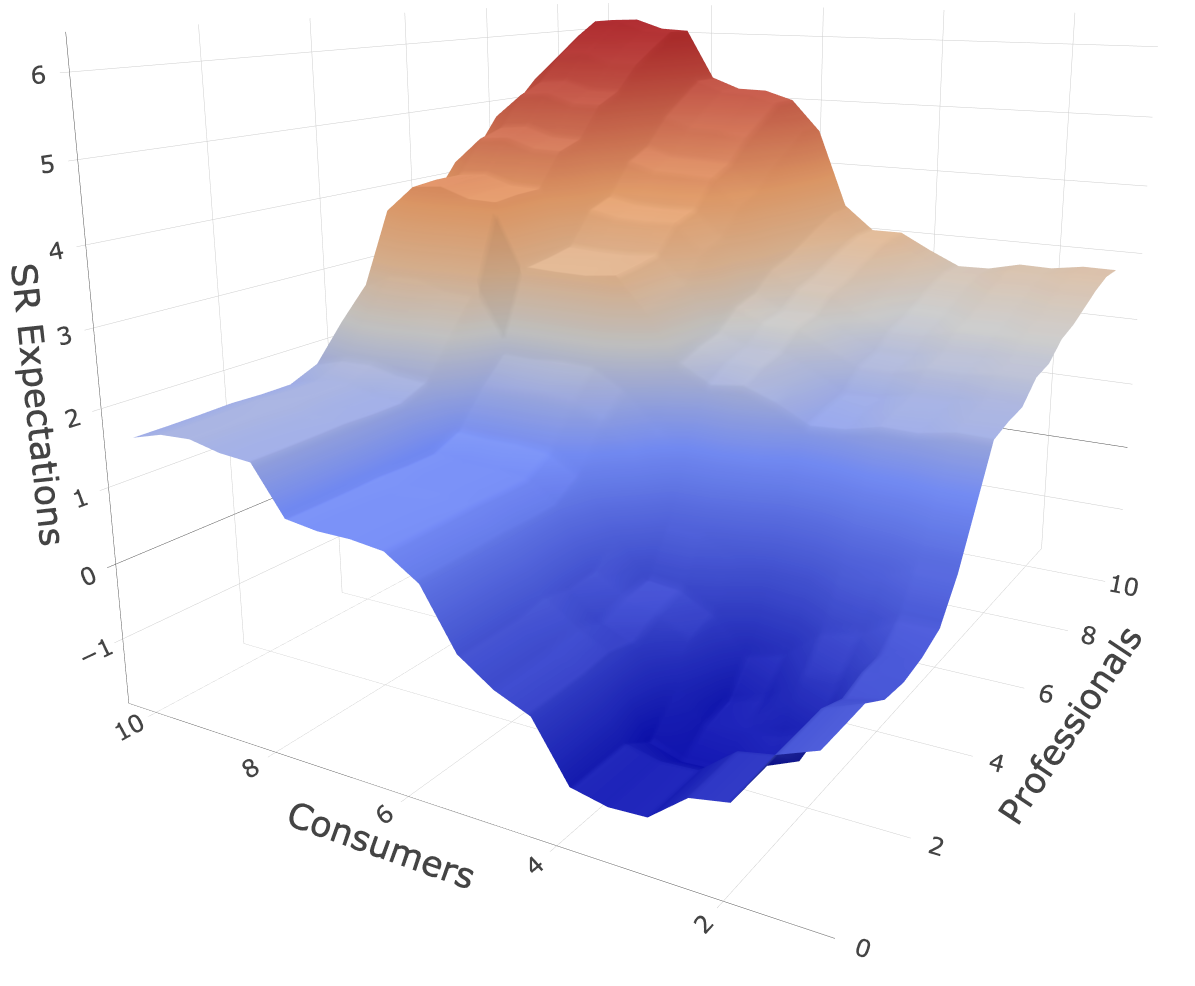}
    \caption{\footnotesize Short-Run Expectations ($\mathcal{E}_t^\text{SR}$)}
  \end{subfigure}
  \caption{\footnotesize Surrogate Model Results for benchmark HNN-F specification, with training ending in 2019Q4.  Vacancies and Hours Worked are detrended and presented as z-scores.  More surrogate model details are in Appendix \ref{sec:VIdet}. }
  \label{3dplot}
\end{figure}

The VI analysis above suggests that key drivers of $g_t$ are vacancies and hours worked, and those of $\mathcal{E}_t^{\text{SR}}$ are professional forecasters' (\texttt{spf\_cpi\_currentYrs}) and consumers expectations (\texttt{inf\_mich}).\footnote{While \texttt{Ylag} comes in before \texttt{spf\_cpi\_currentYrs},  the combined importance of the two SPF indicators is greater than the lags of CPI.   Results are mostly unchanged from using \texttt{spf\_cpih1} instead of  \texttt{spf\_cpi\_currentYrs}. }  Figure \ref{3dplot} reports results on how the two selected indicators are associated with $g_t$  and $\mathcal{E}_t^{\text{SR}}$.  The correlations of the surrogate models' fitted values with the original series are 0.80 for $g_t$ and its reconstruction, and 0.77 for $\mathcal{E}_t^{\text{SR}}$ and its reconstruction, suggesting that they capture a significant part of the relevant variation in $g_t$ and $\mathcal{E}_t$.

 
  
The real activity relationships show significant nonlinearity, particularly with respect to vacancies. There is a notable kink in the behavior of vacancies: the effect is moderately positive below a certain threshold, but beyond this threshold, the derivative increases substantially, causing the curve to become very steep until another plateau is reached. The effect of hours worked is linearly positive when vacancies are low but becomes quite nonlinear at higher vacancy levels, showing a major increase from low to mid-range values. When both hours worked and vacancies are high, the effect is milder than for mid-range values. However, it is important to note that there are very few data points in the region where both vacancies and hours worked are at the top of their range, so the second segment of the observed U-shape should be interpreted with caution. This kink behavior, or hockey stick relationship, between vacancies and $g_t$ is reminiscent of the nonlinear Phillips curve proposed by \cite{benigno2023s}, who propose an economic rationale for the phenomenon. While less transparent than a tightly specified theoretical model, HNN identifies this behavior among many possible nonlinearities from data through 2019Q4.

Overall, the relationships between expectations series and $\mathcal{E}_t^\text{SR}$ show some nonlinearities. However, these can be reasonably well approximated with non-interactive linear relationships. This suggests that while there are deviations from linearity, especially for lower values (see ablation study in Appendix \ref{sec:abla2}), the link between survey-based expectations and inflation appears less complex in terms of functional form than that of real activity.

The highly nonlinear relationship behind $g_t$ and the semi-linear ones behind $\mathcal{E}_t^\text{SR}$ provide further backing for this paper's core finding about the importance of real activity. Much of the prior evidence is based on linear models, which, by construction, attribute larger explanatory power to relationships they can capture well—in this case, expectations rather than real activity. In a more general nonlinear model, $\mathcal{E}_t^\text{SR}$ remains very important. However, $g_t$ becomes equally significant, and even more so for the post-2020 sample, once the model accounts for the nonlinearity in how a diverse basket of real activity indicators can drive inflation.

In Appendix \ref{sec:abla2}, I report results for a linear version of HNN where ReLU activation functions have been replaced with linear ones. The linear model struggles to capture a plausible path for $g_t$, whereas results for short-run expectations are closer to those obtained by the baseline HNN. This reinforces the view that allowing for nonlinearities is of greater importance when modeling real activity. Additionally, the forecasting performance of the linear version (included as a benchmark in Section \ref{sec:fcast}) is substantially inferior to that of HNN and HNN-F, especially in recent years.

\section{Forecasting }\label{sec:fcast}

Current evidence in favor of PC-based forecasting is weak, with minor or non-existent improvements over simpler benchmarks like plain autoregressions \citep{atkeson2001phillips,stock2008phillips,faust2013forecasting}.  Recent extensive evaluations for the Euro area \citep{banbura2020does} suggest there is a case for some cautious hope with specifications allowing for flexible trend inflation and an endogenously estimated gap.  Despite all the evidence on its uneven empirical potency,  PCs are still widely used to forecast and understand inflation  \citep{yellen2017inflation}, mostly because they are rooted in some basic form of macroeconomic theory.  This section provides a validation of the new methodology by comparing its predictive performance to a representative set of traditional econometric approaches and machine learning tools. 

{\noindent \sc \textbf{Setup.}} The pseudo-out-of-sample period starts in 2008Q1 and ends in 2024Q1.  I use expanding window estimation from 1961Q3.  HNNs are re-estimated and tuned every 4 quarters.  Following standard practice, the quality of point forecasts is evaluated using the Root Mean Square Error (RMSE).   The forecasting target is CPI($s=1$), which is the supervisor in benchmark HNN specifications.   Additionally,  results for two alternative supervisors are studied in Appendix \ref{sec:altsup} -- CPI average inflation from $t$ to $t+4$ (${\pi}_{t:(t+4)}=\sum_{s'=1}^4 \pi_{t+s'}$) and Core CPI($s=1$).  Performance results are reported for four subsamples.  First,  we have all the samples including and excluding 2020 observations.\footnote{The exclusion zone is extended to 2021Q1-Q2 for 4 quarters ahead forecasts for the simple reason that they were made during the depth of 2020Q1-Q2 and the models propagate a year later what it thinks is an unusually large negative (yet typical in composition) demand shock.} Additionally,  separate RMSEs for the pre-2020 (up to 2019Q4) and post-2020 (starting in 2021Q1) sample are provided.

{\noindent \sc \textbf{Models.}} A few obvious benchmarks from both sides of the aisle are considered.   On the ML side, there is a fully connected neural network with the same hyperparameters as HNN (\textbf{DNN}) and a Random Forest (\textbf{RF}) with default tuning parameters (typically hard to beat,  \citealt{MRFjae}).  Additionally,  I consider a \textit{linear} network version of HNN-F (\textbf{HNN-F-LR}) where ReLU activation functions have been replaced by linear ones for gaps hemispheres.  Using linear layers in the network makes it some kind of linear groupwise supervised principal components model,  a linear data-rich benchmark.  The latter model allows to get a sense of the contribution of nonlinearities in HNN's overall performance by ablating them.  All these benchmarks use the exact information set as HNN (variables and aforementioned transformations).  

Then, there are inflation-specialized econometric benchmarks of increasing sophistication.  First,  we have the AR(4) which will stand as the generic numeraire of reported MSEs.  Also,  two rolling means are considered,  the one-year mean à   la \cite{atkeson2001phillips} (\textbf{1y Avg}) and a longer-run one (\textbf{10y Avg}).  Bringing in real activity information,  I consider a PC regression (\textbf{PC}, two lags of $\pi_t$ and the CBO gap) estimated on a rolling window of 15 years to allow for time-varying parameters.  Note that this PC regression is given a handicap by using the latest CBO gap which may have been substantially revised ex-post.  Additionally,  an identical PC regression augmented with two lags of oil prices and survey expectations (\textbf{PC+}) is considered to match some of the information set in HNN, and more generally specifications inspired from \cite{coibion2015phillips}.  We also consider \cite{chan2016}'s time-varying bounded Phillips curve model (\textbf{CKP}) where $g_t$ is extracted in a supervised fashion from unemployment by assuming the natural rate of unemployment to follow a random walk.  Key coefficients also follow random walks.  This approach was reported to have sporadic success in forecasting Euro area inflation \citep{banbura2020does}.  All those non-NN methods are re-estimated every quarter.  



\begin{figure}[t!]
  \begin{subfigure}[b]{0.5\textwidth}
\includegraphics[trim={0cm 0cm 0cm 0cm},clip,width=0.995\textwidth]{{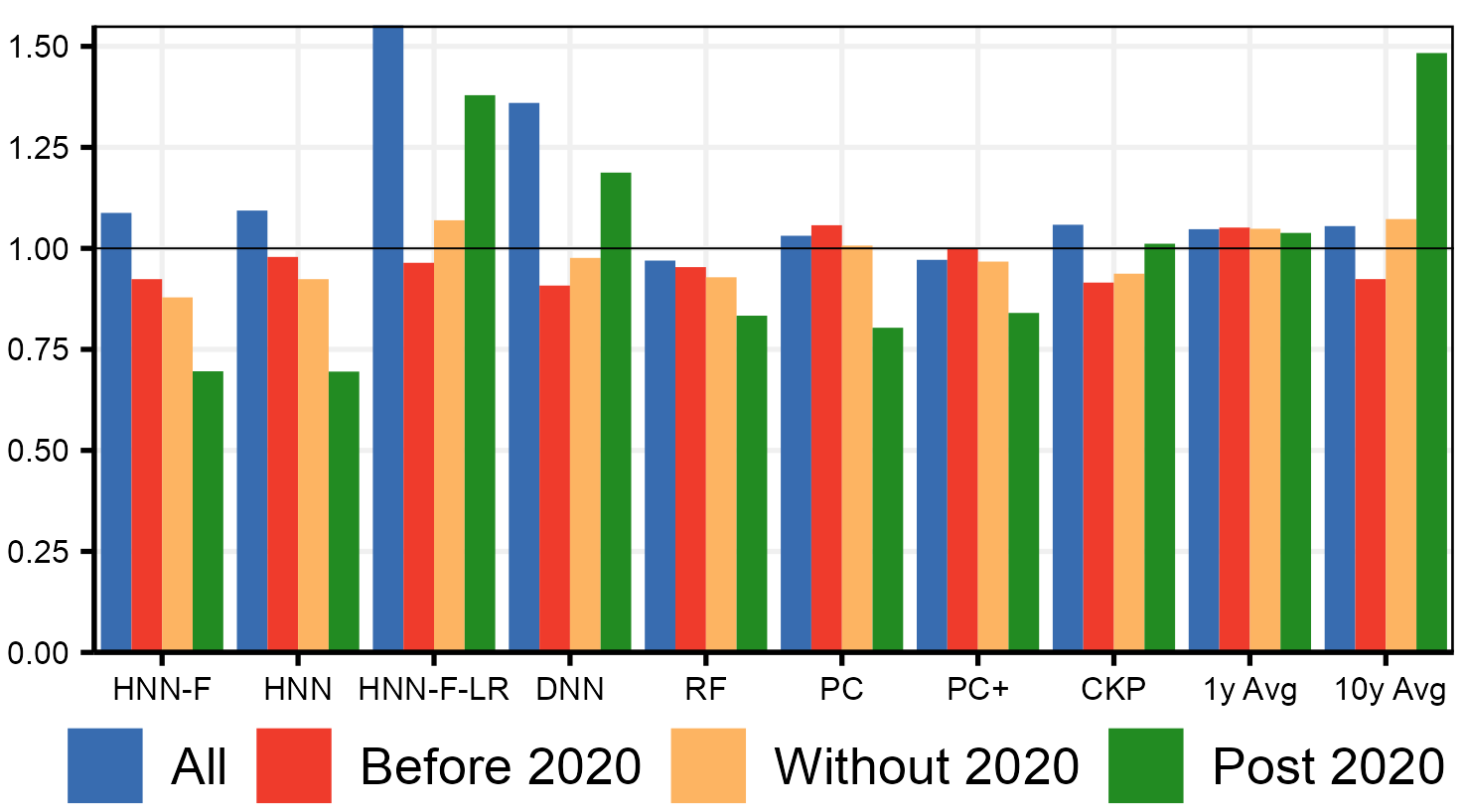}}
\caption{\footnotesize MSE wrt AR(4),  CPI }\label{mse_q_2007}
      \end{subfigure}
  \begin{subfigure}[b]{0.5\textwidth}
\includegraphics[trim={0cm 0cm 0cm 0cm},clip,width=0.995\textwidth]{{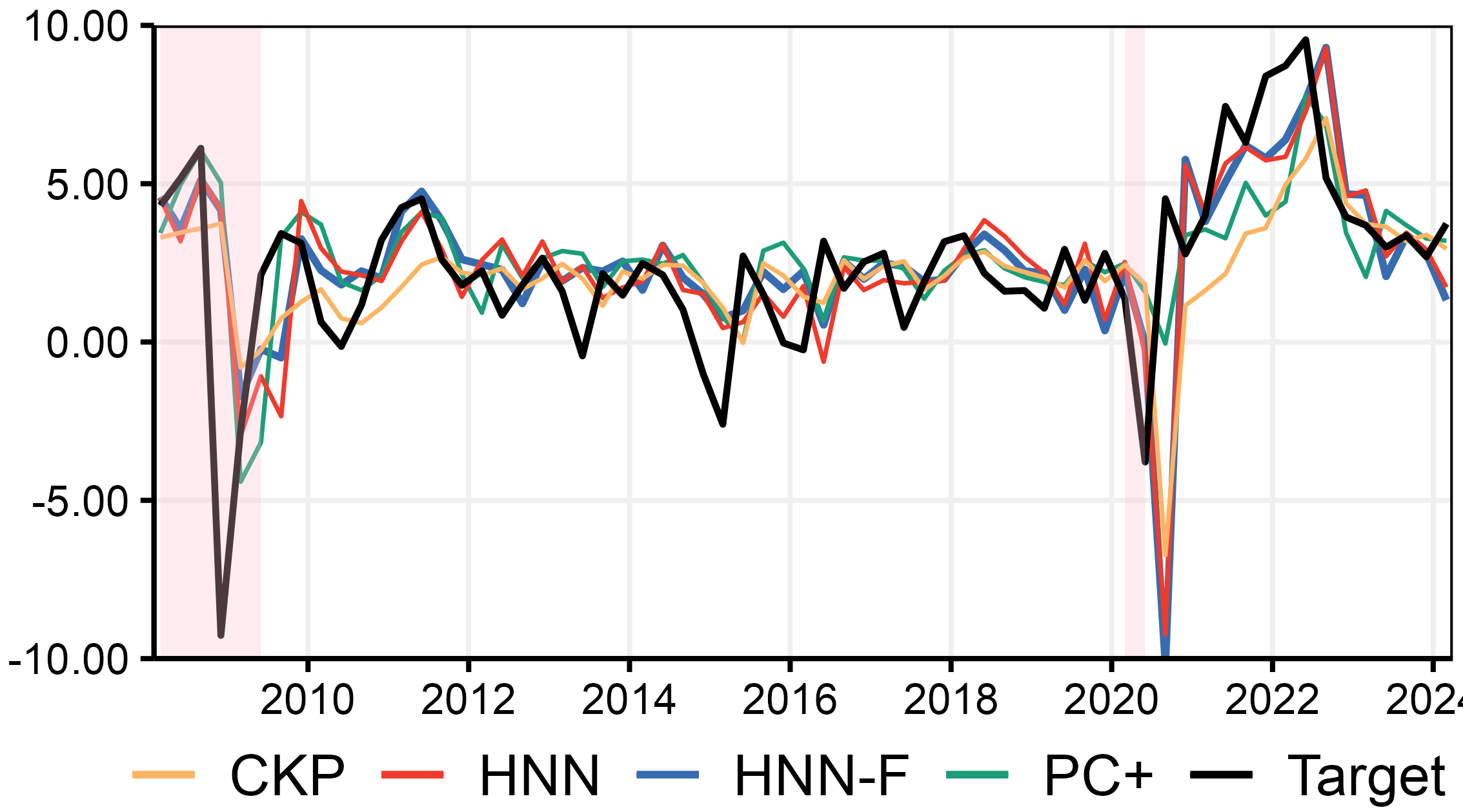}}
\caption{\footnotesize Forecasts,  CPI }\label{fcast_q_2007}
      \end{subfigure}
  \caption{\footnotesize Forecasting Results the main target with test set 2007Q2-2024Q1 and re-estimating NNs every year.  Notes:  Pink shading is NBER recessions.  Corresponding Diebold-Mariano tests can be found in Table \ref{dmtest} (Appendix \ref{addifig}).}\label{forecast_fig}
\end{figure}

{\noindent \sc \textbf{RMSEs and Forecasts Narratives Comparison.}} I now report the forecasting performance of HNNs for the three targets and examine their forecasts. In Figure \ref{mse_q_2007}, HNN and HNN-F perform well for CPI ($s=1$), excluding the aberrant 2020 observations (to which we will come back to below).  Figure \ref{fcast_q_2007} shows that HNN's relative success is partly due to its reasonable accuracy in capturing the recent upswing in inflation. Overall, the bar plot in Figure \ref{mse_q_2007} show improvements ranging from 10\% to 25\% depending on the subsample, with HNN-F and HNN consistently delivering comparable RMSEs.\footnote{Diebold-Mariano test statistics are reported in Table \ref{dmtest} (Appendix \ref{addifig}) and show that in the main case of CPI ($s=1$), these improvements over the benchmark are significant at the 5\% level for the "Without 2020" and "Post 2020" samples. Table \ref{tab:rt1} and Figure \ref{tab:rt2} show that forecasts using real-time  data (and resulting raw RMSEs) are extremely similar to the main ones in Figure \ref{fcast_q_2007} using the 2024Q1 data.} The closest competitors are RF, PC+ (which includes the survey of professional forecasters' forecasts as a predictor), and CKP.  The mild improvement of HNNs over (or tie with) benchmarks for the pre-2020 sample can be understood from, e.g., Figure \ref{cbo_compa}, as all models agree on a near-zero \textit{contribution} of real activity from 2010 to 2020.   However, traditional models are less apt at capturing what crucially matters for policy: when inflation starts to persistently get out of its target range.  For the three years following 2020 (Post 2020 in green), HNN, and particularly HNN-F, perform best for CPI ($s=1$) and the two additional targets in Appendix \ref{sec:altsup}.  Finally,  the \textit{linear} supervised data-rich model (HNN-F-LR) struggles in most samples and highlights the importance of nonlinearities to truly harness the benefits of the additional data—an argument made more generally for ML models in macroeconomic forecasting by \cite{GCLSS2018,GCMS}.

Figure \ref{fcast_q_2007} helps differentiate the narratives. CKP, based on a Bayesian bivariate state-space model of trend inflation and the gap, faces major difficulties during two important historical episodes. First, its forecasts are consistently too low for most of 2008-2012, exemplifying the missing disinflation puzzle—a classic issue with regression models predicting inflation using traditional gaps. HNN mostly avoids this predicament with a rapidly closing gap in the aftermath of the GR. Second, CKP forecasts are significantly below realizations for the two-year period from mid-2020 to mid-2022 (when inflation starts to slow down). The reason for this is visible in Figure \ref{og_coef}: the de-trended unemployment rate, the forcing variable, is negative for most of 2021. Consequently, if it influences inflation, it does so downward, not upward. When the unemployment-based forcing variable finally lands in positive territory, the Phillips curve coefficient makes its contribution minimal.  PC+ also struggles, but less so than CKP, to swiftly and consistently capture the sudden rise in inflation in 2021, which settled around 5\% rather than 2-3\%. HNN delivers more timely forecasts of CPI ($s=1$) in the post-2020 sample by not relying on potentially lagging indicators and leveraging nonlinearities to capture the contribution of real activity in a timely manner.



{\noindent \sc \textbf{On 2020's Performance and Density Forecasting.}}   Unsurprisingly,  yearly results for 2020 and most of 2021 are not great for real-activity-based forecasts,  including HNN.  In a similar fashion to what is reported in Figure \ref{fcast_q_2007}, this is due to HNN and PC regressions not being informed that this is no ordinary recession and that extraordinary governmental programs have been implemented to life support the economy.   This limitation has even stronger consequences when forecasting ${\pi}_{t:(t+4)}$ since the medium-run dynamic transmission mechanism itself is certainly quite different during the pandemic than for previous recessions.  A careful use of the model, combined with the usual external  judgment in forecasting practice,  would have led to discarding the downward spike.   Therefore,  it is justified to focus on performance metrics excluding them.   Nonetheless,  this setback motivated the development of an extended version of HNN in \cite{GCFK} which features an additional hemisphere predicting forecast uncertainty.   The obtained density forecasts are very competitive (including 2020) because the more sophisticated network calls for extreme volatility during 2020 (thus, self-discounting its struggling 2020 forecasts) and more moderate volatility afterwards.

\section{Conclusion}\label{sec:con}

This paper estimates a neural Phillips curve with a deep output gap using the Hemisphere Neural Network (HNN), a machine learning model designed for economically interpretable inflation predictions. HNN avoids various non-innocuous assumptions of traditional econometric methods, leading to significantly different results. A refinement of the plain HNN architecture allows for the separation of contributions from expectations and real activity into coefficients and gaps.

HNNs provide good forecasts, such as capturing part of the CPI upswing in 2021. This success is due to HNN defining an output gap through supervised nonlinear processing of numerous real activity indicators. The most important raw indicators are vacancies and hours worked, suggesting a prominent role for the intensive margin and the demand side of the labor market. Allowing for nonlinearities is crucial in capturing such effects. As a result, the Phillips curve coefficient on HNN's output gap declined in the early 1980s but experienced a significant revival starting from the 2000s.  The role of real activity, found to be significantly stronger than in traditional Phillips curves, explains HNN's ability to foresee mounting inflationary pressures from late 2020 onward.

The HNN framework is versatile and can be applied to various problems and extended in many directions. Overall, this paper demonstrates that deep learning can provide more than just predictions, helping to address key issues in empirical macroeconomic analysis, such as the successful measurement of ambiguously defined yet extremely important latent states.

 \pagebreak
 
 \setlength\bibsep{5pt}
		
\bibliographystyle{apalike}

\setstretch{0.75}
\bibliography{ref_pgc_v181204}

\clearpage

\appendix
\newcounter{saveeqn}
\setcounter{saveeqn}{\value{section}}
\renewcommand{\theequation}{\mbox{\Alph{saveeqn}.\arabic{equation}}} \setcounter{saveeqn}{1}
\setcounter{equation}{0}


\section{Supplementary Material}
\setstretch{1.35}

\subsection{Additional Technical Details}\label{sec:VIdet}
\subsubsection{Additional HNN Implementation Details}

For HNN, we normalize each predictor to have mean 0 and variance 1, which is standard in regression networks.  For HNN-F,  since there is no weight sharing,  we ought to be more careful in order not to give implicitly some hemisphere a higher prior weight in the network. This could occur, for instance, if some $\mathcal{H}$ has a much larger number of inputs than another.  With early stopping performing a type of ridge regularization,  it entails the prior that \textit{each} variable should contribute but in a mild way.  If the real activity group contains 40 times more regressors than the commodities one,  then going for the standard normalization gives a much larger prior weight to its resulting component by construction.  To avoid this scenario,  and give equal a priori importance to $h_t$'s,  we divide each standardized ${X}_{t,k}\in\mathcal{H}_j$ by $\sqrt{\mathbf{card}\left(\mathcal{H}_j\right)}$ (the square root of the number of variables in that hemisphere).  The intuition for using such a denominator comes from the fact that if all variables are mutually uncorrelated and each given a weight of one or minus one (i.e., no learning beyond what ridge prescribed has taken place), then the variance of the simplistic (linear) component $h_{t,j}$ is $\mathbf{card}\left({\mathcal{H}_j}\right)$.  Thus,  dividing each member of that group by the square root of it sets each $h_{t,j}$'s a priori variance to be 1.  

Since early stopping is set optimally for each run in isolation (without considering their ultimate ensembling), the early-stopping points may be overly aggressive and sub-optimally early for the final average. While this phenomenon is not observed in the empirical data for this paper (most runs with very high in-sample \(R^2\) thanks the use of large networks), it can be an issue when applying the model to smaller data sets (such as data starting in the 2000s), especially with a subsampling rate of 65\%. In such cases, immediate improvement can be achieved by regressing the in-sample out-of-bag predictions on the original target in a univariate regression. This acts as a global early-stopping (or later-stopping) step by scaling up or down the fitted values. The scaling coefficient (usually near 1) can then be applied to all draws to ensure coherent uncertainty quantification.


\subsubsection{Variable Importance and Surrogate Model Details}

I cover here the details behind the generation of results in Section \ref{sec:VI}.  For the VI calculations,  I focus on groups of variable $k$,  meaning we will evaluate the overall effect of all transformations and lags of variable $k$ (as mentioned in Section \ref{data}, we include 4 lags of each and moving averages of order 2, 4 and 8).  The variable importance procedure to evaluate the relevance of variable $k$ to $h_{t,j}$ can be summarized as follows.  $\text{VI}_k^j$, for a variable $k \in \mathcal{H}_j$,  works in three steps.  First,  we shuffle randomly variable $k$ (and all its attached transformations, i.e., lags and MARXs).  Second,  we recompute (but do not re-estimate) the component $h_j(\tilde{X}_t)$ (using the shuffled data for $k$ and the original data for all other variables).  Third,  we calculate its distance to the real component estimate $h_j(X_t)$.
Formally,  the standardized $\text{VI}_k^j$,  in terms of \% of increase in MSE, is
\begin{align}
 \text{VI}_k^j= 100 \times \left(\frac{\tfrac{1}{T}\sum_{t=1}^T (h_j(\tilde{X}_t)-h_j(X_t))^2}{Var(h_j(X_t))} \right).
\end{align}
Intuitively,  randomizing important variables that is highly utilized by the model will push $h_{t,j}$ further from its original estimate than randomizing ones that are not utilized at all or are given very small effective weights.  In terms of operations, we increase the subsampling rate for VI calculations (85\%), which allows HNN to identify sparser functions of raw inputs while keeping point estimates unchanged.



The surrogate RF model is set with a subsampling rate of 0.75 (a standard value in RF) and the minimum node size is set to 20, which limits gently the complexity of the underlying trees.  This is recommended to avoid overfitting when RF has little possibilities for diversification by shuffling predictors \citep{MSoRF}.  The expectations data is utilized as is.  "Vacancies" is the linearly detrended log HWIx (originally in units of vacancies).  Hours worked in the manufacturing sector (AWHMAN) is also detrended.  Both are rescaled as z-scores.  The log operation for HWIx  helps with visualization by squashing down the "extreme" post-2020 values.  The kink behavior reported in the main text is still visible without it.  Detrending puts the focus on cycles by taking out a mild trend in both which is implicitly partialized out in HNN anyway through the long-run component.  In both the expectations and real activity cases, the most recent lag available (i.e.,  at time $t$) for each indicator is used in the analysis. 

\subsection{HNN as an Evaluator of Empirically Ambiguous Theories}\label{sec:ext}

With typical PC regressions often being only mildly supported by the data, there has been a business of proposing augmented PCs.  Often times,  the newly proposed component is either suggested from formal theory or common sense economic arguments.  In both cases,  there can be a disconnect between what is in the database and what comes out of the theory, again compromising the proper evaluation of the potency of such augmentations.  By adding new hemispheres dedicated to the newcomers,  HNN can palliate this problem. 


Here is a short summary of algorithmic developments and corresponding results reported below.  Firstly,  \textbf{HNN-F-4NK} extends the latter PC's to include additional hemispheres for "credit conditions" and the central bank's balance sheet, as suggested in \cite{sims2019four}'s 4 equations NK model.  HNN-F-4NK reports that, as derived in \cite{sims2019four},  favorable credit conditions have a negative marginal impact \textit{conditional on other components}.  In sharp contrast,  a simpler approach with time-varying coefficients including the apparently suitable Chicago Fed National Financial Conditions Credit Subindex would suggest no such effect exists, or has the opposite sign.  The second extension,  \textbf{HNN-F-IKS}, creates, among other things, a supervised composite from a panel of international GDP growth data. It is found that, overall, and except for a few spikes (like some during the pandemic), the international "gap" has limited explanatory power for US inflation.  HNN-F-IKS also includes a kitchen sink hemisphere whose variable importance analysis reports extended use of complementary variables that are all forward-looking in nature -- in accord with theory suggesting inflation is an expected discounted stream of future marginal costs.

\subsubsection{An Investigation of the 4-Equation NK Model}\label{sec:4k}

\cite{sims2019four}  introduce a 4-equations New Keynesian model that skillfully blends the tractability (and the derivation of an explicit Phillips curve) of the canonical 3 equations model and relevance for analyzing the effects of quantitative easing (QE).  As a result of incorporating, among other things, financial intermediaries,  bonds of different terms, and credit market shocks,  their Phillips curve includes two additional variables beyond the usual gap and expectations: the real market value of the monetary authority's long-term bond portfolio and credit conditions. While the former is rather clearly defined in terms of observed variables, the latter needs to be proxied, and ambiguity reigns as to which financial market variable will adequately proxy for "credit conditions".  The HNN solution is now obvious: create a $\mathcal{H}$ with a myriad of indicators containing information on the health of credit markets.  

\begin{figure}[t!]
\begin{center} 
\includegraphics[trim={0.7cm 1cm 0cm 0cm},clip,width=\textwidth]{{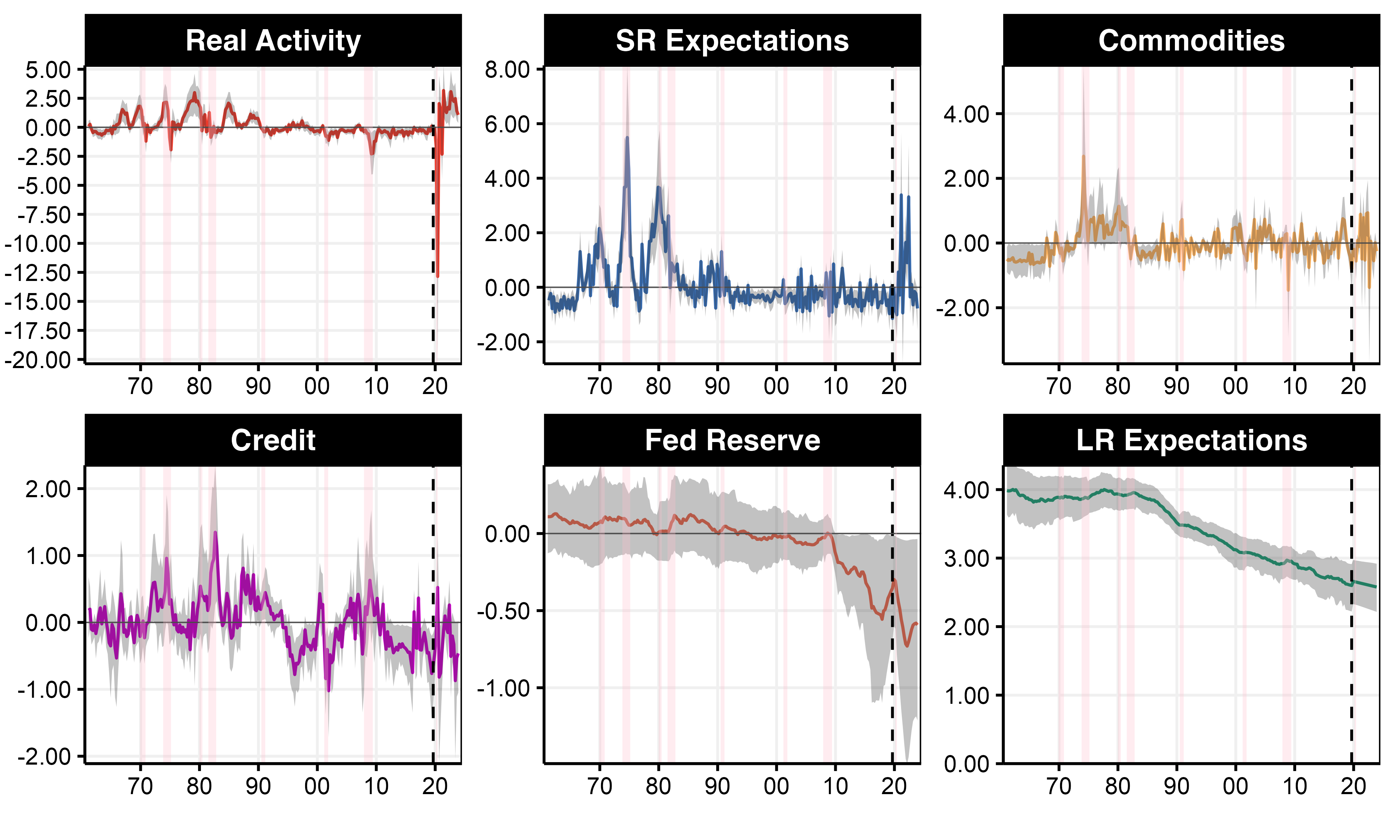}}
\caption{\footnotesize Contributions ($h_{t}$'s) from \textbf{HNN-F-4NK}.  Notes: Dashed line is the beginning of the out-of-sample.  NBER recessions are in pink shadowing.  Gray shading represents the 14-86\% quantiles from the out-of-bag ensemble. }\label{hs_2019_4nk}
\end{center}
\end{figure}

The expected signs for coefficients, as derived from theory, is that \textit{keeping the output gap fixed},  favorable credit conditions bring inflation down, and so does an expanding positive central bank balance sheet.  Those signs are obviously those of marginal effects, i.e.,  when controlling for the output gap.  In this section,  I augment HNN-F with two additional hemispheres inspired from \cite{sims2019four}'s model.  Then, results regarding the effect of credit conditions are compared to a much simpler model -- a PC regression with time-varying parameters that is augmented with oil prices,  the reserves of depository institutions (total and non-borrowed),  and, most importantly, the Chicago Fed National Financial Conditions Credit Subindex. 



\begin{figure}[ht!]
\begin{center} 
\hspace*{-0.25cm}\includegraphics[trim={0cm 0cm 0cm 0cm},clip,width=0.83\textwidth]{{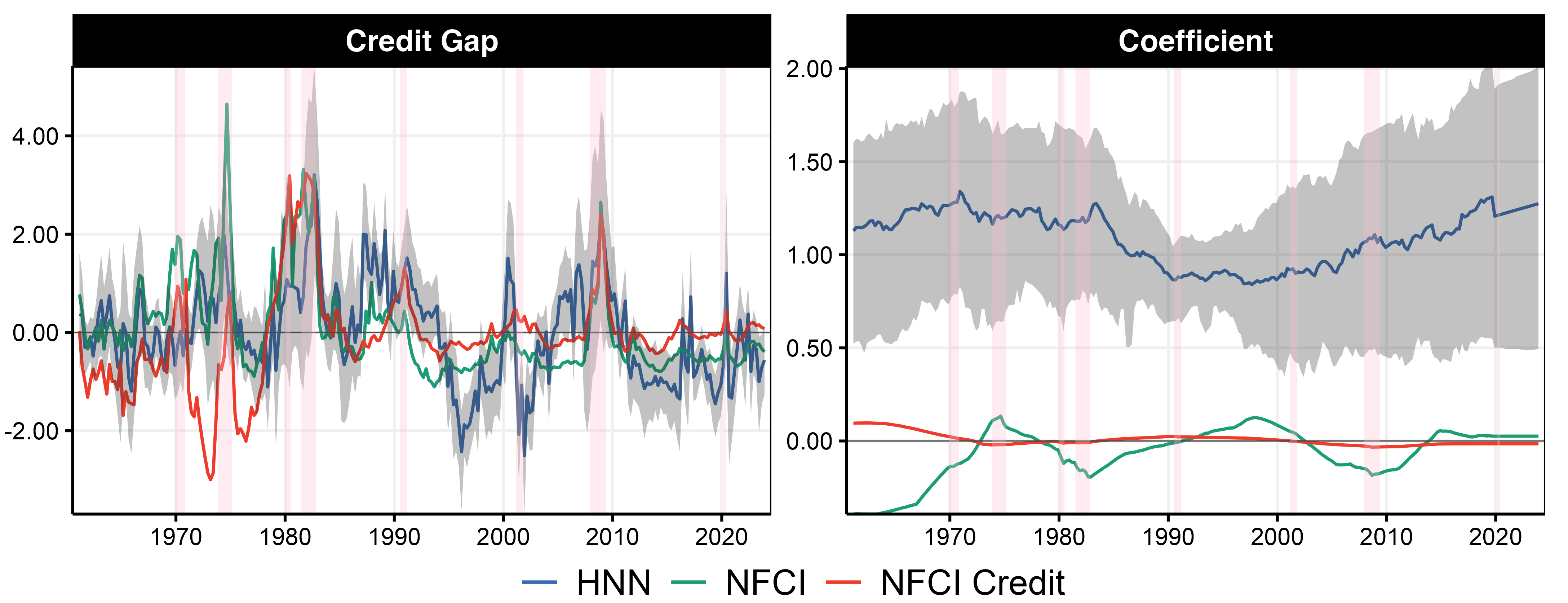}}
\caption{\footnotesize HNN-F-4NK's "credit conditions" and associated time-varying coefficient.  Notes: Dashed line is the beginning of the out-of-sample.  NBER recessions are in pink shadowing.  Gray shading represents the 14-86\% quantiles from the out-of-bag ensemble. NFCI indices prior to 1971 are patched using \cite{SW2002}'s EM algorithm applied on the whole FRED-QD dataset.  Coefficients for the last two specifications are obtained from time-varying parameter PC regressions (see Figure \ref{cbo_compa} notes for details) augmented with the Fed's balance sheet each credit conditions index in turns.}\label{4nk_compare}
\vspace*{-0.5cm}
\end{center}
\end{figure}

Figure \ref{hs_2019_4nk} reports, among other things, the contribution of credit conditions and the Fed expanding balance sheet to $\pi_{t+1}$ as estimated from HNN-F.   The four original components are largely unchanged, mainly because the additional two are of limited relative importance.   Figure \ref{4nk_compare} reports results from augmented PC regressions with time-varying parameters. The NFCI is found to have a negligible impact on $\pi_{t+1}$, whereas the credit index created endogenously by HNN-F from the credit group of variables in FRED-QD (see \cite{mccracken2020fred} for the complete list) has an appreciable effect during certain historical episodes.  For instance,  there is mild upward pressure on prices due to tightening credit conditions before and after the early 1990s recession, as well in running up to the GR.  Also, loose credit conditions and an ever-expanding Fed balance sheet are credited for very light (direct,  not indirectly through the gap) downward pressure on prices during the mid 2010s.  This is, obviously,  the direct marginal effect, keeping the gap fixed.

In Figure \ref{4nk_compare}, the HNN credit conditions index shares some peaks and troughs with NFCI-Credit and mostly overall NFCI, but, all in all, they are only mildly correlated.  As a result,  compared to a more traditional test of the 4-NK model,  we get a much larger (and correctly signed\footnote{This cannot be assessed by looking at the coefficient since it is forced to be positive for identification purposes.  Rather, the statement comes from observing that HNN's index is positively correlated with a known measure of credit stress and that its ups and down are consistent with the kind of cyclical variation we expect from it.}) coefficient for credit conditions in HNN.  This is explained by HNN's index being active during certain periods while either NFCI-Credit is essentially flat (from the early 1980s on, excluding the GR) or has the opposite sign (for almost all of the 1970s).   Thus,  unlike classical methods,  HNN finds a mild positive contribution of tightening credit conditions from the mid 1980s until the early 1990s, an era punctuated by the 1987 stock market crash and a general credit slowdown from 1989-1992.  Additionally,  HNN finds easy credit conditions from 1995 until 2005, with the exception of a small peak following the collapse of the Dotcom bubble.  Overall,  the credit conditions index created by "inflation supervision" is suggestive of a much less persistent behavior and much more action during the Great Moderation than what can be seen from the NFCI-Credit.  Finally,  the coefficient on credit index is found to be declining exogenously through time starting from 1980s but then experiencing a revival in the 2010s.  However,  there is wide uncertainty surrounding the coefficient estimates of the last decade. 


From a methodological standpoint, the takeaway message is the following.  If one chooses the NFCI-Credit, arguably a very legitimate proxy for credit conditions as they enter \cite{sims2019four}'s PC, literally no empirical support is found for the new model.  In contrast, HNN, by constructing a credit index supervised by $\pi_{t+1}$,  finds some evidence for the PC as derived by \cite{sims2019four}. This contribution of credit conditions -- albeit light when compared to that of the original four components -- is nontrivial. The same cannot be said of the Fed's reserves, which have a limited direct effect on $\pi_{t+1}$.  But this could be due to the limited length of the "QE sample".

\subsubsection{Adding an International Component and a Kitchen Sink}\label{sec:int}

The connectedness of the world economy suggests inflation can be influenced by non-domestic factors, like the vigor of the trading partners' economy.  There is cross-sectional and time series evidence on the matter \citep{borio2007globalisation,laseen2016did,bobeica2019missing} and it is not infrequent to see proxies of international economic or inflation conditions enter PC regressions \citep{blanchard2015inflation}.  The 2021 inflation experience,  with many countries reporting historically high YoY inflation rates simultaneously,  does not negate the importance of a global component either.  As always,  the question is how to properly construct a global measure of slack that may or may not influence US inflation, when controlling for its own gap.

To create a global gap (excluding the US),  I construct a hemisphere where the inputs are quarterly GDP growth rates from 1970 for OECD members and potential member states, which data is available \href{https://stats.oecd.org/index.aspx?queryid=350}{here}.  Country aggregations (like G7,  to avoid overlap with domestic variables) are excluded, and so are countries which data starts post-1960.\footnote{Exceptions are made for China and India.  China's data is replaced with that available from FRED that starts in the mid-1990s (with residual seasonality filtered out with dummies) and the interpolated yearly series of the World Bank is spliced in before that.  OECD data is kept for India post-1996 and interpolated yearly data from the World Bank is used prior to that. Transformations mentioned in Section \ref{data} are carried out with the new data.} 


Since this specification is not motivated from any tight theory (as it was the case in Section \ref{sec:4k}), I also indulge in adding a Kitchen Sink hemisphere, which, as the name suggests, will include all the variables in FRED-QD that are not already included in our four benchmark hemispheres.  This will provide yet  another robustness check on the path estimated for key components like $g_t$ and ${\mathcal{E}_t^\text{SR}}$. This can also point, via the VI analysis, to variables that could eventually deserve their own hemispheres in extensions of this work.  The resulting specification is referred to as \textbf{HNN-F-IKS}, namely,  HNN-F with an \textbf{i}nternational component and a \textbf{k}itchen \textbf{s}ink.

\begin{figure}[t!]
\begin{center} 
\hspace{-0.5em}\includegraphics[trim={0.3cm 1cm 0cm 0cm},clip,width=\textwidth]{{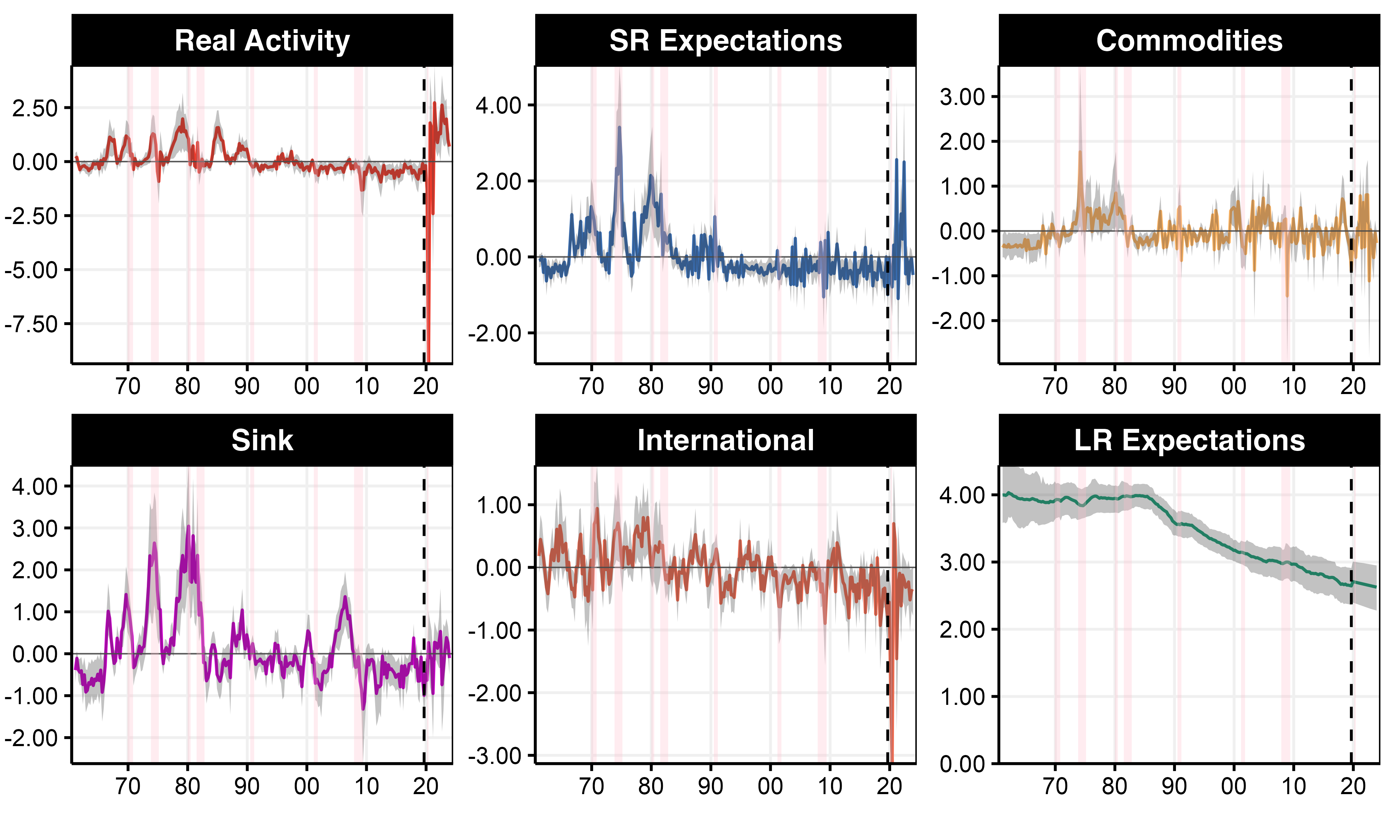}}
\caption{\footnotesize Contributions ($h_{t}$'s) from \textbf{HNN-F-IKS}.  Notes: Dashed line is the beginning of the out-of-sample.  NBER recessions are in pink shadowing.  Gray shading represents the 14-86\% quantiles from the out-of-bag ensemble. }\label{hs_2019_iks}
\end{center}
\end{figure}

The international output gap seems to be of limited importance compared to other components ---  its contribution is typically contained within the -0.5 to 0.5 range and the bands often times include 0.  This statement, of course,  does not apply to the pandemic era where massive swings similar to those of $g_t$ are observed.  Notable recent episodes are a flash negative contribution circa the GR and a gently negative one in the mid-2010s, corresponding to the missing inflation period.\footnote{\cite{laseen2016did} also report on the informativeness of external factors for the 2008-2015 period in a conditional BVAR exercise.  However, results from HNN, which dispenses with many assumptions from BVAR and related methods (but comes with its own,  in all fairness),  points out this effect to be mild. }

In Figure \ref{hs_2019_iks},  it is found that $g_t$ and ${\mathcal{E}_t^\text{SR}}$ are qualitatively unchanged,  but one can notice an overall weakened effect (with respect to Figure \ref{hs_2019}) of both components especially in the 1970s.  The major reason for that last observation is arguably the commanding presence of the kitchen sink, which contribution entertains some important highs in the 1970s, as well as three intriguing bumps before the 1990, 2001, and 2008 recessions.  Importantly,  it is worth remembering that its very inclusion changes the definition of $g_t$ and ${\mathcal{E}_t^\text{SR}}$ as per the network structure.  Thus,  their reported dampening should be taken with a grain of salt.

\begin{figure} 
\centering
\vspace{-0.45cm}
\hspace{-0.25cm}\includegraphics[trim={0cm 1.5cm 0.2cm 0cm},clip,width=0.53\textwidth]{{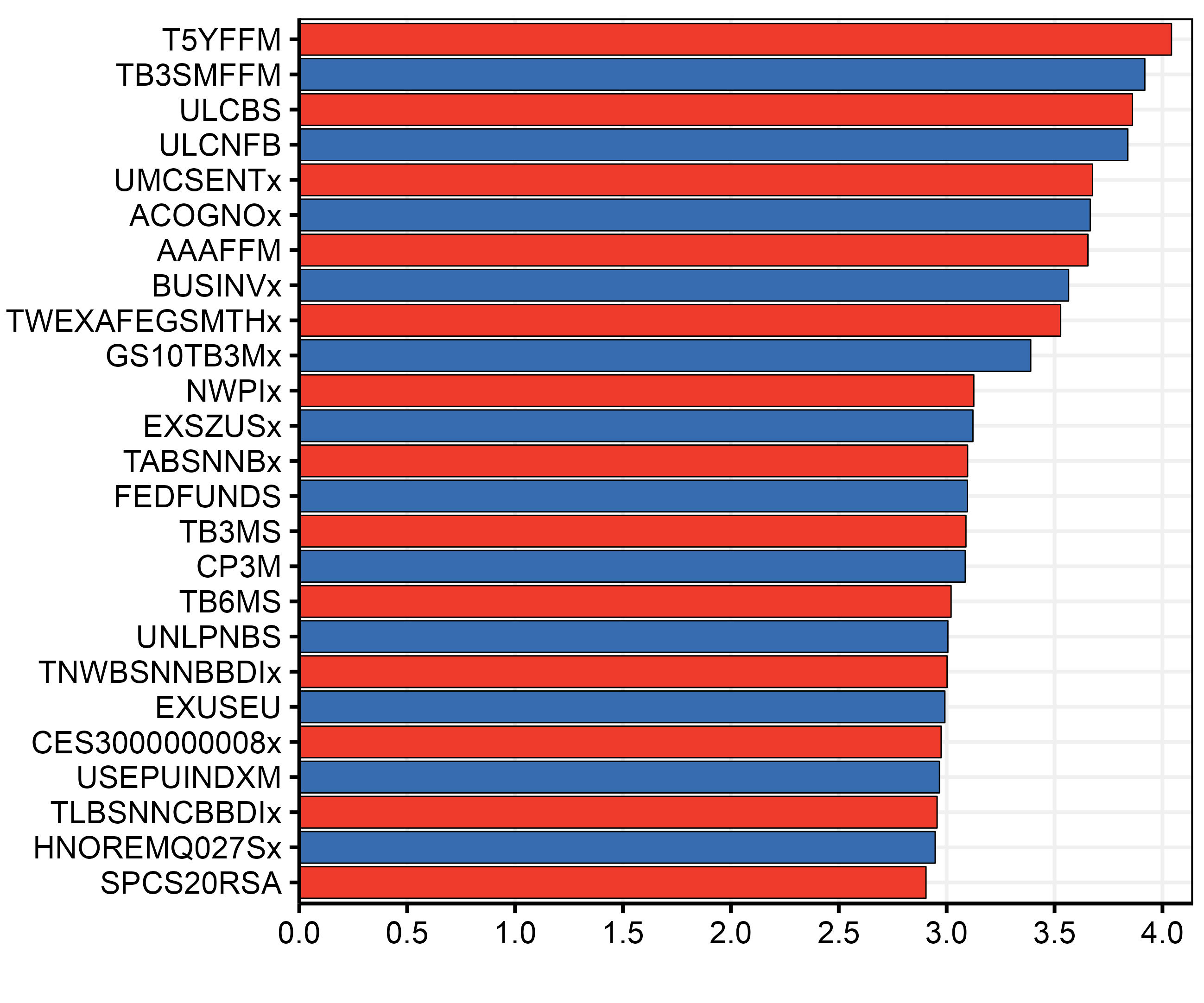}}
  \caption{\footnotesize VI Results for the  Kitchen Sink component of HNN-F-IKS specification, with training ending in 2019Q4.  Mnemonics are those of FRED-QD.}\label{vi_sink}
  \vspace{-0.45cm}
\end{figure}

Nonetheless, understanding what is in the sink could clearly prove valuable. Figure \ref{vi_sink}, by reporting the VI for the sink component, strongly suggests that information about future economic outcomes is key: the great majority of indicators populating the top 10 are considered leading indicators. First, we have spreads, which are market-based forward-looking variables (5-year treasury bill minus federal funds rate (FFR), 3-month commercial paper minus FFR, 3-month treasury bill minus FFR). The literature documenting the potency of spreads as predictors of business cycle turning points is vast \citep{stock1989new, estrella1998predicting}. Their link to inflation seems thinner \citep{stock2008phillips} in linear PC regressions, but \cite{MRFjae} finds that this link appears to be highly nonlinear (using a newly developed Random Forest approach). HNN can also deal with the necessary nonlinearities. Second, the consumer survey variant of such expectations comes up in the 5th position (\texttt{UMCSENTx}).

Overall, there is a clear push from forward-looking variables during the periods that precede economic downturns. This large weight accorded to variables inputting information about future real activity conditions and/or future prices is not surprising, as the latter is directly related to expectations about future marginal costs (and so are unit labor costs\footnote{In fact, in a well-known paper, \cite{gali1999inflation} showed that proxying for marginal costs directly with the labor share gives a significant Phillips curve slope coefficient, whereas using some form of output gap does not. }, entering at positions 3 and 4 in Figure \ref{vi_sink}) -- solving forward the NKPC yields that $\pi_t$ is a function of expected future marginal costs.\footnote{In unreported results, unit labor costs were included in the baseline HNN-F specification, which is a legitimate enterprise in itself if we wish to extract ${\text{mc}}_t$ directly rather than $g_t$. The estimates of the gap (or ${\text{mc}}_t$) did not budge, but unit labor costs ranked highly in VI. This suggests that, while unit labor costs carry pertinent information, it was already proxied for by a nonlinear combination of real activity variables  contained in $\mathcal{H}_g$.} The empirical importance of considering forward-looking expectations about the marginal costs has been highlighted before, mostly from a structural model perspective \citep{del2015inflation}. Nonetheless, it is worth remembering that VI results for the kitchen sink are more dispersed than those of Figure \ref{vi_bench}, and it is clear that a plethora of regressors contribute to the component beyond those at the top.




\subsection{Additional Figures and Tables}\label{addifig}

  \begin{figure}[h!]
\begin{center} 
\hspace*{-0.05cm}\includegraphics[trim={0cm 1.75cm 0cm 0cm},clip,width=0.8\textwidth]{{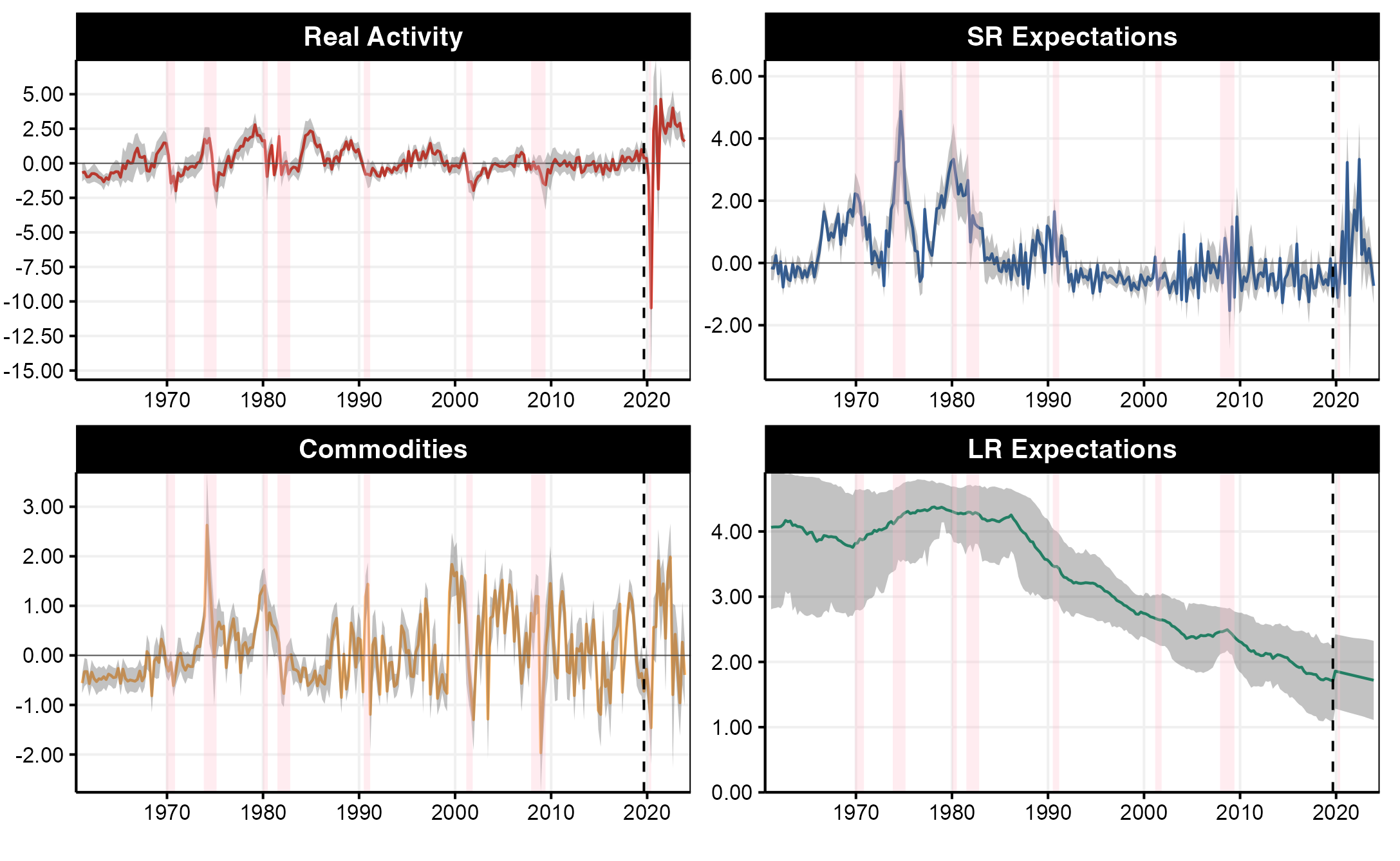}}
\caption{\footnotesize Contributions ($h_{t}$'s) from \textbf{HNN}.  Notes: Dashed line is the beginning of the out-of-sample.  NBER recessions are in pink shadowing.  Gray shading represents the 14-86\% quantiles from the out-of-bag ensemble. }\label{hs_2019_hnn}
\end{center}
\end{figure} 
     \vspace{-0.8cm}                        
  \begin{figure}[h!]
\begin{center} 
\hspace*{-0.05cm}\includegraphics[trim={0cm 1.75cm 0cm 0cm},clip,width=0.8\textwidth]{{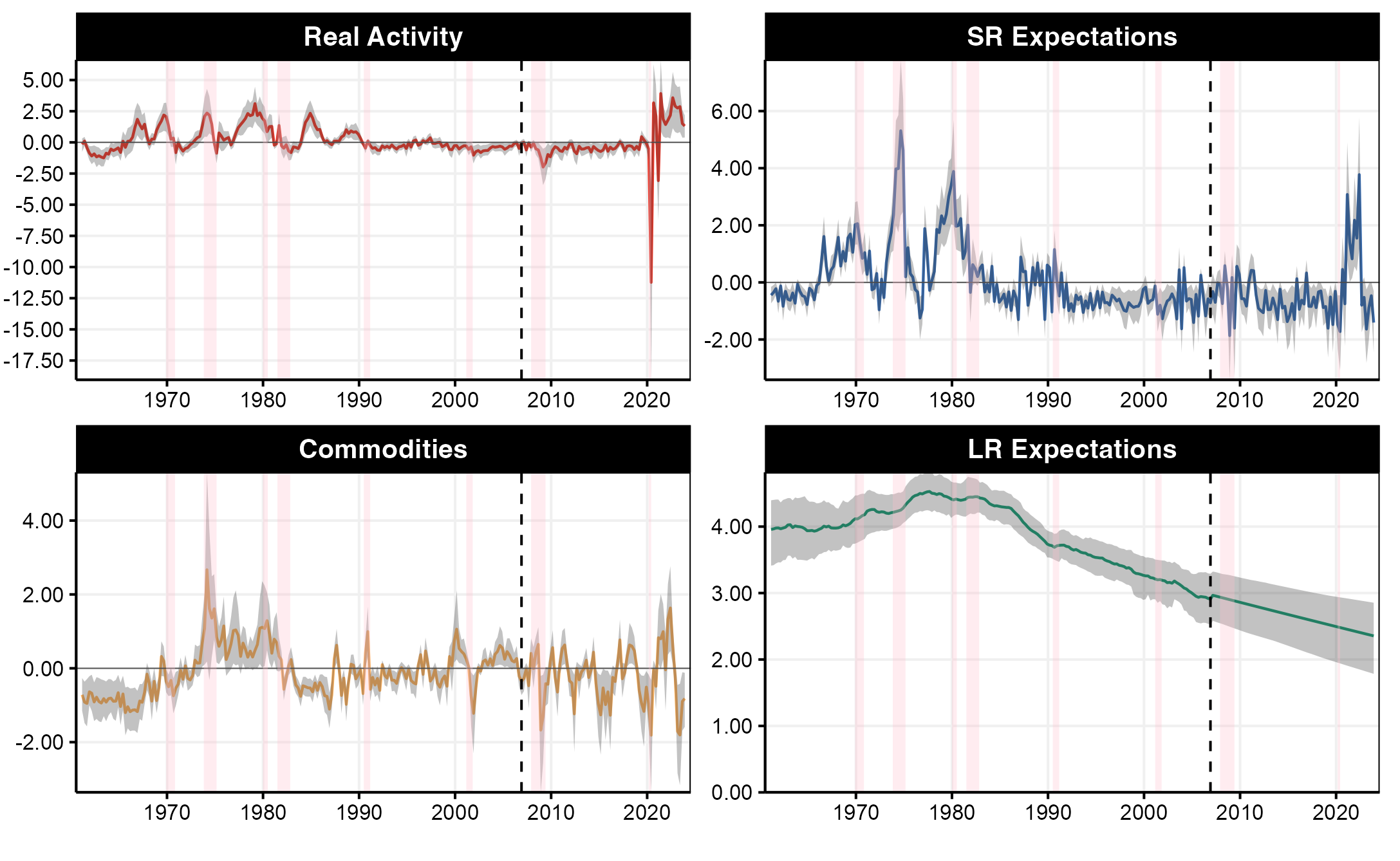}}
\caption{\footnotesize Contributions ($h_{t}$'s) from HNN-F with training ending in 2007.  Notes: Dashed line is the beginning of the out-of-sample.  NBER recessions are in pink shadowing.  Gray shading represents the 14-86\% quantiles from the out-of-bag ensemble. }\label{hs_2007}
\end{center}
\end{figure} 
     
\begin{figure}[h!]
\begin{center} 
\hspace*{-0.05cm}\includegraphics[trim={0cm 1.65cm 0cm 0cm},clip,width=0.75\textwidth]{{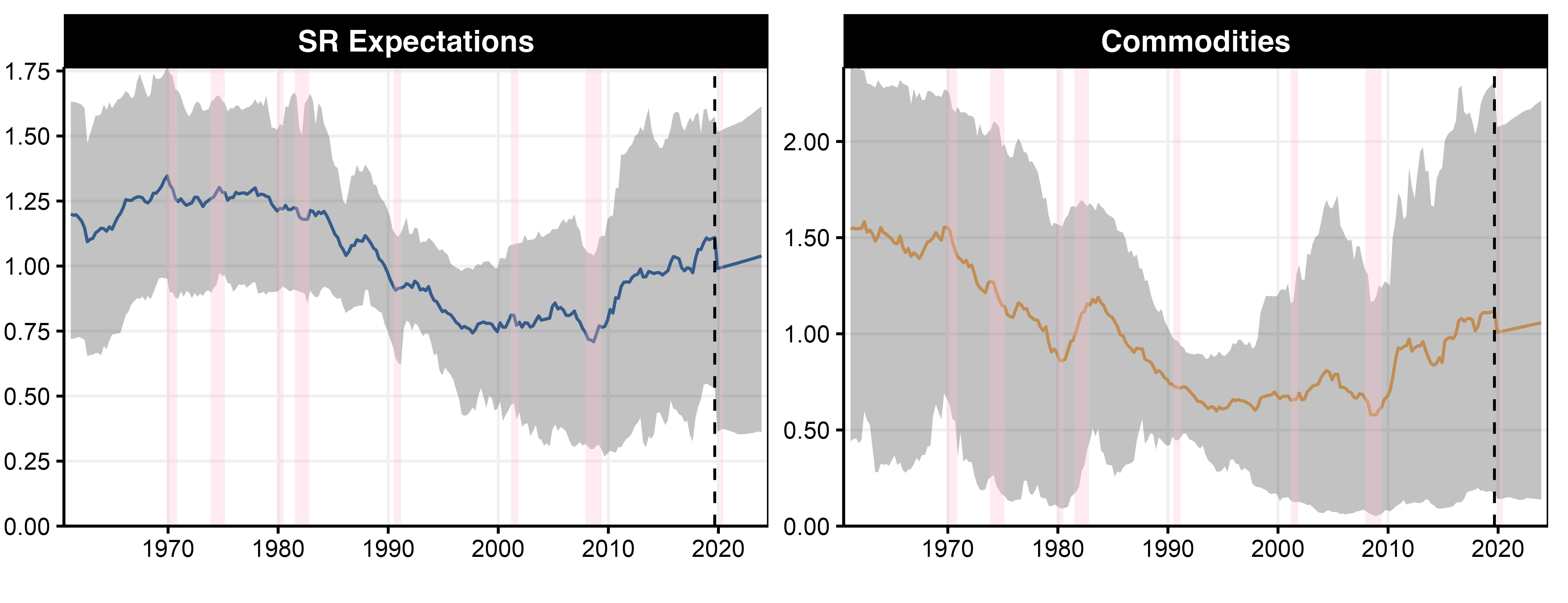
}}
\caption{\footnotesize HNN-F's remaining two time-varying coefficients.  Notes: Dashed line is the beginning of the out-of-sample.  NBER recessions are in pink shadowing.  Gray shading represents the 14-86\% quantiles from the out-of-bag ensemble. }\label{leftover_coefs}
\end{center}
\end{figure} 

\begin{figure}[h!]
\begin{center} 
\hspace*{-0.05cm}\includegraphics[trim={0cm 0cm 0cm 0cm},clip,width=0.65\textwidth]{{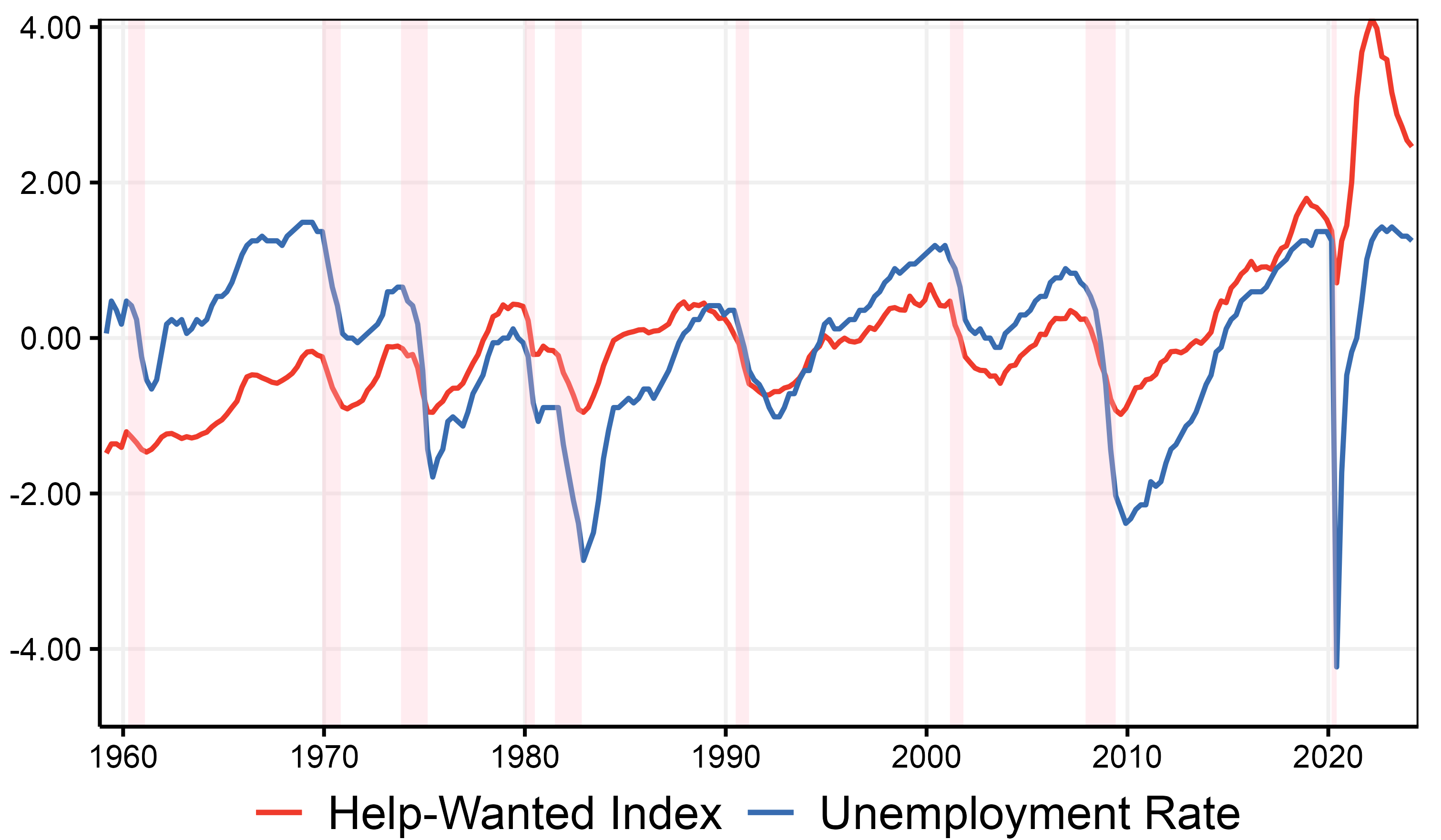
}}
\caption{\footnotesize Compared (standardized) measures of slack.  Unemployment was multiplied by -1. }\label{ur_hwi}
\end{center}
\end{figure} 

\begin{table}[h!]
\center
\scriptsize
\footnotesize
\setlength{\tabcolsep}{0.2em}
 \caption{\normalsize Diebold-Mariano Test P-values for Figure \ref{forecast_fig}'s Forecasts versus AR(4)} \label{dmtest}
\begin{tabular}{l|llllll}
\addlinespace[2pt]
\toprule \toprule
\addlinespace[4pt]
\multicolumn{7}{c}{CPI ($s=1$) } \\
\cmidrule(lr){1-7}\addlinespace[2pt]
             & HNN   & HNN-F & DNN   & RF    & CKP   & PC+ \\
             \addlinespace[2pt]
             \toprule
All          & 0.60 & 0.67 & 0.36 & 0.65 & 0.62 & 0.64 \\
Before 2020  & 0.67 & 0.11 & 0.13 & 0.35 & 0.13 & 1.00 \\
Without 2020 & 0.19 & 0.04 & 0.81 & 0.18 & 0.17 & 0.63 \\
Post 2020    & 0.01 & 0.01 & 0.03 & 0.14 & 0.50 & 0.02  \\
\bottomrule
\addlinespace[4pt]
\multicolumn{7}{c}{CPI Core ($s=1$) } \\
\cmidrule(lr){1-7}
             & HNN   & HNN-F & DNN   & RF    & CKP   & PC+ \\
\addlinespace[2pt]
             \toprule
All          & 0.36 & 0.24 & 0.08 & 0.11 & 0.20 & 0.27 \\
Before 2020  & 0.05 & 0.01 & 0.08 & 0.30 & 0.69 & 0.03 \\
Without 2020 & 0.72 & 0.28 & 0.06 & 0.37 & 0.08 & 0.10 \\
Post 2020    & 0.17 & 0.93 & 0.08 & 0.19 & 0.14 & 0.74  \\
\bottomrule
\addlinespace[4pt]
\multicolumn{7}{c}{CPI ($s=4$) } \\
\cmidrule(lr){1-7}
             & HNN   & HNN-F & DNN   & RF    & CKP   & PC+ \\
\addlinespace[2pt]
             \toprule
All          & 0.92 & 0.41 & 0.41 & 0.58 & 0.31 & 0.45 \\
Before 2020  & 0.21 & 0.03 & 0.30 & 0.34 & 0.69 & 0.20 \\
Without 2020 & 0.65 & 0.84 & 0.03 & 0.64 & 0.64 & 0.43 \\
Post 2020    & 0.71 & 0.95 & 0.01 & 0.39 & 0.12 & 0.37  \\
\bottomrule \bottomrule
\end{tabular}
\end{table}

\vspace{1cm}
\hspace*{-1.25cm}
\begin{minipage}{1.05\linewidth}
		
\begin{minipage}{0.5\linewidth}
	\centering
\small
\hspace*{0.25cm}
	\begin{threeparttable}
	\captionof{table}{Raw RMSEs from 2018Q1 to 2024Q1} \label{tab:rt1}
 \begin{tabular}{l| c c c c} 
\toprule \toprule
\addlinespace[1pt]
& &  HNN-F &  & HNN-F,  Real-Time  \\
\midrule
\addlinespace[5pt] 

All & & 1.40  &  & 1.49   \\ \addlinespace[2pt]
Before 2020 & &  1.50  &  & 1.28    \\ \addlinespace[2pt]
Without 2020 & & 0.78  &  & 0.75  \\ \addlinespace[2pt]
Post 2020 & & 0.70  &  & 0.70    \\ \addlinespace[2pt]

\bottomrule \bottomrule
	\end{tabular}
\begin{tablenotes}[para,flushleft]
	\scriptsize 
	\end{tablenotes}
\end{threeparttable}
 \end{minipage}\hfill
 \begin{minipage}{0.5\linewidth}
 	\centering
 	\captionof{figure}{Forecasts In Real-Time}\label{tab:rt2}
	\hspace*{-0.5cm}	\includegraphics[width=0.98\textwidth, trim = 0mm 0mm 0mm 2mm, clip]{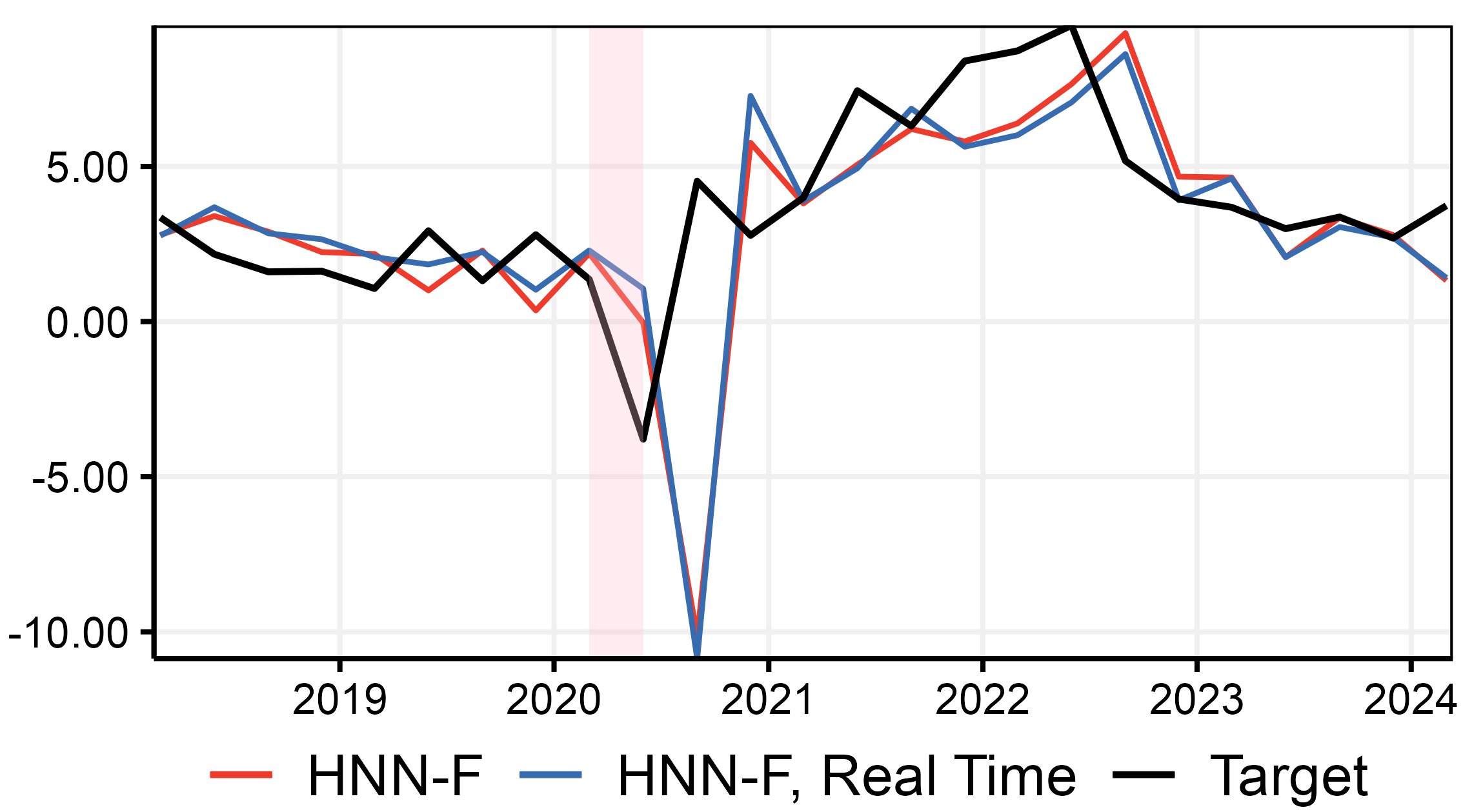}
		
 \end{minipage}          
\end{minipage}\ \\
\vspace*{0.25cm}

\newpage

\subsection{Reliability and Sensitivity to Re-estimation} \label{app:rewrite}

\begin{figure}[ht!]
\begin{center} 
\includegraphics[width=0.8\textwidth]{{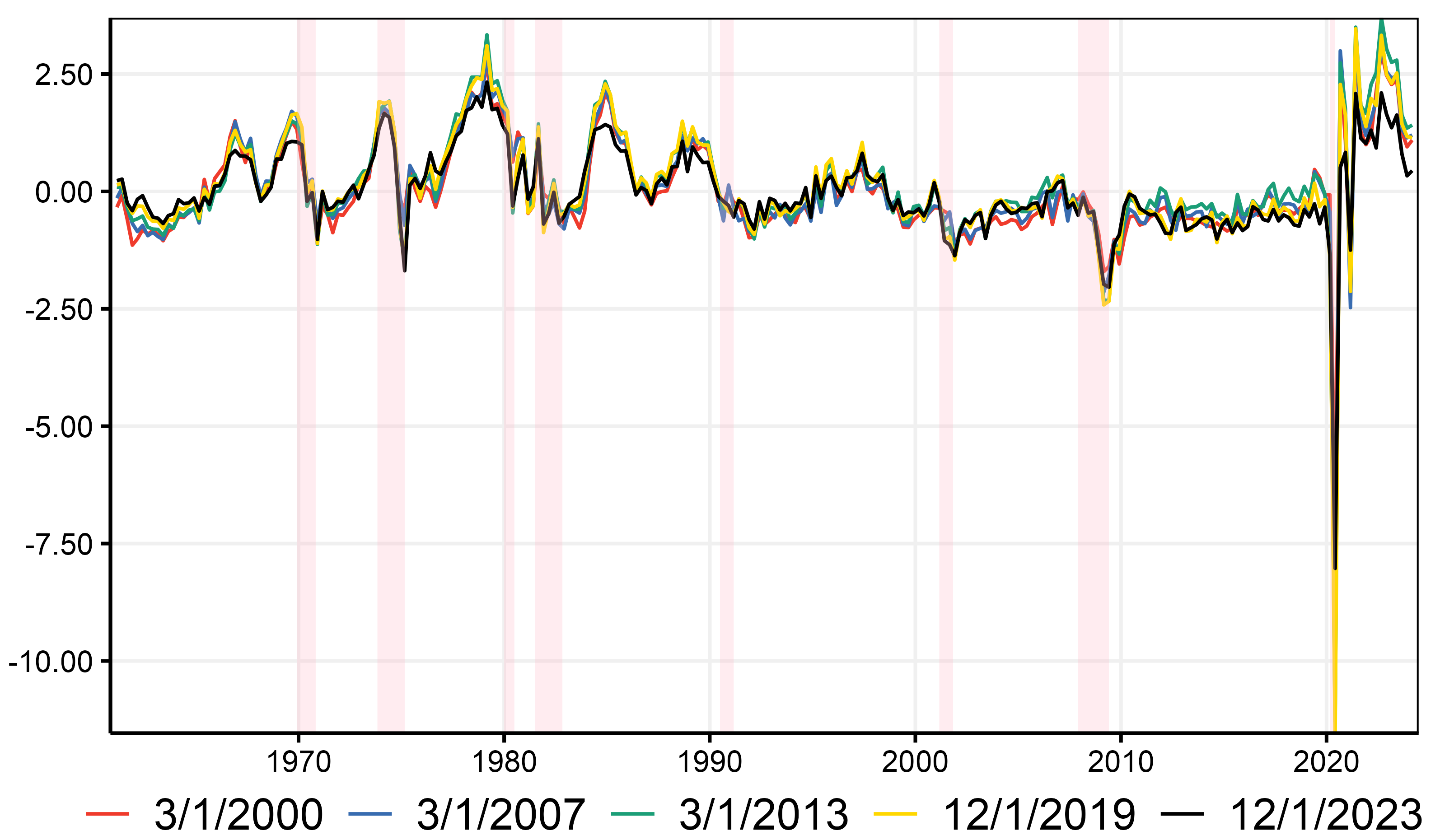}}
\caption{\small  $g_t$'s estimated on various training sets. }\label{g_revisions}  
\end{center}
\end{figure}

A relevant statistical question is whether HNN could be prone to rewriting history --- because many of the gap estimation methods based on plain filtering are \citep{orphanides2002unreliability, guay2005hodrick}. Figure \ref{g_revisions} suggests that HNN-F's estimation of $g_t$ is rather stable, with the qualitative patterns observed in Figure \ref{og_coef} being completely intact. There are some mild quantitative disagreements between the 2000 version and the remaining four, especially for the positive $h_{t,g}$ preceding the crisis. As for the aftermath of the crisis and the 2010s, there are some mild quantitative disagreements, but the pattern -- strikingly different from those of traditional methods -- is the same across specifications. That is, we get a major but short-lived dip following the crisis, a brief comeback to 0, then a long mildly negative phase up until 2018. All estimations agree on economic pressures on inflation increasing from the mid-2010s up until the Pandemic, with a slight disagreement on the overall level of $g_t$. 

Historical results are robust to the inclusion/exclusion of extreme pandemic observations, and $g_t$ movements are rather similar whether they are projected out-of-sample from 2019 or using all the data up until today. The quantitative discrepancy between the 2019 and the full-sample versions is obviously larger during 2020, but so is estimation uncertainty.  For the last estimation,  2020 data is excluded from estimation for obvious reasons \citep{lenza2020estimate, schorfheide2020real}, but the early 2021 data is kept in and has the effect of dampening the gap's movement in the last 2 years. Nonetheless, the re-estimated gap is still within the bands of its former self. Overall, results with training ending in 2019Q4 allow to evaluate whether a statistical model that has not seen 2021-2023 could capture well the inflation surge.


\subsection{Robustness Checks}\label{sec:rob}

HNN might be a very flexible nonparametric model, but it still features various compositional assumptions that can influence forecasts and the interpretation of the resulting components. In this section, we conduct many one-shot deviations from the baseline HNN specification to see, among other things, whether the key finding that the effect of real activity is stronger than what is captured by standard Phillips curves when allowing for nonlinearities still holds.

\begin{figure}[h!]
\begin{center} 
\hspace*{-0.05cm}\includegraphics[trim={0cm 0cm 0cm 0cm},clip,width=0.85\textwidth]{{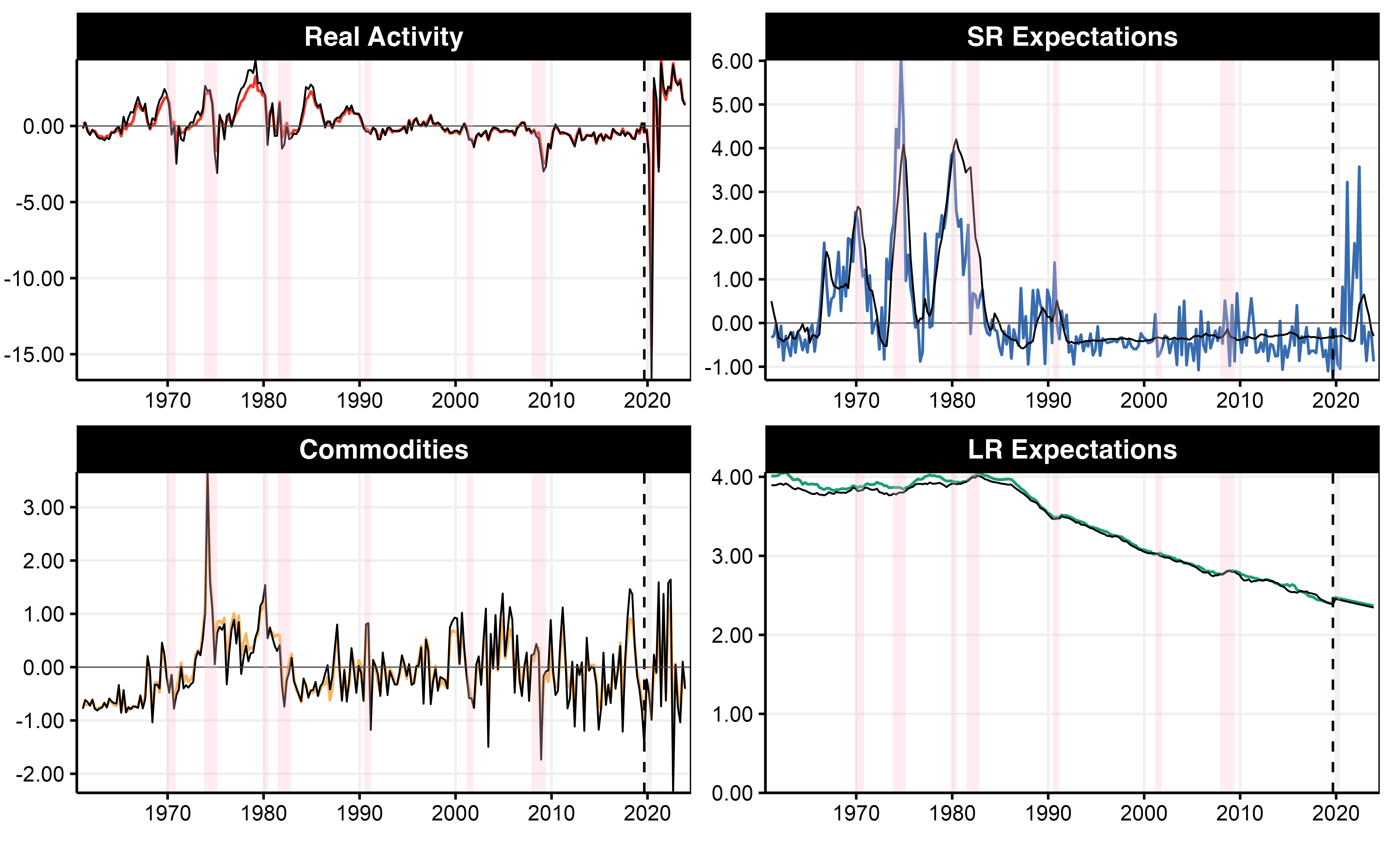
}}
\caption{\footnotesize Contributions ($h_{t}$'s) from baseline HNN (in colors) and alternative specification (in black) using only survey information in the short-run expectations hemisphere.  Notes: Dashed line is the beginning of the out-of-sample.  NBER recessions are in pink shadowing.  }\label{expect}
\end{center}
\end{figure}

{\noindent \sc \textbf{Tighter Specification of Expectations. }} The baseline HNN's short-run expectations hemisphere features lags of CPI as well as lags from many price sub-indices featured in the FRED-QD database. This can blur the line between the role of real activity and that of expectations, and possibly lead to an overstatement of the role of real activity in driving inflation. The cleanest possible separation is to use data that is already defined as a (possibly noisy) measurement of true expectations from actual economic agents. In Figure \ref{expect}, we report results when taking out all the lagged price data from the short-run expectations hemisphere.

We see that this removes high-frequency movements in the SR hemisphere but leaves its overall path intact, except for the post-2020 data. This is not entirely surprising given that price sub-indices feature a sizable amount of high-frequency variation, whereas survey expectations mostly do not. In that specification, the effect of short-run expectations is very quaint from the mid-1990s onward. One can also notice that the blue line (full specification) seems to lead the black one during Volcker's disinflation and during the Pandemic. Therefore, price sub-indices appear to feature relevant forward-looking signals about future inflation that survey-based expectations may lack. \textit{Most importantly},  the real activity hemisphere results are arguably unchanged. Thus, having more cleanly defined short-run expectations does not alter the core finding on the importance of real activity.

{\noindent \sc \textbf{Alternative Timing of the Target Variable.}}  While HNN's design is motivated as a generalization of \cite{stock2008phillips} forecasting PCs or approaches looking at dynamic effects of real activity on inflation \citep{del2020s}, the PC coming from New Keynesian theory (and utilized in, e.g., \cite{blanchard2015inflation} and \cite{coibion2015phillips}) features a contemporaneous target. This can also be done within HNN to evaluate whether the baseline forward-looking specification overstates the importance of real activity (or has a different shape) compared to the specification directly derived from theory.

\begin{figure}[h!]
\begin{center} 
\hspace*{-0.05cm}\includegraphics[trim={0cm 0cm 0cm 0cm},clip,width=0.85\textwidth]{{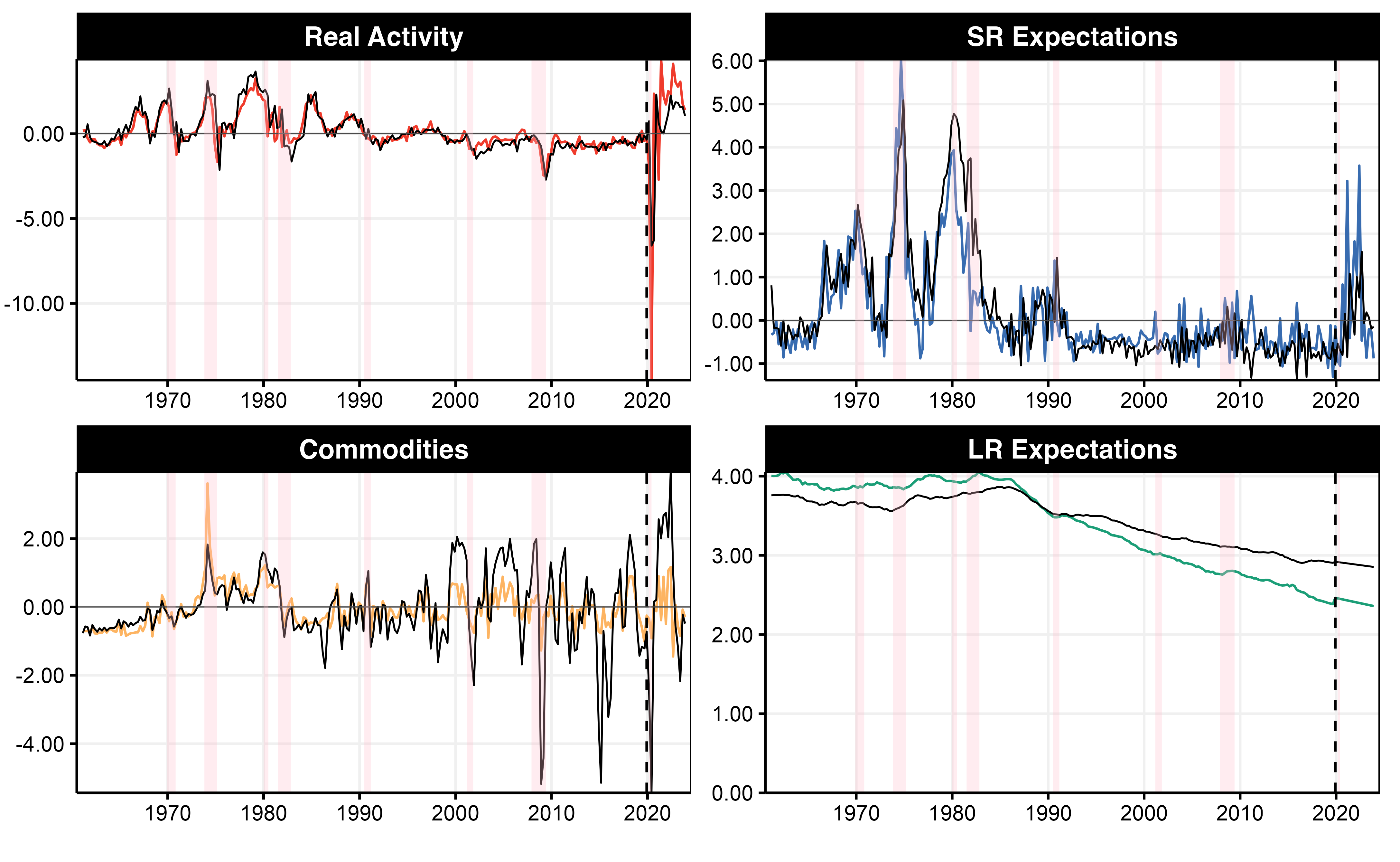
}}
\caption{\footnotesize Contributions ($h_{t}$'s) from baseline HNN (in colors) and alternative specification (in black) using a contemporaneous target ($s=0$) and only survey information and lags of $\pi_t$ in the short-run expectations hemisphere.   Notes: Dashed line is the beginning of the out-of-sample.  NBER recessions are in pink shadowing. }\label{compta}
\end{center}
\end{figure} 

Results are reported in Figure \ref{compta}. Timing of target and predictor variables has been set as theory would directly suggest. To get an even clearer identification of expectations and a mapping to classic specifications, only expectations survey data and lags of $\pi_t$ are utilized in the short-run expectations hemisphere. I find that key historical results about real activity are unchanged. More prolonged quantitative differences can be seen from 2001 to 2007, where the black line sits marginally below the red one for six years. In-sample results for real activity being mostly unchanged can be attributed to real activity having a slow and gradual effect on inflation, but also that the effect of commodities (which is inevitably magnified at $s=0$) has much less persistence. Thus, while $s=0$ vs $s=1$ can make a sizable difference for commodities, it is more muted for other hemispheres. For instance, we see how the black line now clearly captures the three large post-2020 downward oil price shocks which were arguably impossible to capture at $s=1$. By linearity of the final layer in HNN, this also means the implicit target onto which the remaining hemispheres are fitted is some kind of core inflation (that would keep food in the index).

We also see that $g_t$ is less forward-looking during the Pandemic by taking longer to firmly settle into positive territory and persistently generating values akin to those in the 1970-1980s. It is not extracted from a forecasting model anymore, a very unusual setup for machine learning regression with macroeconomic "big data," and as such may have a hard time generalizing well out-of-sample. This change can be partly attributed to the Commodities hemisphere getting a much more prominent role through its accounting relationship to CPI at $s=0$. During the Pandemic, the commodity price cycle's contribution nearly doubles in the baseline specification. From the perspective of obtaining a gap to forecast inflation developments for the next quarter or over the coming year, there is the natural question of which timing is optimal to extract the relevant latent states. By construction, if the objective is to be looking ahead, the baseline specification is the most natural setup, which distills non-persistent commodity price cycles and leaves relevant variation in $\pi_t$ for other states to be extracted.



{\noindent \sc \textbf{Redefining Long-run Expectations.}} The specification of long-run expectations in HNN is relatively simple and aligns with various papers in the time-varying parameters literature that define long-run objects (expectations or otherwise) and the time-varying intercept of the regression. HNN approximates this by featuring a nonparametric trend component defined as long-run expectations. Michigan Survey expectations (for the next 12 months) were allocated to the short-run expectations hemisphere in the baseline specification. However, it is worthwhile to ask what happens if those are instead included in the long-run hemisphere, either in replacement of or in combination with the time trend.

\begin{figure}[h!]
\begin{center} 
\hspace*{-0.05cm}\includegraphics[trim={0cm 0cm 0cm 0cm},clip,width=0.85\textwidth]{{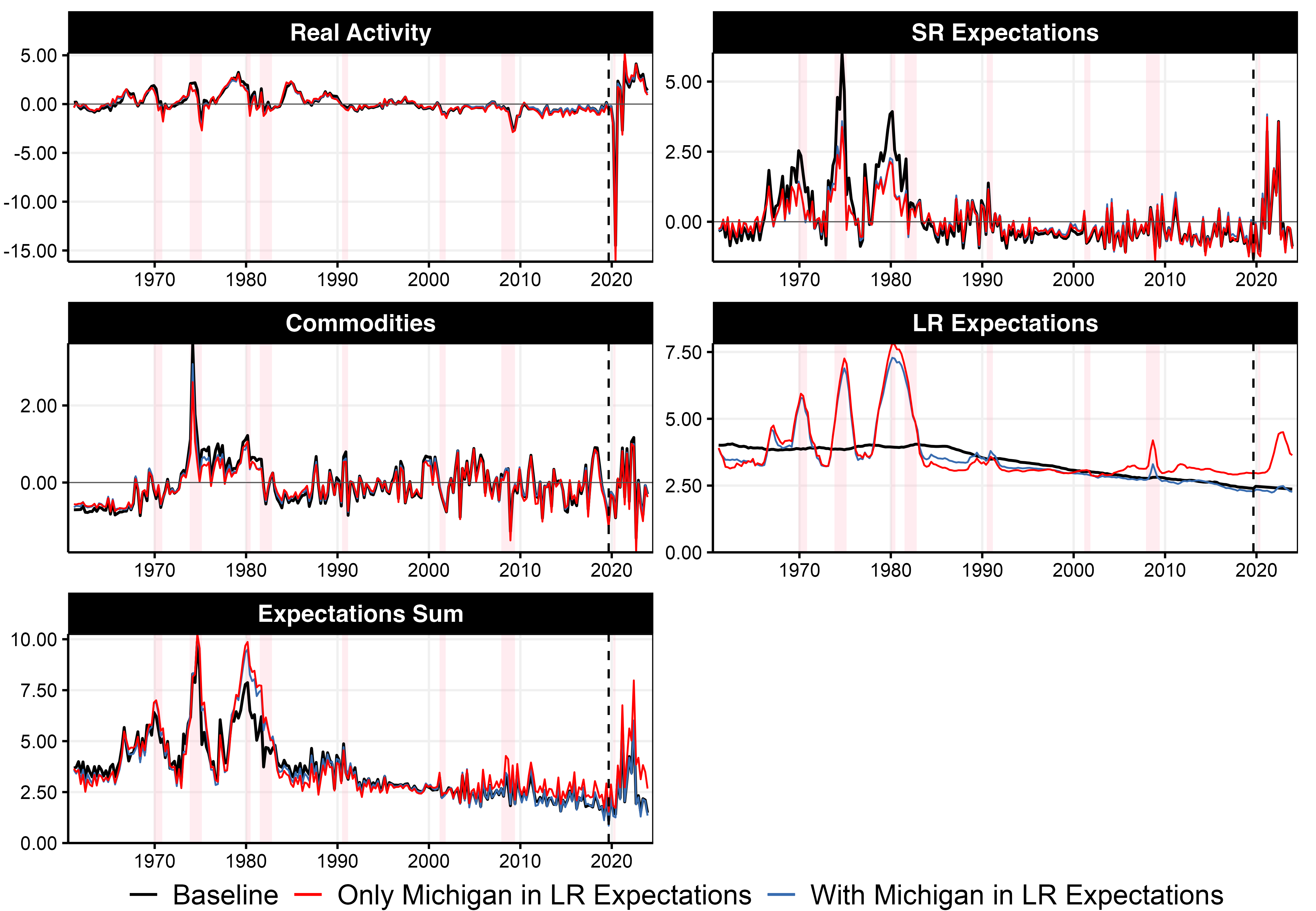}}
\caption{\footnotesize Contributions ($h_{t}$'s) from baseline HNN and alternative specifications using survey-based long-run inflation expectations data in the LR Expectations hemisphere.    Notes: Dashed line is the beginning of the out-of-sample.  NBER recessions are in pink shadowing. }\label{lrstuff}
\end{center}
\end{figure} 

Results are reported in Figure \ref{lrstuff}. We see that including the Michigan survey data in the LR hemisphere generates a reallocation in the 1970s of short-run expectations' role towards that of the long-run ones, as expected. This is true for both specifications, with or without the trend. However, the version without the trend seems unable to capture the lower LR values from 2000 onward. Bringing the trend back into the mix solves that issue as we see that the blue line overlaps almost perfectly with the black line from 2000 onward. Thus, when it comes to the separation of long-run versus short-run expectations effects on inflation, it depends on what one wishes to qualify as long-run or short-run, at least for the 1970s. The \textit{sum} of long-run and short-run expectations is what ultimately drives inflation. That sum is reported in the last panel of Figure \ref{lrstuff}. We see that all lines significantly overlap except in the late 1970s where the role of expectations is more substantial in the alternative specifications. A noticeable discrepancy is evident in the out-of-sample data, especially during the last two years. The "Michigan only" path diverges significantly, standing alone, whereas the two specifications that include the trend are in agreement. This underscores the importance of incorporating the trend in the long-run expectations hemisphere.  Most importantly, the path of real activity contribution remains almost completely unchanged for all specifications of long-run expectations as one would expect given that the sum of both expectations hemispheres remains somewhat stable across specifications.


\subsection{Changing Supervisors}\label{sec:altsup}

The deep output gap and associated results from HNN have been learned through supervision with headline inflation.  Changing supervisors could alter results.  For instance,  it has been often been documented that alternative measures of inflation -- typically stripped-down version of the CPI designed to be less volatile -- can deliver different results,  for instance,  about the strength of the PC \citep{ball2019nonpuzzling,stock2019slack,luciani2020common}.  Investigating such alternatives can be informative on the robustness of $g_t$ and the mechanics of HNN.  In this section, I report $g_t$'s and $\gamma_t$'s obtained from HNN-F with two alternative supervisors.   The first is Core CPI (headline minus food and energy).  The second is the  average inflation rate over the forthcoming year.  Tuning and architecture details remain intact from Section \ref{sec:archi}, except that dropout is turned off for these two less noisy targets.



\begin{figure}[h!]
\begin{center} 
\hspace*{-0.25cm}\includegraphics[trim={0cm 0cm 0cm 0cm},clip,width=0.83\textwidth]{{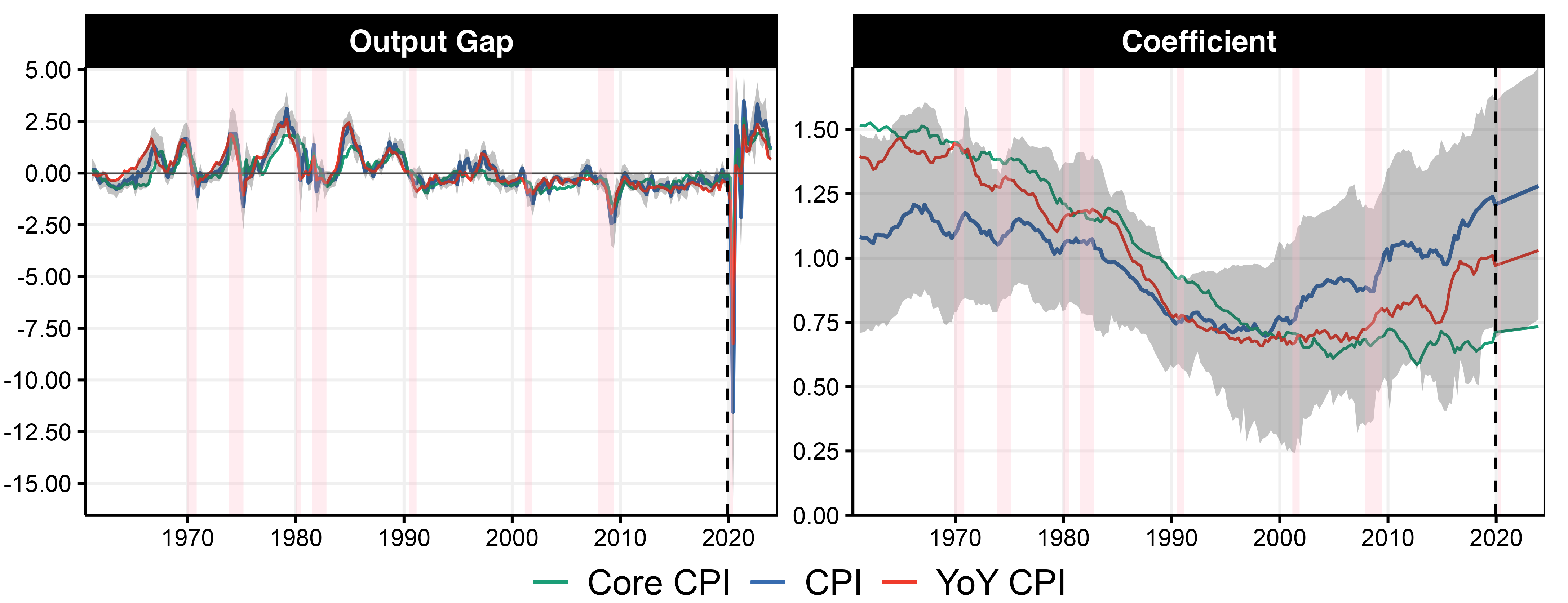}}
\caption{\footnotesize HNN-F gaps and associated coefficients with alternative supervisors.  Notes: Dashed line is the beginning of the out-of-sample.  NBER recessions are in pink shadowing.  Gray shading represents the 14-86\% quantiles from the out-of-bag ensemble. }\label{altsups}
\vspace*{-0.5cm}
\end{center}
\end{figure} 

Estimation results reported in Figure \ref{altsups} are suggestive that there is such a thing as a unique real activity latent state driving future inflation for various horizons.\footnote{A natural extension of this paper's framework would be to consider multiple targets -- namely all horizons from 1 to, say, 8 quarters -- and extract a common gap.} All ${g}_t$'s follow a clear common pattern, with that of yearly inflation showing larger amplitudes than one quarter ahead $\pi_t$ up to 1990, and Core CPI being slightly less. All gaps close relatively slowly following the early 2000s recession (with Core CPI's gap closing the slowest of the 3) and all close extremely fast following the GR. They also share a common arc-shaped mildly negative gap from 2011 to the onset of the COVID-19 era. During the Pandemic, the general pattern is again common to all three, but magnitudes seldom differ, in line with the wide uncertainty of the last two years. For instance, ${g}_t$ obtained from CPI (benchmark) is dissenting from those of CPI Core and YoY CPI by calling for a negative gap in 2021Q2. The other two ${g}_t$'s remain (very) positive from the end of 2020, which is the basis for their respective upward forecasts in 2021 reported below.

In terms of coefficients ($\gamma_t$'s), those are typically lower for both alternative supervisors, and so is their revival in the 2000s, with that of ${\gamma}_t^{\text{Core}}$ being practically non-existent. Credible regions suggest slack's contribution to inflation is similar for the CPI at $s=1$ and $s=4$. ${\gamma}_t^{\text{Core}}$'s overall level suggests a slightly lower passthrough from real activity to core inflation. Nevertheless, the main message from the previous section remains: there is a nonlinear measure of real activity which still impacts inflation greatly, and does so more significantly than classic gaps.

In Appendix \ref{sec:fedsup}, a more radical departure from the benchmark specification is conducted with the federal funds rate replacing inflation as supervisor. Accordingly, this last gap is extracted from a Taylor rule rather than a PC and will represent the $g_t$ the Fed "has in mind". Interestingly, the resulting gap looks more like a traditional filtered one, suggesting there may be a gap between the monetary authority's view of economic slack (in line with typical econometric estimates used by economists) and what can rationalize the inflation record. This also suggests that HNN's preference for a faster moving $g_t$ when modeling \textit{inflation} is not a mechanical result due to data transformation choices and other technicalities as the model can deliver smoother gaps if the dependent variable calls for it.

\begin{figure}[t!]
   \begin{subfigure}[b]{0.5\textwidth}\hspace{-0.25cm}\includegraphics[trim={0cm 0cm 0cm 0cm},clip,width=0.995\textwidth]{{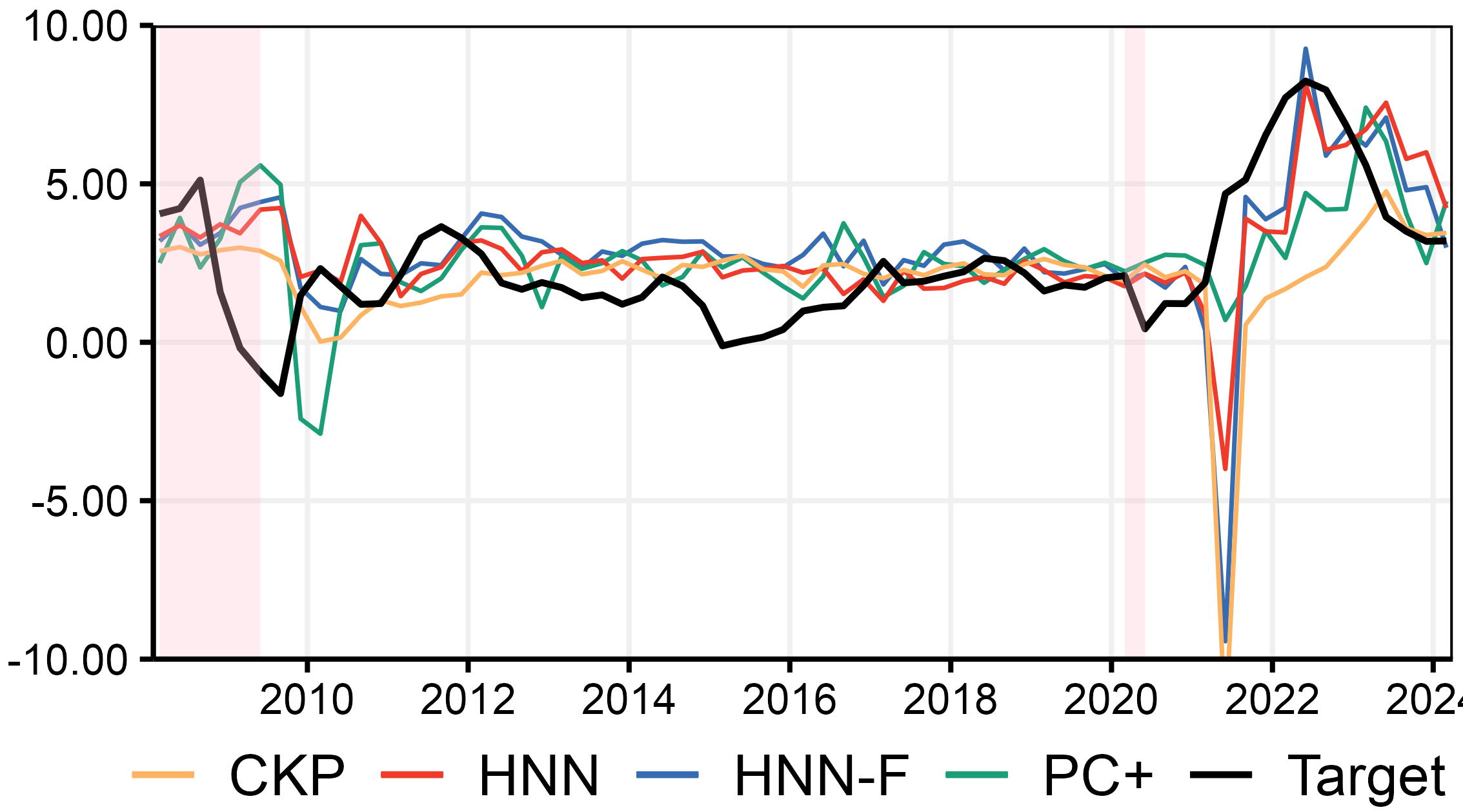}}
    \caption{\footnotesize Forecasts, CPI (s = 4)}\label{fcast_q_2007_h4}
   \end{subfigure}
   \begin{subfigure}[b]{0.5\textwidth} \includegraphics[trim={0cm 0cm 0cm 0cm},clip,width=0.995\textwidth]{{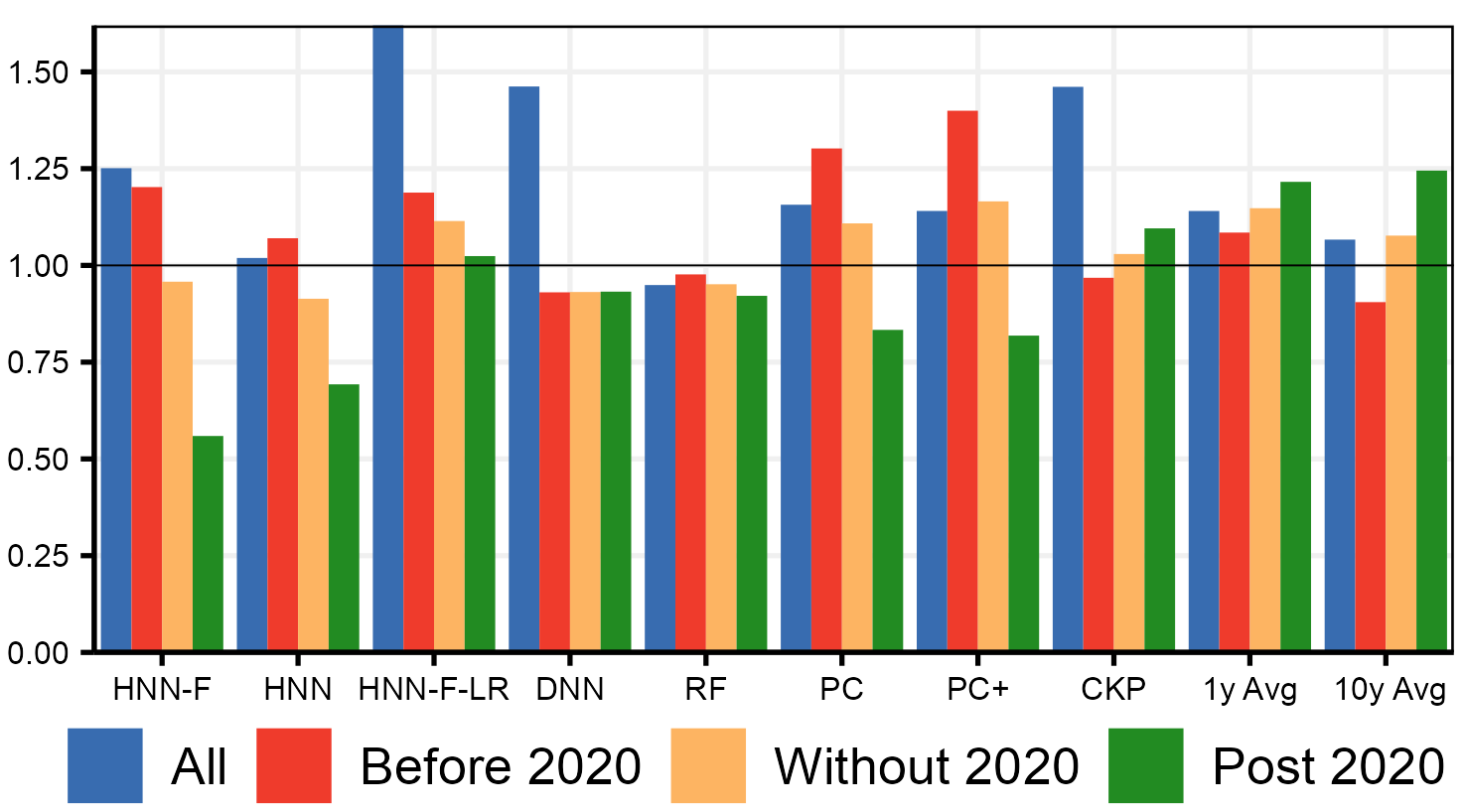}}
   	\caption{\footnotesize  RMSE wrt AR(4), CPI (s = 4)}\label{mse_q_2007_h4}
   \end{subfigure}
       \par\bigskip             
   \begin{subfigure}[b]{0.5\textwidth} \includegraphics[trim={0cm 0cm 0cm 0cm},clip,width=0.995\textwidth]{{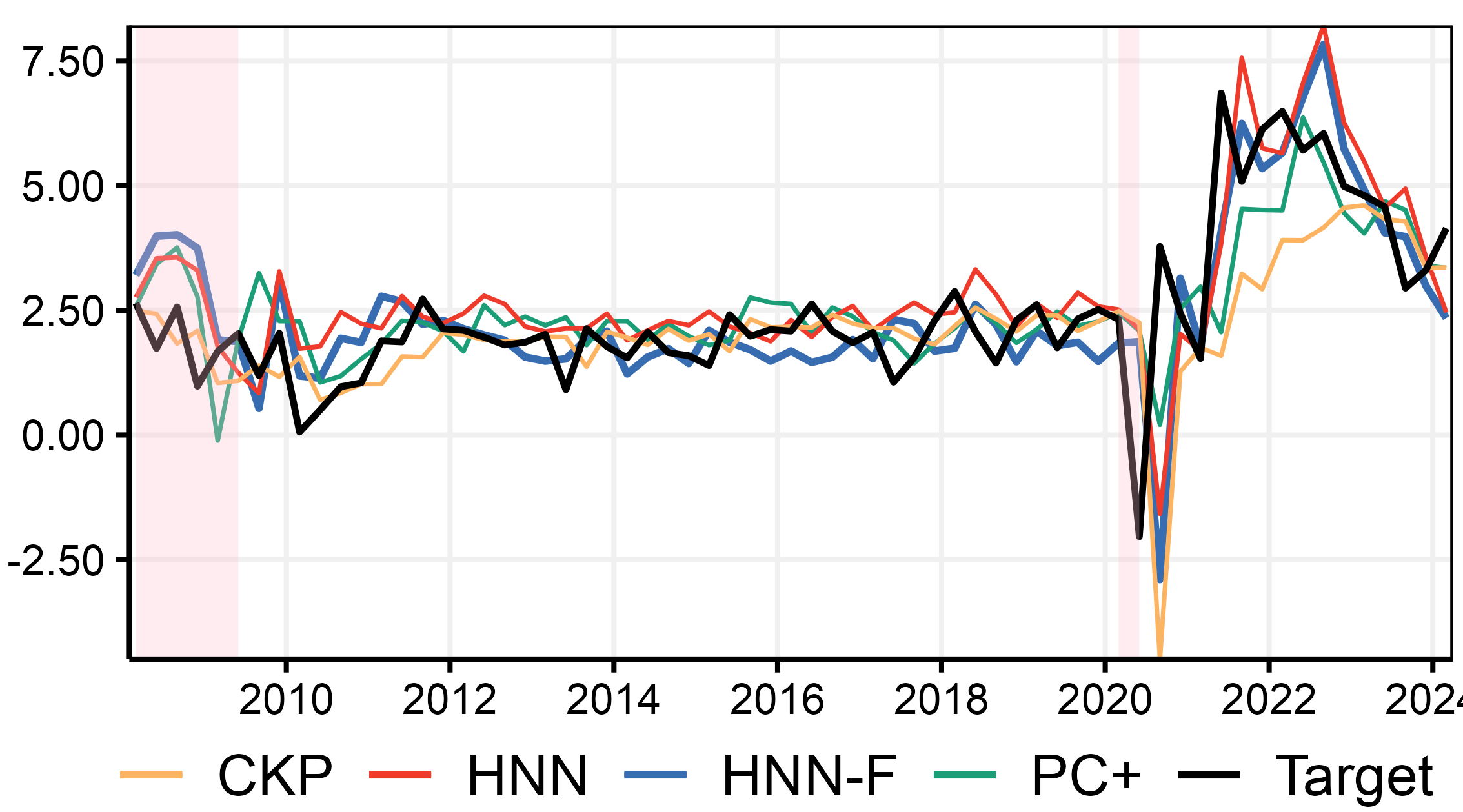}}
	\caption{\footnotesize Forecasts, Core CPI}
   \end{subfigure} 
   \begin{subfigure}[b]{0.5\textwidth} \includegraphics[trim={0cm 0cm 0cm 0cm},clip,width=0.995\textwidth]{{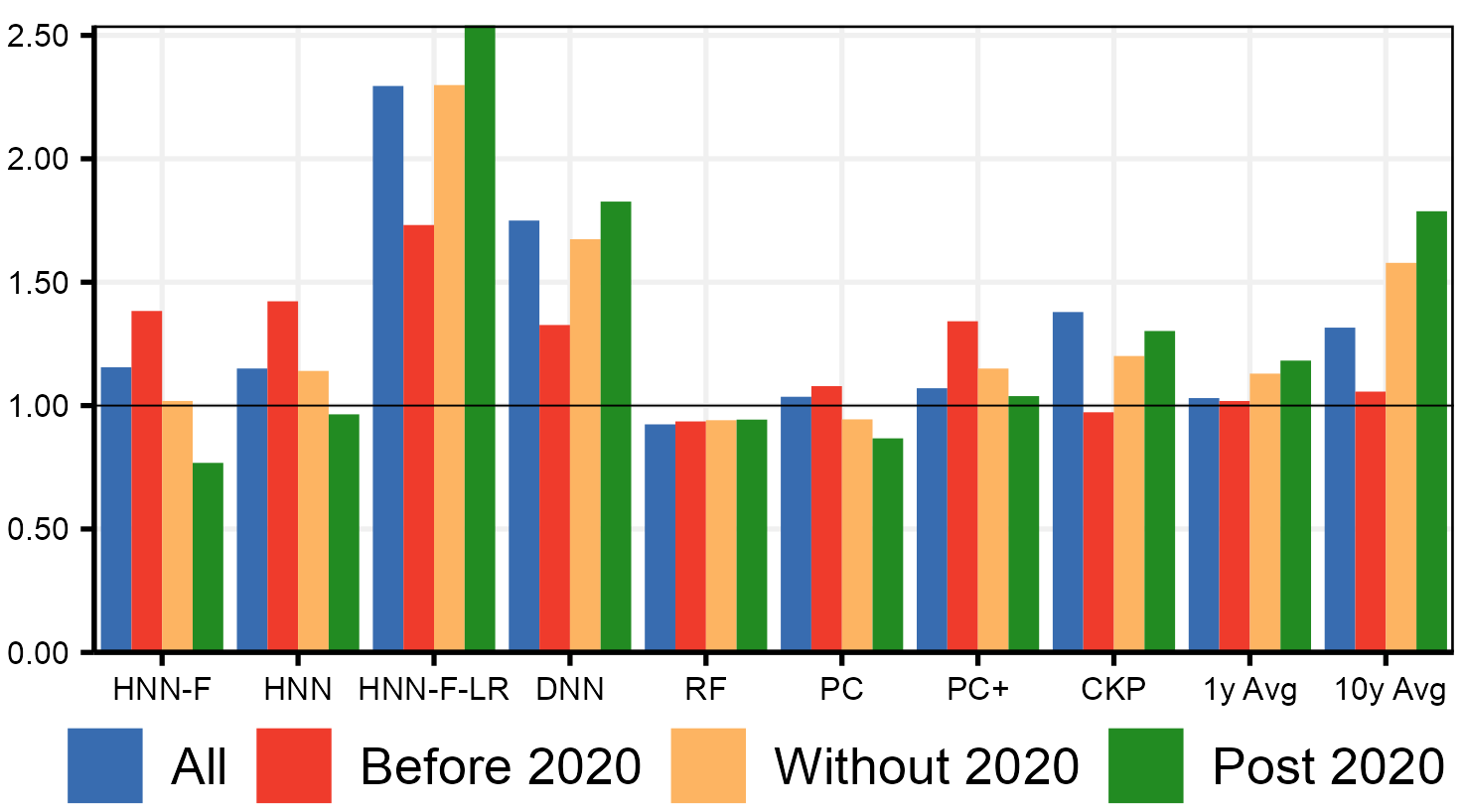}}
	    \caption{\footnotesize RMSE wrt AR(4), Core CPI}
   \end{subfigure}
  \vspace{0.0001em}
  \caption{\footnotesize Forecasting Results for 2 targets with test set 2007Q2-2024Q1 and re-estimating NNs every year.  Notes:  Pink shading is NBER recessions.  Corresponding Diebold-Mariano tests can be found in Table \ref{dmtest} (Appendix \ref{addifig}).}\label{forecast_fig2}
\end{figure}

{\noindent \sc \textbf{Forecasting Results.}}   For \textit{one-year ahead forecasts}, Figures \ref{mse_q_2007_h4} and \ref{fcast_q_2007_h4} reveal that HNN and HNN-F provide the best PC-based forecasts, particularly when excluding 2020. Performance-wise, they are on par with DNN and RF (not natively interpretable) for $s=4$ steps forecasts but have a definitive edge in the post-2020 sample. Unlike PC+ in Figure \ref{fcast_q_2007_h4}, HNN-F and HNN are not misled into predicting long-lasting disinflation (or even deflation) following the  GR because HNN-F's gap is closing as fast as that of the benchmark CPI($s=1$) estimation and $\gamma_t$ is moderately small (see Section \ref{sec:altsup}). For the Pandemic era, HNNs predict an average CPI inflation of 4\% from 2020Q4 to 2021Q3 inclusively, which is well above target. In contrast, PC+ calls for a modest 2.5\% and CKP expects inflation to be below target. For these two alternative targets, HNN cannot beat the benchmarks in the more stable pre-2020 era, and gains are localized to the return of inflation in the post-2020 sample. Core inflation appears harder to be captured efficiently by HNN (or any model, almost none of which surpass the AR(4) for any subsample). Still, HNNs are the only class of models delivering outperformance through early warning signals during the crucial post-2020 sample, where core inflation rose to a level unseen since at least the early 1990s.

\subsection{Ablation Studies} \label{sec:abla}

HNN involves various ingredients,  like the use of many economic indicators and nonlinear supervised processing.  In this section,  I conduct a brief inspection of what happens when dispensing with one or the other.  

\begin{table}[h!]
\vspace*{0.95em}
\small
\centering
\rowcolors{2}{white}{gray!15}
\setlength{\tabcolsep}{0.96em}
\caption{Data-Poor $\mathcal{H}$'s}
\label{hdefine_poor}
\vspace*{-0.75em}
\begin{tabularx}{0.7\textwidth}{|l|X|}
\toprule
\toprule

$\mathcal{H}$  & Content   \\ 
\midrule
\textbf{$\mathcal{E}_t^{\text{LR}}$}   &  $t$ (exogenous time trend) \\
\textbf{$\mathcal{E}_t^{\text{SR}}$}   &  Inflation expectations from SPF, and Michigan Survey,  lags of $\pi_t$, $t$  \\
\textbf{$g_t$}   & Unemployment, $t$   \\
\textbf{$c_t$}   & Oil price,  $t$  \\

\bottomrule
\bottomrule
\end{tabularx}
\end{table}

\begin{figure}[h!]
\begin{center} 
\hspace*{-0.05cm}\includegraphics[trim={0cm 1.65cm 0cm 0cm},clip,width=0.85\textwidth]{{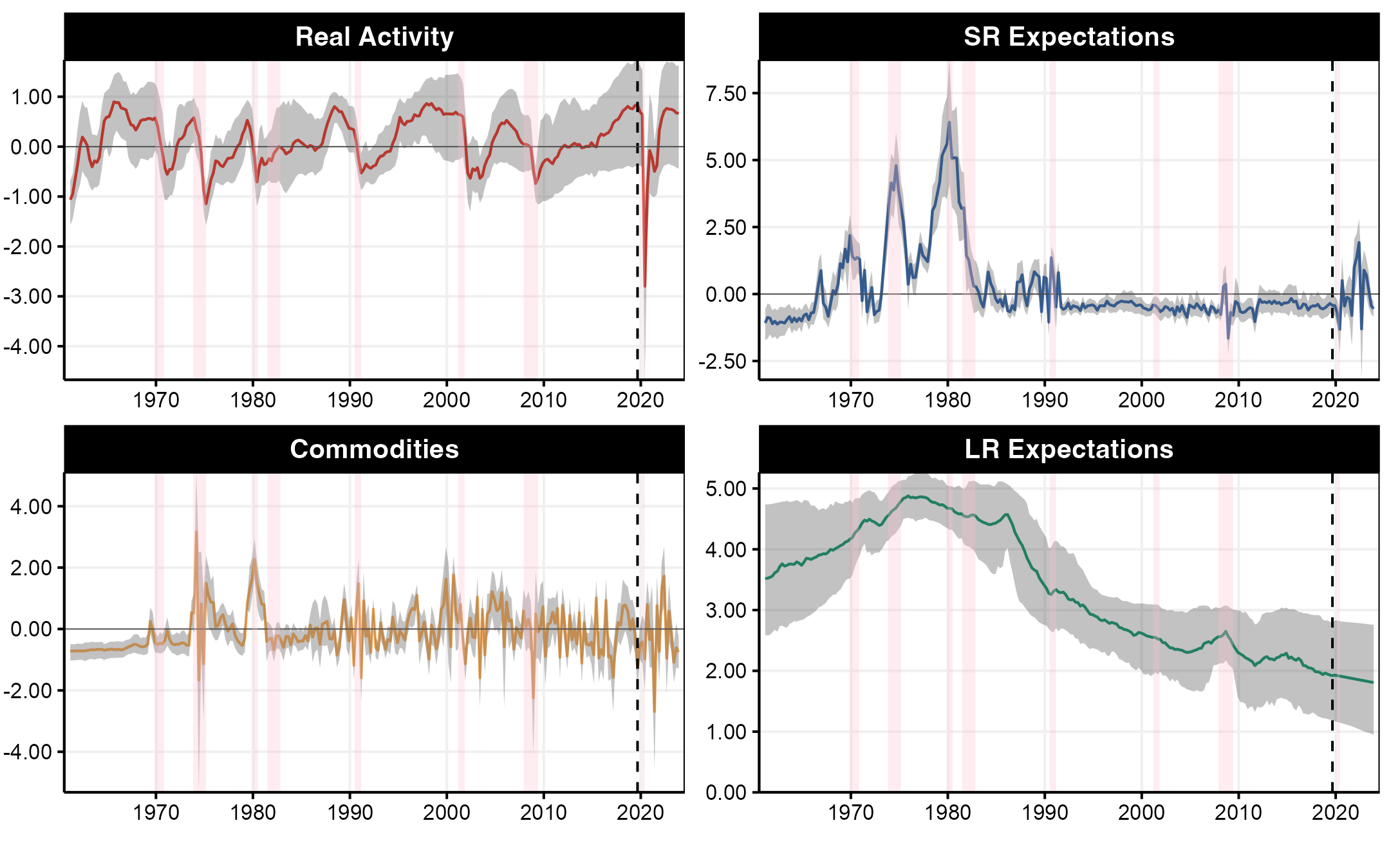}}
\vspace*{-0.1cm}
\caption{\footnotesize A Look at HNN components ($h_{t}$'s) with limited data in hemispheres.  Notes: Dashed line is the beginning of the out-of-sample.  NBER recessions are in pink shadowing.}\label{hs_URonly}
\end{center}
\end{figure}

First, I consider a specification where key hemispheres only contain what would enter in a typical modern PC OLS-based regression, as detailed in Table \ref{hdefine_poor}. Essentially, more fine-grained data on prices and any real activity indicator except for unemployment have been excluded (with respect to Table \ref{hdefine}). Unemployment is now in levels, and the aforementioned transformations (lags and moving averages) are kept. The idea is to have the neural network filter unemployment itself by nonlinearly interacting it with $t$, analogously to what unsupervised filtering does. In this context, for identification reasons (detrending unemployment and estimating a trending coefficient on it), HNN (not HNN-F) results are reported in Figure \ref{hs_URonly}, and the focus is kept on contributions.

Here are key observations from Figure \ref{hs_URonly}. The contribution of real activity is much smaller in absolute terms throughout the sample than it is for baseline specifications, highlighting the importance of diversified real activity indicators. The bands include 0 much more often starting from the 2000s, in line with traditional results using filtered unemployment. Speaking of, the extracted $h_{t,g}$ looks much more like filtered unemployment, albeit with a smaller $\lambda$ (Hodrick-Prescott smoothing parameter) than what is typically used. Lastly, and rather not unexpectedly, $h_{t,g}$ is negative for a long time during the Pandemic, making it rather unequipped to force its inflation forecast upward during the last three quarters. All in all, the inclusion of many real indicators in $\mathcal{H}_g$ seems vital for a more proactive characterization of real activity. This is not an unfamiliar conclusion \citep{SW2002}.

The expectations component is more aligned with what is reported throughout the paper (e.g., Figures \ref{hs_2019} and \ref{hs_2019_hnn}). This is not surprising given the high importance accorded to survey expectations and lags of the CPI by HNN, as shown by VI calculations in Figure \ref{vi_p}. However, in Figure \ref{hs_URonly}, their nonlinear processing is even more evident: the component is drastically shut down starting from 1990 and only wakes up for one obvious spike during the Great Recession. The contribution, however, excludes the very noticeable early peak in 2021Q2. Thus, dispensing with the vast number of price series originally included in HNN leads to missing key abrupt changes in the short-run expectations component that go under the radar of traditional aggregated measures.

\begin{figure}[ht!]
  \begin{subfigure}[b]{0.5\textwidth}
\includegraphics[trim={0cm 0cm 0cm 0cm},clip,width=0.995\textwidth]{{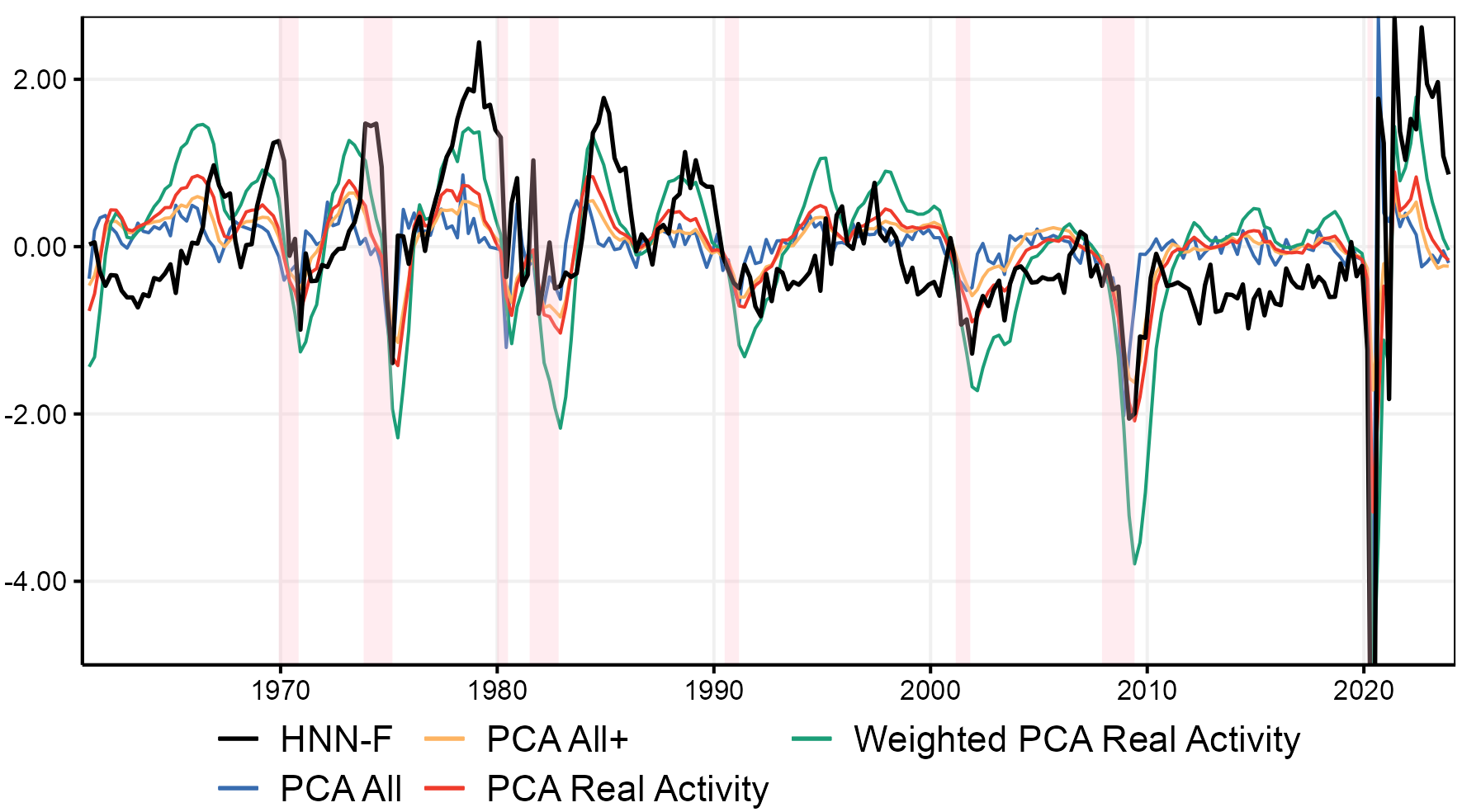}}
\caption{\footnotesize Alternative Data-Rich $g_t$ Extraction}\label{pca_gap}
      \end{subfigure}
  \hspace{-0.2em}
  \begin{subfigure}[b]{0.5\textwidth}
\includegraphics[trim={0cm 0cm 0cm 0cm},clip,width=0.995\textwidth]{{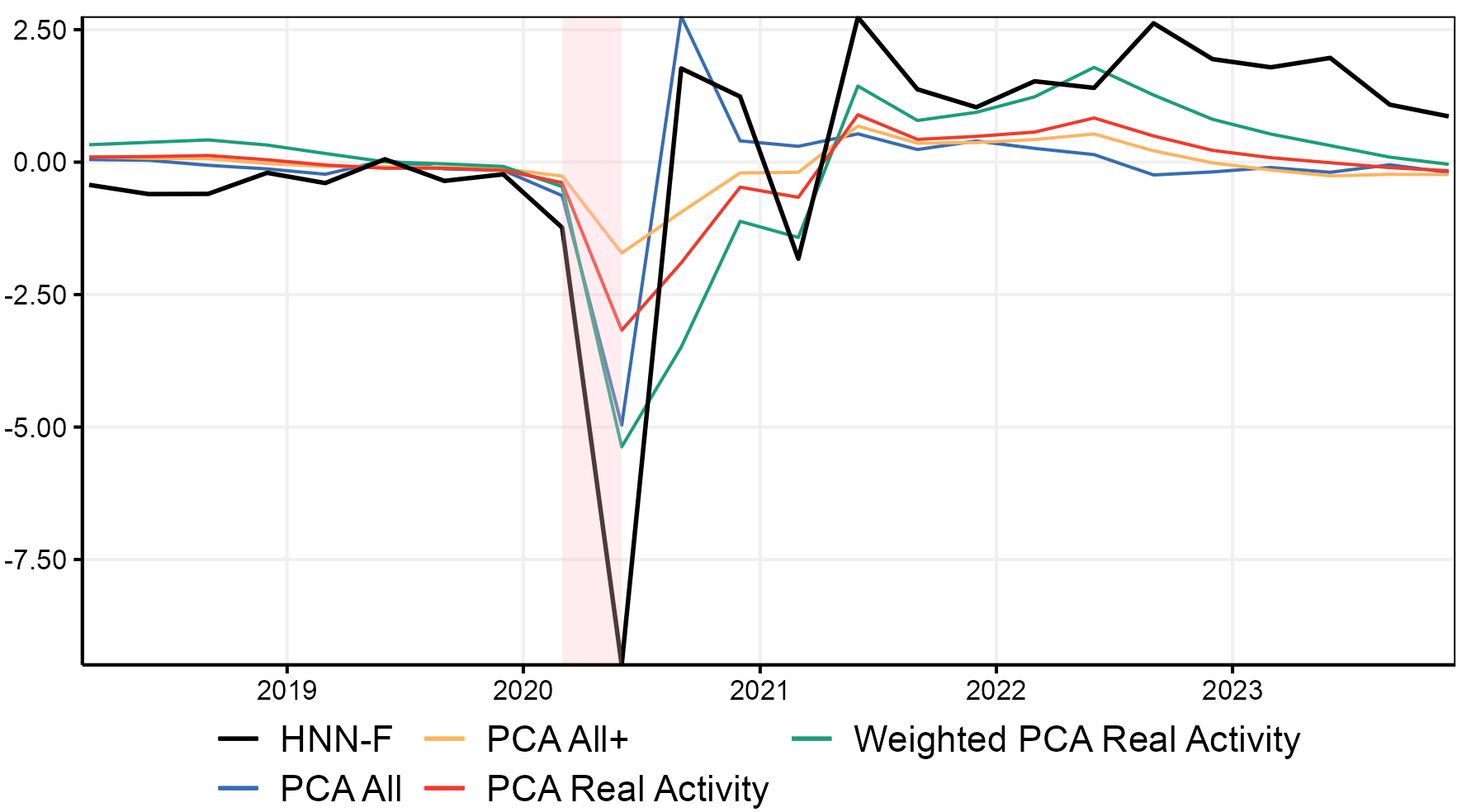}}
\caption{\footnotesize Alternative Data-Rich $g_t$ Extraction -- Last 3 years}\label{pca_gap_zoom}
      \end{subfigure}
      \par\bigskip            
        \begin{subfigure}[b]{0.5\textwidth}
\includegraphics[trim={0cm 0cm 0cm 0cm},clip,width=0.995\textwidth]{{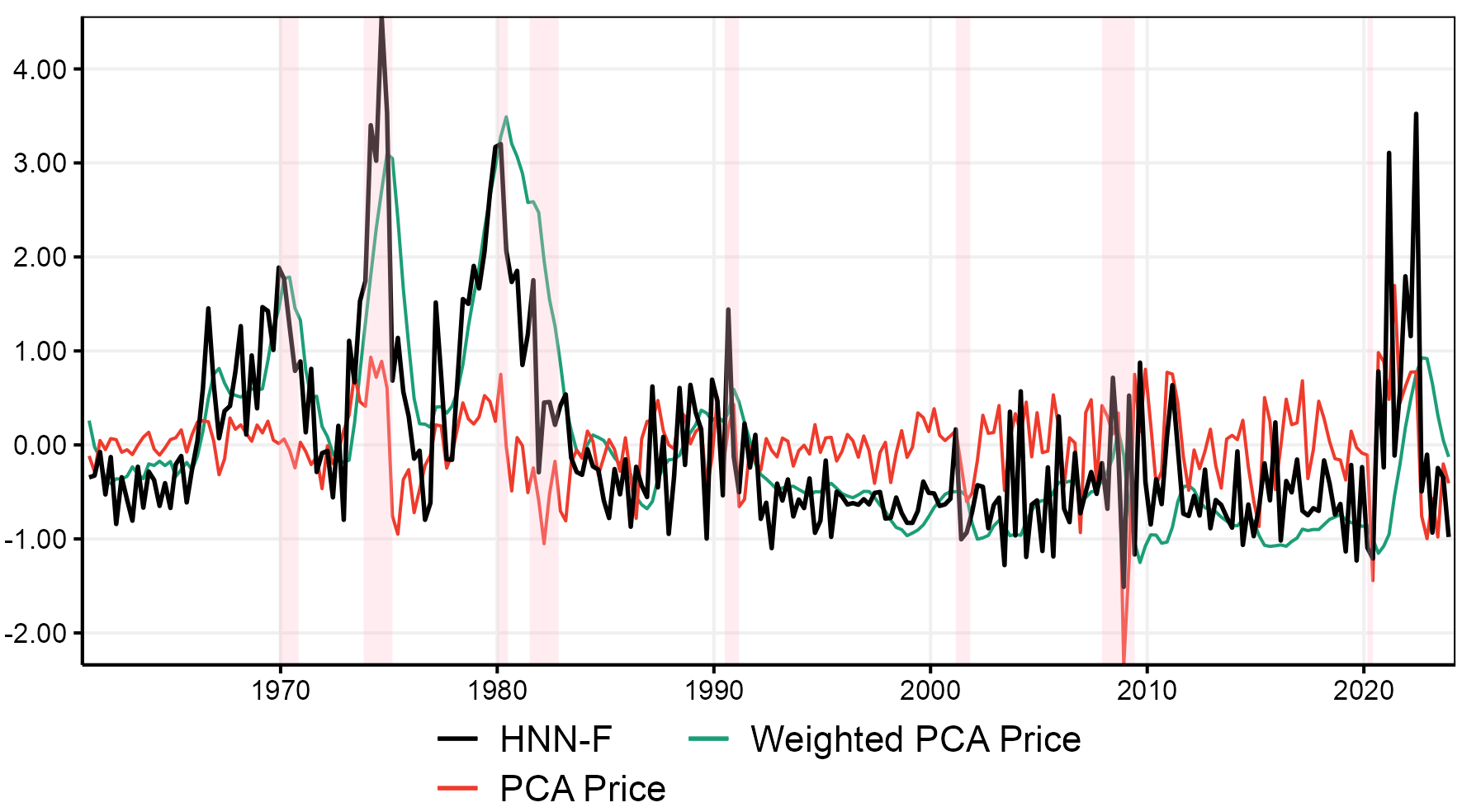}}
\caption{\footnotesize Alternative Data-Rich $\mathcal{E}_t^{\text{SR}}$ Extraction}\label{pca_price}
      \end{subfigure}
        \hspace{-0.2em}
              \begin{subfigure}[b]{0.5\textwidth}
\includegraphics[trim={0cm 0cm 0cm 0cm},clip,width=0.995\textwidth]{{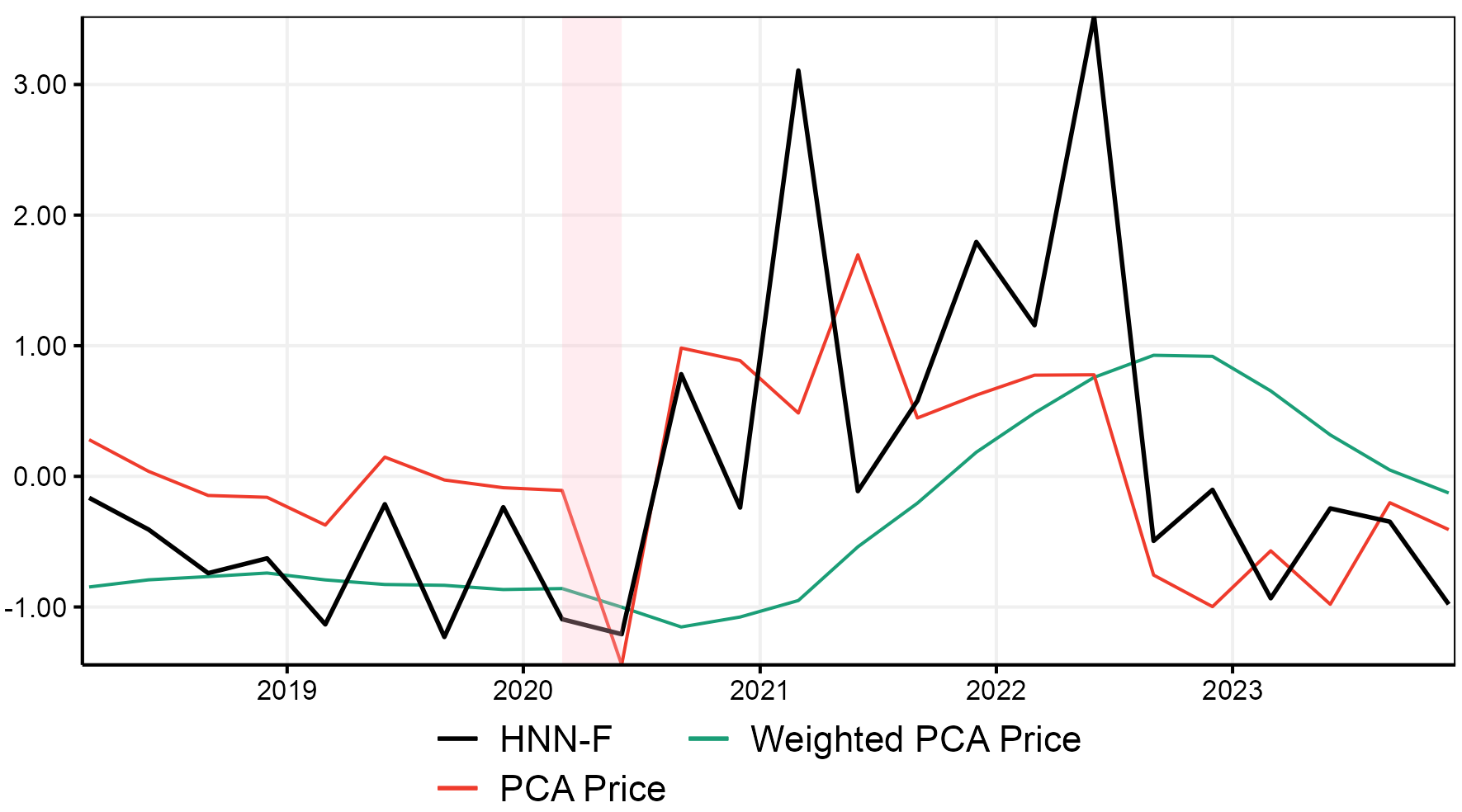}}
\caption{\footnotesize Alternative Data-Rich $\mathcal{E}_t^{\text{SR}}$ Extraction -- Last 3 years}\label{pca_price_zoom}
      \end{subfigure}
  \caption{\footnotesize "Gaps" of the four $\mathcal{H}$'s through time according to different extraction techniques.  Notes: all NN estimations end in 2019 and are projected out-of-sample from there.   Pink shading is NBER recessions.  }\label{pca_checks}
\end{figure}

The second ablation study maintains the data-rich environment but dispenses with nonlinearities and supervision (in part). Figure \ref{pca_checks} conducts different PCA-based extractions of the data contained in $\mathcal{H}_g$ and $\mathcal{H}_{\mathcal{E}^\text{SR}}$. A few alternatives are reported. \textbf{PCA Real Activity} means reporting the first factor of $\mathcal{H}_g$. \textbf{Weighted PCA Real Activity} means that, after standardization, variables were given weights according to VI estimates from Figure \ref{vi_bench}. This brings back some of the supervision from HNN. The real activity part of Figure \ref{pca_checks} contains two additional extractions. \textbf{PCA All} is the first factor of all the contemporaneous data included in HNN, whereas \textbf{PCA All+} includes lags and MARX's as well. The rationale for including the last two (in the top part of the panel only) stems from the often-reported finding that the first factor extracted from broad macroeconomic panels looks very much like a real activity factor \citep{mccracken2020fred}.

Findings are as follows. In Figure \ref{pca_gap}, PCA extractions, except for the weighted version, form a cluster throughout. For the period spanning from 2000 to the Pandemic era, that cluster seemingly includes $g_t^{\text{HNN}}$. However, during periods of overheating, differences are manifest and no linear method seems to approximate $g_t^{\text{HNN}}$. This is true of the 1970s and also of the current period (Figure \ref{pca_gap_zoom}), with two linear extractions signaling no overheating at all, and two others showing a rather quaint or short-lived one. Weighted PCA Real Activity is mostly far away from $g_t^{\text{HNN}}$, suggesting that nonlinear processing of important variables (such as the Help-Wanted Index, whose trend is clearly visible in Weighted PCA Real Activity) cannot be dispensed with. Nevertheless, this "mildly supervised" PCA extraction is the only one pointing to a widening positive gap in 2021â€”as does $g_t^{\text{HNN}}$. From mid-2021 until the end of the sample, only $g_t^{\text{HNN}}$ remains elevated.

Notable differences between PCA extracts and HNN-F's own $\mathcal{E}^{\text{SR}}$ are  apparent in Figure \ref{pca_price}.  While Weighted PCA is substantially correlated with $\mathcal{E}^{\text{SR}}$ in-sample,  the former lags behind the latter starting from 2021.  Interestingly,   HNN's $\mathcal{E}^{\text{SR}}$' similarity with plain PCA only occurs during localized episodes such as during the Great Recession,  and more obviously so,  from mid-2020 onward.  From these observations,  it is obvious that something more sophisticated must be at work within HNN, and nonparametric nonlinear processing seems to be vital in extracting $\mathcal{E}^{\text{SR}}$ from price and expectations data that is actually forward-looking. 


\subsection{Ablating ReLU : Another Test For The Relevance of Nonlinearities}\label{sec:abla2}

\begin{figure}[h!]
\begin{center} 
\hspace*{-0.05cm}\includegraphics[trim={0cm 0cm 0cm 0cm},clip,width=0.85\textwidth]{{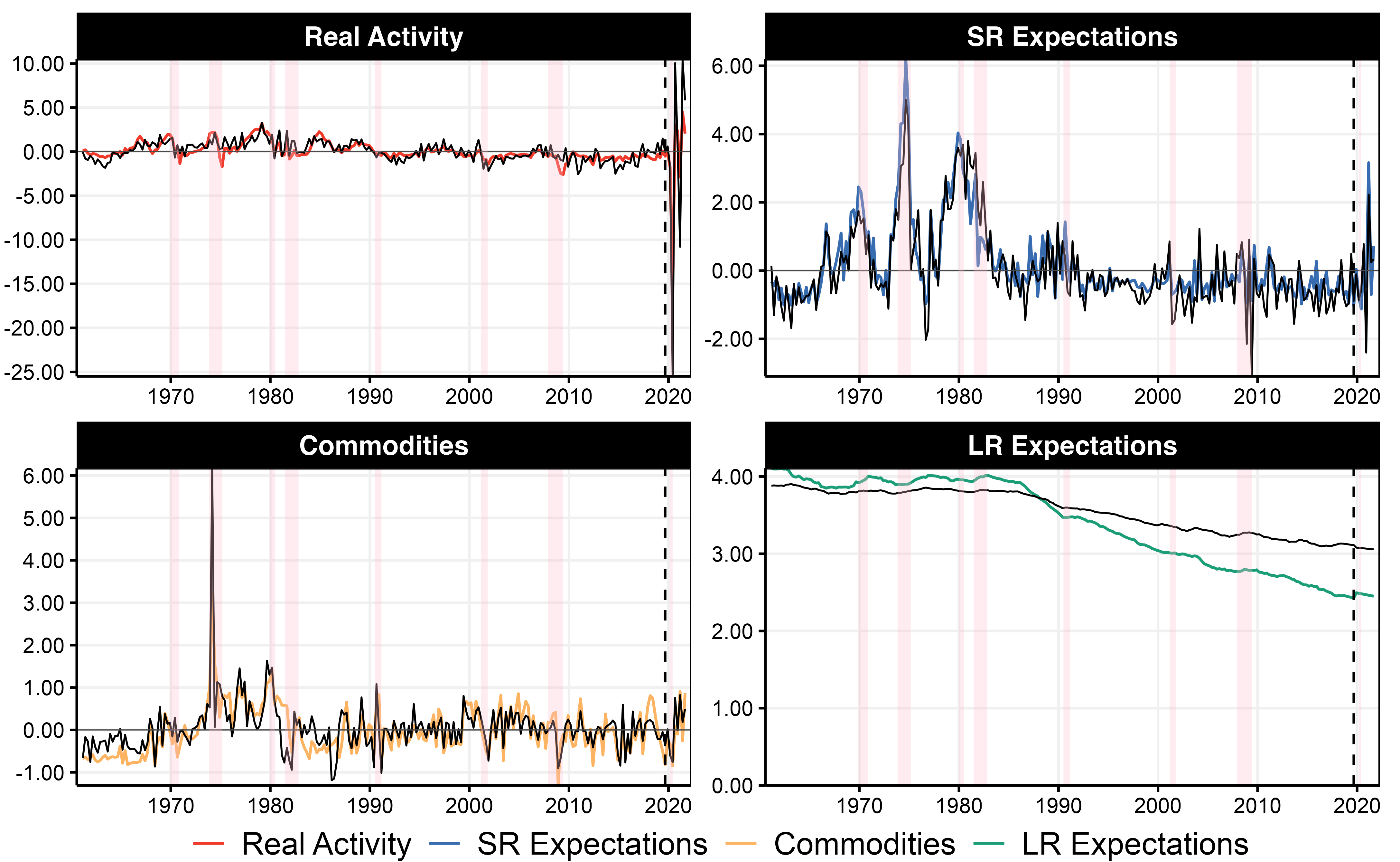
}}
\caption{\footnotesize Contributions ($h_{t}$'s) from baseline HNN (in colors) and alternative specification (in black) using a contemporaneous target ($s=0$) and only survey information and lags of $\pi_t$ in the short-run expectations hemisphere.   Notes: Dashed line is the beginning of the out-of-sample.  NBER recessions are in pink shadowing. }\label{ablarelu}
\end{center}
\end{figure}

One can alternatively turn off nonlinearities in HNN by changing the ReLU activation function to a linear one.  That way,  HNN becomes some kind of supervised groupwise PCA (or ridge regression).  In Figure \ref{ablarelu},  I report a  version where the 3 gaps hemispheres are linear but all coefficient hemispheres are nonlinear (allowing for more flexible time-variation than a simple linear trend).

There are two important findings. First, the real activity component is the most impacted by this ablation and features a less economically plausible path.  Its cyclical variation is not nearly as evident in nonlinear results. For instance, the linear gap near-completely brushes off the Great Recession. Most of the 1970s peaks and troughs are barely discernible from high-frequency noise that is more prevalent in the linear specification. In contrast, the business cycle is clearly visible in the nonlinear gap. There is more harmony between the linear and nonlinear specifications from 1990 to 2007, where the nonlinear specification appears like a smoother version of the linear one. Finally, the linear real activity path provides implausible estimates of the gap in the post-2020 sample. Indeed, while the baseline HNN features a contribution of real activity mostly aligned with the highest highs of the 1970s, the linear one goes off the grid by providing a contribution that is almost four times that which is estimated in the 1970s. As discussed in \cite{GCMS}, ML nonlinearities play various roles, especially in times of crisis, where linear extrapolation may easily trigger substantial outliers.

The second observation is that short-run expectations are far less impacted by ablating nonlinearities. Key inflationary episodes are somewhat similar, in the 1970s and from 2020 onward. The only obvious difference is that the nonlinear specification filters out a series of \textit{downward} spikes: there is a clear one in the mid-1970s, another in the mid-1980s, during the Great Recession, and finally, in 2020. While the role of nonlinearities is not as pronounced as for real activity, the deletion of those downward spikes helps avoid sporadic negative forecasts of inflation. Furthermore, it is again the case that the nonlinear specification delivers results that are economically more plausible than the linear one: survey inflation expectations from professional forecasters rarely feature these rapid downward movements, and LR expectations of the fully nonlinear specification are more aligned with conventional wisdom.

The specification with linear gap hemispheres and time-varying coefficients is evaluated as a data-rich benchmark in forecasting results of Section \ref{sec:fcast} as HNN-F-LR. Its performance is far inferior to its nonlinear counterpart. Despite showing noticeable action in the real activity hemisphere post-2020, the linear gap appears to show an overstated elevation of real activity and thereby RMSEs suffer from it. Overall, results from the ablation studies suggest that both using vast amounts of data and nonlinear supervised processing are essential to obtain the desirable $g_t$ and $\mathcal{E}^{\text{SR}}$ delivered by HNN-F.

  
 
 \subsection{Taylor Rule Supervision} \label{sec:fedsup}

This section explores a curiosity which can be understood as a more radical change of supervisors than what was reported in Section \ref{sec:altsup}. From an econometric point of view, it showcases yet another potential application of HNN beyond inflation and Phillips curves.

Inflation is retired in favor of the federal funds rate, and the supervision relationship becomes an empirical Taylor rule. That is, we are extracting the contribution of the gap and inflation to the monetary policy instrument values. An interesting economic question is whether $h_{t,g}^{\text{FFR}}$ looks remotely like $h_{t,g}^{\text{CPI}}$. In other words, does the "Fed view" of the gap assuming the Taylor rule is a valid approximation of its behavior coincide with what the inflation record suggests?

There are two important changes with respect to the baseline specification. First, $\pi_{t+1}$ is replaced by the federal funds rate next period ($r_{t+1}$). Second, the energy/commodities group is replaced by the "Smoothing" group, which includes lags of $r_{t+1}$. This inclusion is typical of empirical Taylor rules and statistically accommodates the fact that the monetary authority avoids drastic changes in $r_t$.

\begin{figure}[h!]
\begin{center} 
\hspace*{-0.05cm}\includegraphics[trim={0cm 1.65cm 0cm 0cm},clip,width=0.85\textwidth]{{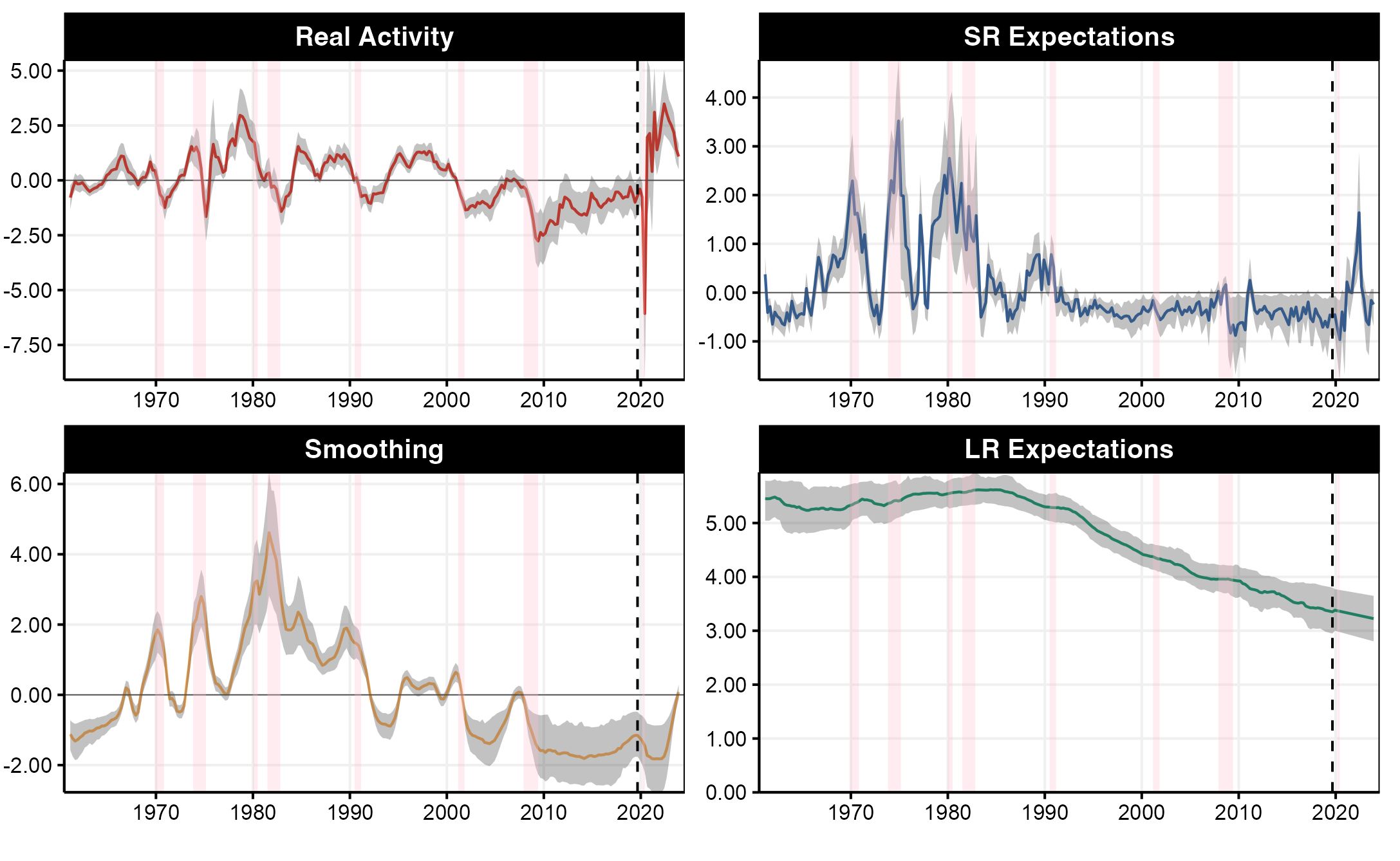}}
\vspace*{-0.1cm}
\caption{\footnotesize Contributions ($h_{t}$'s) based on a Taylor Rule,  with training ending in 2019.  Notes: Dashed line is the beginning of the out-of-sample.  NBER recessions are in pink shadowing.}\label{hs_fedview}
\end{center}
\end{figure} 

In Figure \ref{hs_fedview}, $h_{t,g}^{\text{FFR}}$ looks much more like what one would obtain from a traditional filtering-based gap, except for the Pandemic episode. First, there is a certain persistence to it that is characteristic of specifications assuming autoregressive laws of motion. Second, albeit remaining cyclical, the contribution is mostly negative starting from the 2000s, whereas it was roughly symmetric around 0 beforehand and is reminiscent of many unemployment-related measures of slack \citep{stock2019slack}. Third, it mostly exhibits a slow climb back to 0 starting from 2008, similar to what one would obtain from, e.g., the CBO gap.

There are three episodes where predictions (unreported) from this model can be off for a few quarters. First, the two ZLB episodes, where this Taylor rule prescribes interest rates going below zero is not inconsistent with the deployment of quantitative easing following the GR and during the Pandemic period. Interestingly, the third episode is in late 2020 and 2021, with $h_{t,g}^{\text{FFR}}$ calling for much higher rates than those currently in effect. In other words, if the Fed were consistent with how it responded to slack/overheating (as extracted by HNN-F) during the last decades, rates would have been hiked sooner. Grains of salt on this statement are that (i) the Fed changed its approach to inflation targeting in 2020, and (ii) pandemic-era slack featured many aspects different from previous recessions, which entered the Fed's information set and not that of HNN.


The evident wedge between $h_{t,g}^{\text{FFR}}$ and $h_{t,g}^{\text{CPI}}$ reported throughout the paper hints that there may be a gap between the monetary authority's view of economic slack and what matches the inflation record.  Nonetheless, this application is meant as illustrative about HNN's versatility,  and to understand further how supervision affects $g_t$.  A comprehensive assessment of "neural Taylor rules" is material for future work.

\subsection{Simulation Study}\label{sec:simul}

The simulation study seeks to establish three things.  First, that HNN can actually recover true latent states under a known (and representative) DGP.  Second,  that a large overparametrized architecture performs well at  recovering nonlinearities,  and relatively better in this environment than smaller architectures -- justifying the choice of a larger architecture in the empirical work.  Third, we want to check that the ensemble-based approach to quantify uncertainty delivers plausibly sized bands that can be used to complement the analysis of point estimates obtained by HNN.

\subsubsection{Data Generation Process (DGP)}

The simulation study involves generating data through three main steps.  Its design is such to retain some kind of simplicity and yet mimic key features of the time series data onto which HNN is trained in the empirical section.  It uses workhorse DGPs for gaps and coefficients, and the input data is simulated such as to resemble  macroeconomic data. 

\begin{enumerate}
\item \textbf{Simulating Input Data from Dynamic Factor Models (DFM).}  The input data is simulated through
   \[
   \begin{aligned}
   & \text{factors}_{t, j} = \rho \cdot  \text{factors}_{t-1, j} + u_{t, j}, \quad u_{t, j} \sim \mathcal{N}(0, 1), \\
   & \text{loadings}_{k,j} \sim \mathcal{N}(0, 1), \\
   & X_{t, k} = \sum_{j}^J \text{factors}_{t, j} \cdot \text{loadings}_{k, j} + e_{t, k}, \quad e_{t, k} \sim \mathcal{N}(0, 1),
   \end{aligned}
   \]
   where \( \rho = 0.75 \) is the autocorrelation coefficient, and \( t, j, k\) are indices for time, factors, and loadings, respectively. The observed variables are then normalized to the range [0, 1].  We will have $K=20$ and $K=100$.   The number of factors $J$  is set to 5.  The choice of the DFM is to mimic two important features of the macroeconomic data utilized by HNN: persistence and cross-correlation of predictors.  The 0-1 normalization is chosen so to match that of $X$'s traditionally used to enter the ubiquitous Friedman 1 DGP,  which is widely used in statistics and econometrics to evaluate nonlinear tools \citep{MARS}.

\item  \textbf{Generating "gaps" from Friedman DGP.} We takes the simulated datasets from the previous step and nonlinearly transform them into two distinct gaps according to
   \begin{align*}
  g_{t,1} &= 10 \sin(\pi \cdot X_{t,1} X_{t,2}) + 20 (X_{t,3} - 0.5)^2 + 10 X_{t,4} + 5 X_{t,5}\\
  g_{t,2} &= 10 \sin(\pi  \cdot X_{t,11} X_{t,12}) + 20 (X_{t,13} - 0.5)^2 + 10 X_{t,14} + 5 X_{t,15}
   \end{align*}

   in the $K=20$ case.  Note that $\pi$ in this section means the irrational number,  not inflation.   As is standard in the Friedman 1 DGP,  not all regressors are utilized,  but they will still be given to the estimation algorithm.   Thus, in the $K=20$ case,  there will be 5 uninformative predictors per hemisphere (and thus a ratio of 1-1 in terms of informative/uninformative predictors).  In the high-dimensional ($K=100$) case,   each gap is now a sum of 5 Friedman DGPs using a total of 25 distinct predictors each.  Thus, the ratio of informative to uninformative predictors is the same, but now HNN must constitute a gap from estimating 5 times more nonlinear relationships than in the $K=20$ setup.

\item  \textbf{Generating the time-varying coefficients,  contributions, and the target.} The final target is generated through
\[
\begin{aligned}
    h_{t, 1} &= \beta_t g_{t, 1} \\
    h_{t, 2} &= \beta_t g_{t, 2} \\
    Y_t &= h_{t, 1} + h_{t, 2} + \epsilon_t
\end{aligned}
\]

   where \( \beta_t = 1 + 0.6 \times \sin (3 \cdot \frac{ \pi \cdot  t}{T - 1}) \) is a time-varying coefficient smoothly oscillating between 0.4  and 1.6,  and completing two full cycles between $t=1$ and $t=T$.   \( \epsilon_t \sim \mathcal{N}(0, \sigma^2) \) is the noise.   $\sigma^2$ is set so to have a signal-to-noise ratio (SNR) of 1 and 2,  thus having a model with the "true" $R^2$ being 0.5 and 0.66.  For reference,  the benchmark HNN on the inflation data has an out-of-bag $R^2$ of approximately 62\%.  Note that both $\beta_t$'s follow a similar pattern for simplicity, but HNN is not provided that information in estimation.  Simulations are conducted with sample sizes of $T=400$ and $T=1000$.  The latter helps in verifying that HNN indeed gets estimates that are very close to the true DGP under close-to-ideal estimation conditions.  
\end{enumerate}

\subsubsection{Three HNN Configurations}

We use HNN-F,  which is the main model utilized in the analysis and choose hyperparameters corresponding to the empirical section with minor modifications. The block size is set to 10, a horizon at which persistence in the $X_{t,k}$ remains but is negligible ($\rho^{10} = 0.75^{10} \approx 0.05$ ).  The dropout  rate,  which had limited effects for larger networks in our application,  seemed to impede medium and small networks from optimizing smoothly,  and thus has been turned off.  All other tuning parameters remain unchanged.  Thus, the \textbf{large} network is given $\texttt{neurons} = 400$ and $\texttt{layers} = 3$  for gaps,  while the coefficients hemispheres (with the only input being $t$) have $\texttt{neurons} = 100$ and $\texttt{layers} = 3$.  The \textbf{medium} networks keep the same number of layers but shrinks the number of neurons by a factor of 10 (resulting in 40 and 10 per layer).  The \textbf{small} network shrinks this further to have 20 neurons for each gap and 5 for each coefficient hemisphere.  Thus,  the ratio of total neurons allocated to gaps versus coefficients is the same for the three networks sizes.
 
\subsubsection{Evaluation  Metrics}

I consider three performance metrics that elicit different kinds of information useful to answer the questions raised above.   First,  we have a measure of how close point estimates for the various outputs (gaps, coefficients, contributions) are \textit{to the truth}.  The $R^2$ with respect to the true DGP is reported so that units are comparable across output types.  It can be understood as "how much" of the variance of the true object can be captured by the estimated one or a squared correlation (without pre-demeaning).  The measure is calculated using out-of-bag estimates,  as is done in the empirical analysis.  This implies that for an estimate that would be worse than simply setting, e.g.,  $g_t$,  to its mean, the measure can be negative (like is often found with out-of-sample $R^2$ for return prediction in finance).  It is important to note that, by construction,  the performance metrics reported here directly tell us which model (large, medium, small) would fare better if we were to use them in a recursive out-of-sample forecasting experiment.   Indeed,  the $R^2$ metrics for contributions   inform us about which model's estimates are closest to \textit{the true conditional mean} (rather than the training target), the latter defining the upper bound of predictive performance for a squared error loss.

The other two metrics are designed to evaluate the soundness of the uncertainty quantification methodology.   I report 68\% and 95\% nominal coverage.  Note that in general,  the Bayesian Bootstrap (from which HNN's uncertainty quantification strategy is inspired) is not necessarily expected to deliver results hitting nominal coverage targets,  which is a  frequentist concept. Nonetheless,  such metrics are informative about how those methods will behave when using them on real data.  For instance,  it could be worrisome if coverage is always very low and that bands are always too tight with respect to what seems to be a reasonable assessment of uncertainty.  In all cases,  I use the relevant percentiles from the draws coming out of the procedure and calculate the frequency (averaged over $t$,  20 simulations, and the pair of similar objects) at which the truth lies within the interval.  



\begin{table}[t!]
\footnotesize
\centering
\caption{\normalsize {Results of the Simulation Study } \vspace*{-0.3cm}} \label{tab:simul}
\begin{threeparttable}
\setlength{\tabcolsep}{0.50em}
\setlength\extrarowheight{3.95pt}
\begin{tabular}{l | ccc | lccc | lccc}
\toprule \toprule
& \multicolumn{3}{c}{ $R^2$ (with true components) }    & & \multicolumn{3}{c}{ 68\% coverage  } &&\multicolumn{3}{c}{ 95\% coverage }  \\
\cmidrule(lr){2-4} \cmidrule(lr){5-8} \cmidrule(lr){9-12}
& Large & Medium & Small & & Large & Medium & Small && Large & Medium & Small \\
\midrule
\addlinespace[5pt]
\rowcolor{gray!15}
\multicolumn{12}{l}{\textbf{ Contributions }} \\
\addlinespace[4pt]
\text{SNR} = 2, $T = 1000$, $K = 20$& 0.88 & 0.76 & 0.63 && 0.79 & 0.31 & 0.26 && 0.99 & 0.67 & 0.55 \\
\addlinespace[2pt]
\text{SNR} = 2, $T = 400$,\phantom{0} $K = 20$& 0.82 & 0.72 & 0.63 && 0.91 & 0.40 & 0.31 && 1.00 & 0.81 & 0.69 \\
\addlinespace[2pt]
\text{SNR} = 1, $T = 1000$, $K = 20$& 0.77 & 0.68 & 0.59 && 0.72 & 0.39 & 0.32 && 0.97 & 0.71 & 0.61 \\
\addlinespace[2pt]
\text{SNR} = 1, $T = 400$,\phantom{0} $K = 20$& 0.69 & 0.64 & 0.61 && 0.86 & 0.42 & 0.30 && 1.00 & 0.83 & 0.64 \\
\addlinespace[2pt]
\text{SNR} = 1, $T = 400$,\phantom{0} $K = 100$& 0.60 & 0.56 & 0.48 && 0.88 & 0.49 & 0.30 && 1.00 & 0.88 & 0.67 \\
\addlinespace[2pt]
\text{SNR} = 2, $T = 400$,\phantom{0} $K = 100$& 0.70 & 0.62 & 0.54 && 0.88 & 0.48 & 0.29 && 1.00 & 0.87 & 0.66 \\
\addlinespace[2pt]
\midrule
\addlinespace[5pt]
\rowcolor{gray!15}
\multicolumn{12}{l}{\textbf{ Gaps }} \\
\addlinespace[4pt]
\text{SNR} = 2, $T = 1000$, $K = 20$& 0.90 & 0.85 & 0.74 && 0.69 & 0.94 & 0.98 && 0.94 & 1.00 & 1.00 \\
\addlinespace[2pt]
\text{SNR} = 2, $T = 400$,\phantom{0} $K = 20$& 0.85 & 0.82 & 0.74 && 0.68 & 0.92 & 0.98 && 0.94 & 1.00 & 1.00 \\
\addlinespace[2pt]
\text{SNR} = 1, $T = 1000$, $K = 20$& 0.81 & 0.77 & 0.68 && 0.61 & 0.86 & 0.98 && 0.89 & 0.99 & 1.00 \\
\addlinespace[2pt]
\text{SNR} = 1, $T = 400$,\phantom{0} $K = 20$& 0.78 & 0.76 & 0.70 && 0.67 & 0.90 & 0.99 && 0.94 & 1.00 & 1.00 \\
\addlinespace[2pt]
\text{SNR} = 1, $T = 400$,\phantom{0} $K = 100$& 0.66 & 0.66 & 0.57 && 0.66 & 0.87 & 0.98 && 0.93 & 0.99 & 1.00 \\
\addlinespace[2pt]
\text{SNR} = 2, $T = 400$,\phantom{0} $K = 100$& 0.73 & 0.71 & 0.64 && 0.67 & 0.89 & 0.98 && 0.93 & 0.99 & 1.00 \\
\addlinespace[2pt]
\midrule
\addlinespace[5pt]
\rowcolor{gray!15}
\multicolumn{12}{l}{\textbf{ Coefficients }} \\
\addlinespace[4pt]
\text{SNR} = 2, $T = 1000$, $K = 20$& 0.87 & 0.36 & 0.11 && 0.72 & 0.89 & 0.96 && 0.95 & 0.99 & 1.00 \\
\addlinespace[2pt]
\text{SNR} = 2, $T = 400$,\phantom{0} $K = 20$& 0.79 & 0.21 & 0.01 && 0.77 & 0.88 & 0.96 && 0.96 & 0.99 & 1.00 \\
\addlinespace[2pt]
\text{SNR} = 1, $T = 1000$, $K = 20$& 0.69 & 0.29 & 0.10 && 0.65 & 0.82 & 0.95 && 0.92 & 0.98 & 1.00 \\
\addlinespace[2pt]
\text{SNR} = 1, $T = 400$,\phantom{0} $K = 20$& 0.39 & 0.08 & -0.01 && 0.76 & 0.86 & 0.97 && 0.96 & 0.98 & 1.00 \\
\addlinespace[2pt]
\text{SNR} = 1, $T = 400$,\phantom{0} $K = 100$& 0.20 & 0.09 & 0.01 && 0.78 & 0.84 & 0.96 && 0.96 & 0.98 & 1.00 \\
\addlinespace[2pt]
\text{SNR} = 2, $T = 400$,\phantom{0} $K = 100$& 0.54 & 0.19 & 0.03 && 0.77 & 0.86 & 0.96 && 0.96 & 0.98 & 1.00 \\
\addlinespace[2pt]
\bottomrule \bottomrule
\end{tabular}
\begin{tablenotes}[para,flushleft]
 {\scriptsize \item \textit{Notes: }This table displays results for the simulations under various configurations which are a triple of signal-to-noise ratio (SNR), number of observations for training ($T$),  and total number of regressors $K$ entering HNN for estimation (and $\frac{K}{2}$ contribution to  the DGP).  Three metrics of performance are reported.  First,  we have a measure of recovery of the true states  ($R^2$ between estimated and true components) to evaluate how well larger vs.  smaller networks perform.   Second,  we have two coverage metrics to evaluate if the ensemble technique for uncertainty quantification delivers plausible uncertainty estimates. Results are averaged over both components in the DGP and over 20 simulations.}
\vspace*{-0.2cm}
\end{tablenotes}
\end{threeparttable}
\end{table}

\subsubsection{Results}

{\noindent \sc \textbf{Accuracy of Point Estimates.}} Overall results regarding the accuracy of points estimates are as one would expect: all architectures deliver their best performance when facing a  larger SNR,  a larger sample,  and lower-dimensional true DGP.  This is true for gaps, coefficients, and their product (contributions).   Nonetheless,  even in the most demanding environment (\text{SNR} = 1, $T = 400$,\phantom{0} $K = 100$), the large network of the kind utilized in the empirical analysis still retrieves the true contribution time series with a $R^2$ of 66\%,  which implies a correlation with the true states of at least 80\%.  In the most favorable conditions,  the large network retrieves gaps with a $R^2$ of 90\%,  contributions at 0.88\%,  and coefficients at 87\%.  

The hardest feature to capture is clearly the exogenously evolving time-varying coefficient,  which variability deteriorates faster with a lowering SNR and smaller $T$.  In the most hostile environment (\text{SNR} = 1, $T = 400$,\phantom{0} $K = 100$),  the $R^2$ gets as low as 20\% while that of the gap is 60\%.  Despite this setback,  it is worth remembering the true $\beta_t$ path used for simulation is rather ambitious (on purpose) and features two full sinusoidal cycles between $t=1$ and $t=T$.  Thus,  combining this with a lower SNR than that of the paper and a large $K$ showcase the inevitable limits of the method.


In terms of relative performance, the large networks consistently outperform the medium and small networks when it comes to contributions (and thus the two parts of the conditional mean). The improvement from small to medium is greater than from medium to large, suggesting decreasing marginal returns to adding neurons. In the six environments studied here, the larger networks fare better at recovering the two parts of the conditional mean, with a margin of 8 to 15 percentage points.

Examining gaps and coefficients separately, we see that the larger network's significant advantage over the medium one comes from successfully capturing the nonlinear function of time that represents the coefficient. Both the medium and small networks struggle significantly to recover the coefficient paths. Additionally, the large network is more proficient at recovering gaps (by a margin of up to 16\% over the small one), which are themselves time-invariant nonlinear functions of the inputs.


While those results do not explicitly demonstrate the double descent phenomenon,  they exhibit one of its well-known implications: large networks tend to provide more reliable results.  It is certainly not excluded that there exist a smaller architecture delivering performance similar to that of the large network (one that lives in the classical regimes,  \citealt{belkin2019reconciling}) -- but it appears harder to find it in the simulations conducted here and in the empirical analysis of the paper.

{\noindent \sc \textbf{Reliability of Uncertainty Quantification.}} Here,  the central focus is the evaluation of the \textit{large} network which corresponds to what is utilized in the empirical work.  For gaps,  the quantity which has a fixed definition though time,  the 68\% nominal coverage results range from 61\% to 69\%.   The 95\% coverage results range from 89\% to 94\% with an average over the 6 environments of ~93\%.  For coefficients,  we get more conservative results with an average of 74\% for the 68\% target,  and 95\% for the 95\% target.  Naturally, uncertainty piles up for contributions, which multiplies gaps and coefficients.   The bands for contributions tend to be larger on average (84\% and 99\%) than what frequentist coverage would command for both targets.  Thus, if anything,  the evidence provided in simulations suggest that the 12\%-84\% quantile range reported in Figure \ref{hs_2019} is a conservative assessment of the overall precision of estimates.


 

For medium- and small-sized networks, the uncertainty quantification tool faces significant challenges. First, the nominal coverage is substantially (and consistently) too high for both gaps and coefficients. However, for contributions—which are the product of gaps and coefficients—the coverage is significantly too low. This discrepancy can be explained by the fact that large networks tend to integrate out initialization randomness, whereas smaller networks do not \citep{d2020double}. Consequently, much of the band inflation observed in small and medium networks, relative to their frequentist target, likely arises from excessive inclusion of initialization randomness. This randomness cancels out when considering the product of gaps and coefficients, suggesting difficulties in separating fluctuations related to gaps from those related to coefficients.  In particular, small and medium networks fail to adequately capture the coefficient path, as indicated by the \( R^2 \) values.  However, they show generally decent performance in terms of the $R^2$ of contributions, which blend the gaps and coefficients.  In unreported results for both simulated and empirical data, we observe that the baseline HNN optimization path for a given fixed sample remains extremely consistent across runs. In contrast, smaller networks exhibit substantial variation driven by initialization values. A potential solution for smaller networks would be to pre-ensemble a few networks to average out initialization noise for a single draw. However, given that the larger networks considered in this paper are significantly faster to train than smaller ones—they converge faster despite requiring the estimation of more parameters—this alternative would be substantially more costly to implement.

Hence, the uncertainty quantification method utilized in this paper is not a silver bullet applicable to all networks under all conditions.  Indeed,  simulation results suggest that for small- and medium-sized networks the uncertainty quantification method proposed in Section 2.4 would need some adjustments.  However, it works well for intuitive reasons under conditions that we understand and operate within --i.e.,  large networks.  Thus, it provides researchers with informative (and easily obtainable) uncertainty quantification for the type of architecture proposed in this paper: a large network that is stable in optimization and able to capture time-varying coefficients as well as their attached gaps.

\begin{table}[t!]
\footnotesize
\centering
\caption{\normalsize {In-sample and Out-of-Sample $R^2$'s with respect to true latent series} \vspace*{-0.3cm}} \label{tab:uq2}
\begin{threeparttable}
\setlength{\tabcolsep}{0.50em}
\setlength\extrarowheight{2.95pt}
\begin{tabular}{l | ccc | cccc} 
\toprule \toprule
& \multicolumn{3}{c}{ Train (Out-of-Bag) }   & & \multicolumn{3}{c}{ Test }  \\
\cmidrule(lr){2-4} \cmidrule(lr){6-8}
& Large & Medium & Small & & Large & Medium & Small \\
\midrule
\addlinespace[5pt]
\rowcolor{gray!17}
\multicolumn{8}{l}{\textbf{ Contributions }} \\
\addlinespace[2pt]
\addlinespace[2pt]
\text{SNR} = 2,\phantom{0} $T_{\text{train}} = 300$,  \phantom{0} $T_{\text{test}} = 100$,\phantom{0} $K = 20$&  0.80 &  0.76 &  0.56 &&  -- &  -- &  --  \\
\addlinespace[2pt]
\text{SNR} = 2,\phantom{0}  $T_{\text{train}} = 300$, \phantom{0}  $T_{\text{test}} = 100$,\phantom{0} $K = 100$&  0.78 &  0.75 &  0.65 &&  -- &  -- &  -- \\
\addlinespace[2pt]
\midrule
\addlinespace[5pt]
\rowcolor{gray!17}
\multicolumn{8}{l}{\textbf{ Gaps }} \\
\addlinespace[2pt]
\addlinespace[2pt]
\text{SNR} = 2, \phantom{0} $T_{\text{train}} = 300$, \phantom{0}  $T_{\text{test}} = 100$,\phantom{0} $K = 20$&  0.79 &  0.77 &  0.63 &&  0.80 &  0.78 &  0.68 \\
\addlinespace[2pt]
\text{SNR} = 2, \phantom{0} $T_{\text{train}} = 300$,  \phantom{0} $T_{\text{test}} = 100$,\phantom{0} $K = 100$&  0.74 &  0.75 &  0.69 &&  0.75 &  0.77 &  0.72 \\
\addlinespace[2pt]
\midrule
\addlinespace[5pt]
\rowcolor{gray!17}
\multicolumn{8}{l}{\textbf{ Coefficients }} \\
\addlinespace[2pt]
\text{SNR} = 2, \phantom{0} $T_{\text{train}} = 300$,  \phantom{0} $T_{\text{test}} = 100$,\phantom{0} $K = 20$&  0.52 &  0.69 &  0.35 &&   -- &  -- &  -- \\
\addlinespace[2pt]
\text{SNR} = 2, \phantom{0} $T_{\text{train}} = 300$, \phantom{0}  $T_{\text{test}} = 100$,\phantom{0} $K = 100$&  0.50 &  0.62 &  0.43 &&   -- &  --  & -- \\
\addlinespace[2pt]
\bottomrule \bottomrule
\end{tabular}
\begin{tablenotes}[para,flushleft]
\footnotesize 
\textit{Notes}: Coefficients and contributions results are removed on the test sample because they cannot be evaluated with a fixed train-test split. Results are averaged over 10 simulations and the pair of similar objects.  
\end{tablenotes}
\end{threeparttable}
\end{table}

{\noindent \sc \textbf{A Fixed Train-Test Split Check.}} I take the \( T=400 \) data points and split them into \( T_{\text{train}} = 300 \) and \( T_{\text{test}} = 100 \).  Focusing on \(\text{SNR}=2\), the DGP is simplified to a more realistic single sinusoidal full cycle for \(\beta_t\) between \( t=1 \) and \( T_{\text{train}}=300 \),  and the data is drawn from a single factor model per hemisphere.  Note that  coefficients cannot be evaluated in this manner unless we re-estimate after a certain number of observations, which would be computationally intensive in simulation environments.   Since contributions are the product of the coefficients and the gap, they are also excluded.  Results are reported in Table \ref{tab:uq2}.   Overall,  for test set performance,  there is a tie between large- and medium-sized neural networks.  It is important to remember that the medium-sized network already features many more parameters than observations. The small network consistently delivers results that are below the two larger networks. As expected, since OOB is a pseudo-out-of-sample measure, we find that absolute and relative performance numbers for gaps are very similar on the test sample.   On the out-of-bag sample, the large and medium networks deliver similar results for contributions and gaps, with the medium-sized network having the upper hand for coefficients.




{\noindent \sc \textbf{Summary.}}  We aimed to elucidate two main questions. First, are large networks better at recovering true nonlinear latent states than small ones? Second, for the baseline HNN architecture, a large dense network, is the uncertainty quantification approach adequate? The simulations conducted in this section provide affirmative answers to both questions. In terms of point estimates, the large network is marginally better than the medium one and significantly better than the small one. Regarding the uncertainty quantification technique, we find that the proposed method, which efficiently recycles the ensemble, delivers reasonable coverage results for HNN key outputs, but only when utilizing a large network architecture. Therefore, the use of large networks for the empirical analysis is justified.

\subsection{Mnemonics for Benchmark HNNs}\label{sec:mne}
\begin{lstlisting}[language=R]
#These are for HNN-F.  Add "trend" to the first three hemispheres to get HNN.

real.activity.hemisphere <- c("PAYEMS","USPRIV","MANEMP","SRVPRD",
	"USGOOD" ,"DMANEMP","NDMANEMP","USCONS","USEHS",
                  "USFIRE","USINFO","USPBS","USLAH","USSERV",
                  "USMINE","USTPU","USGOVT","USTRADE",
                  "USWTRADE","CES9091000001","CES9092000001",
                  "CES9093000001","CE16OV","CIVPART",
                  "UNRATE","UNRATESTx","UNRATELTx","LNS14000012",
                  "LNS14000025","LNS14000026",
                  "UEMPLT5","UEMP5TO14","UEMP15T26","UEMP27OV",
                  "LNS13023621","LNS13023557",
                  "LNS13023705","LNS13023569","LNS12032194",
                  "HOABS","HOAMS","HOANBS","AWHMAN",
                  "AWHNONAG","AWOTMAN","HWIx","UEMPMEAN",
                  "CES0600000007", "HWIURATIOx","CLAIMSx","GDPC1",
                  "PCECC96","GPDIC1","OUTNFB","OUTBS","OUTMS",
                  "INDPRO","IPFINAL","IPCONGD","IPMAT","IPDMAT",
                  "IPNMAT","IPDCONGD","IPB51110SQ","IPNCONGD",
                  "IPBUSEQ","IPB51220SQ","TCU","CUMFNS",
                  "IPMANSICS","IPB51222S","IPFUELS") 
                  
SR.expec.hemisphere <- c("Y", "PCECTPI","PCEPILFE",
		"GDPCTPI","GPDICTPI","IPDBS", "CPILFESL","CPIAPPSL",
                  "CPITRNSL","CPIMEDSL","CUSR0000SAC","CUSR0000SAD",
                  "WPSFD49207",    "PPIACO","WPSFD49502","WPSFD4111",
                  "PPIIDC","WPSID61","WPSID62","CUSR0000SAS","CPIULFSL",
                  "CUSR0000SA0L2","CUSR0000SA0L5", "CUSR0000SEHC",
                  "spf_cpih1","spf_cpi_currentYrs","inf_mich")
                  
commodities.hemisphere <- c("WPU0531","WPU0561","OILPRICEx","PPICMM")

LR.expec.hemisphere <- c("trend")

credit.hemisphere <- c("BUSLOANSx","CONSUMERx","NONREVSLx",
                   "REALLNx","REVOLSLx","TOTALSLx","DRIWCIL",  "DTCOLNVHFNM",
                 "DTCTHFNM","INVEST","nfci","nfci_credit","nfci_nonfin")

\end{lstlisting}

\end{document}